\documentclass[aps,
twocolumn,groupedaddress,nofootinbib,nobalancelastpage,nobibnotes]{revtex4}
\pdfoutput=1
\usepackage{amsmath,amsfonts,amssymb,mathrsfs,graphicx,color,epsfig}
\usepackage[squaren]{SIunits}
\usepackage{verbatim}
\usepackage{enumerate}
\usepackage[hyperfootnotes=false]{hyperref}
\usepackage{xcolor}
\usepackage{slashed}
\usepackage[utf8]{inputenc}
\usepackage{rotating}
\usepackage{isotope}

\newcommand{\ie}{{\it i.e.}}

\newcommand{\eg}{{\it e.g.}}

\newcommand{\cf}{{\it cf.}}

\newcommand{\eq}{Eq.}

\newcommand{\fig}{Fig.}
\newcommand{\Fig}{Fig.}

\newcommand{\Figs}{Figures}
\newcommand{\Ref}{Ref.}
\newcommand{\Refs}{Refs.}
\newcommand{\Sec}{Section}

\newcommand{\App}{Appendix}

\newcommand{\Tab}{Tab.}

\newcommand{\deltacp}{\delta_\mathrm{CP}}

\newcommand{\equ}[1]{\eq~(\ref{equ:#1})}
\newcommand{\figu}[1]{\fig~\ref{fig:#1}}

\newcommand{\bi}{\begin{itemize}}
\newcommand{\ei}{\end{itemize}}

\begin{document}


\title{Cosmic-Ray and Neutrino Emission from Gamma-Ray Bursts with a Nuclear Cascade}

\author{Daniel Biehl}
\affiliation{DESY, Platanenallee 6, 15738 Zeuthen, Germany}

\author{Denise Boncioli}
\affiliation{DESY, Platanenallee 6, 15738 Zeuthen, Germany}

\author{Anatoli Fedynitch}
\affiliation{DESY, Platanenallee 6, 15738 Zeuthen, Germany}

\author{Walter Winter}
\affiliation{DESY, Platanenallee 6, 15738 Zeuthen, Germany}

\date{\today}

\begin{abstract}
We discuss neutrino and cosmic-ray emission from Gamma-Ray Bursts (GRBs) with the injection of nuclei, where we take into account that a nuclear cascade from photo-disintegration can fully develop in the source. 
One of our main objectives is to test if recent results from the IceCube and the Pierre Auger Observatory can be accommodated with the paradigm that GRBs are the sources of Ultra-High Energy Cosmic Rays (UHECRs). While our key results are obtained using an  internal shock model, we discuss how the secondary emission from a GRB shell can be interpreted in terms of other astrophysical models.
It is demonstrated that the expected neutrino flux from GRBs weakly depends on the injection composition, which implies that prompt neutrinos from GRBs can efficiently test the GRB-UHECR paradigm even if the UHECRs are nuclei. 
We show that the UHECR spectrum and composition, as measured by the Pierre Auger Observatory, can be self-consistently reproduced in a combined source-propagation model. In an attempt to describe the energy range including the ankle, we find tension with the IceCube bounds from the GRB stacking analyses. In an alternative scenario, where only the UHECRs beyond the ankle originate from GRBs, the requirement for a joint description of cosmic-ray and neutrino observations favors lower luminosities, which does not correspond to the typical expectation from $\gamma$-ray observations.
\end{abstract}

\maketitle

\section{Introduction}
\label{sec:intro}

The origin of the ultra-high energy cosmic rays (UHECRs)  at the highest energies $\gtrsim 10^{18} \, \mathrm{eV}$, which are most likely of extra-galactic origin, is one of the unsolved mysteries in astroparticle physics. Recent results of the Auger Collaboration favor a mixed chemical composition of cosmic-ray particles injected from the sources into the extragalactic medium ~\cite{Aab:2016zth}. One possible candidate source class are Gamma-Ray Bursts (GRBs)~\cite{Piran:2004ba}, which are the most energetic electromagnetic outburst class, and which we consider in this study. 

GRBs are expected to produce a significant flux of high-energy neutrinos~\cite{Waxman:1997ti} if  cosmic ray protons interact with the target photons produced in the prompt gamma-ray emission expected to be created by internal shocks. GRBs, however, are the best-tested source class in stacking searches of the IceCube neutrino telescope using the gamma-ray counterpart, because these analyses make use of timing and direction to almost completely suppress the atmospheric background. Therefore, neutrino telescopes can efficiently test the GRB-UHECR paradigm~\cite{Abbasi:2012zw}. Although earlier prompt neutrino flux predictions~\cite{Guetta:2003wi} have been updated in the meanwhile~\cite{Hummer:2011ms,Li:2011ah,He:2012tq}, it is clear that current IceCube data~\cite{Aartsen:2017wea} exert pressure on the allowed parameter space for conventional neutrino emission models and the GRB-UHECR paradigm~\cite{Baerwald:2014zga}. Regions in parameter space for which the pion production efficiency is high and neutrons are efficiently produced together with the neutrinos, which can easily escape from the source~\cite{Ahlers:2011jj}, are already excluded. If, however, the cosmic ray escape is dominated by directly escaping protons (for which the Larmor radius can reach the size of the shell), fewer neutrinos are expected, and the charged cosmic rays are ejected with a hard spectrum $\propto E^{-1}$, whereas neutrons, which are not magnetically confined, escape with a softer spectrum~\cite{Baerwald:2013pu}; see also discussions in \Refs~\cite{Globus:2014fka,Unger:2015laa} from the source perspective. Note that such hard spectra are indeed found as escape spectra from the sources in current UHECR fits~\cite{Aab:2016zth}. 

There are a few important caveats in this picture. First of all,  other prompt emission models have been considered~\cite{Zhang:2012qy} and tested~\cite{Aartsen:2017wea}; for example, the magnetic reconnection model~\cite{Zhang:2010jt} is less constrained than the internal shock or photospheric scenarios. Second, all of the mentioned constraints have been obtained in one zone collision models, \ie, the parameters have been assumed to be the same for all collisions. Multi-zone models, however, tend to predict lower neutrino fluxes~\cite{Bustamante:2014oka,Globus:2014fka,Bustamante:2016wpu}. The reason is that the pion production efficiency strongly drops with collision radius, which means that few collisions at inner radii dominate the neutrino flux, whereas cosmic rays and gamma-rays (on average) come from larger collision radii~\cite{Bustamante:2014oka}. And third, all of the IceCube constraints have been derived for proton primaries. To test, if they also apply to a heavier composition in the jets, is one of the motivations of this work.

The behavior of UHECR nuclei in GRB jets has been studied in \Refs~\cite{Anchordoqui:2007tn,Wang:2007xj,Murase:2008mr,Globus:2014fka,Joshi:2015gfn,Boncioli:2016lkt}, where in many cases the motivation was to study the necessary conditions for UHECR survival. If, however, the radiation densities are high, a cascade of isotopes lighter than the injected primary emerges due to photo-disintegration in the source~\cite{Globus:2014fka,Boncioli:2016lkt}. In this work, we refer to this phenomenon as "the nuclear cascade". In \Ref~\cite{Boncioli:2016lkt} the nuclear cascade in a GRB shell has been self-consistently computed with a disintegration model at a level of sophistication comparable to CRPropa used for the cosmic ray propagation~\cite{Kampert:2012fi}, and it has been demonstrated that the feed-down to lower masses can have an impact on the ejected cosmic ray (and also the neutrino) flux. It can also be affected by the photo-disintegration model -- especially if a more sophisticated TALYS~\cite{Koning:2007} or FLUKA-based~\cite{Ferrari:2005zk,BOHLEN2014211} disintegration model is used compared to the Puget-Stecker-Bredekamp disintegration chain~\cite{Puget:1976nz}. The same approach will be used in this study, applying it for the first time to systematic parameter scans over the GRB parameters, taking into account the corresponding neutrino constraints.

In order to test the GRB-UHECR paradigm, a combined source-propagation model is needed: the accelerated nuclei are to be injected into the radiation zone, where the secondaries are produced, escape from that zone, and are then propagated through the extragalactic space to Earth; see \Refs~\cite{Baerwald:2014zga,Globus:2014fka,Unger:2015laa} for examples. One of the open questions from the UHECR perspective is, apart from the impact of the composition, where the transition energy between the lower-energy (possibly Galactic) and UHECR contribution occurs. Especially for protons, there are already constraints from cosmogenic neutrino data, for example ~\cite{Heinze:2015hhp}, and extragalactic diffuse gamma-ray data \cite{Supanitsky:2016gke,Berezinsky:2016jys} if the transition occurs below the ankle -- which is the so-called ``dip model'' \cite{Aloisio:2006wv}. In combined source-propagation models, a transition at the ankle has been considered in \Refs~\cite{Globus:2014fka,Globus:2015xga,Globus:2017ehu} for UHECR nuclei. A GRB proton dip model clearly overproduces the prompt neutrinos~\cite{Baerwald:2014zga}, while more generic models can effectively describe the transition to a lighter composition below the ankle by disintegrated nucleons~\cite{Unger:2015laa} -- which however does not consider the source-class dependent  pion production efficiency in the GRBs. 

We test both hypotheses in this study: GRBs as the sources of UHECR nuclei both covering the ankle (``Mixed Composition Dip Model'') and above the ankle (``Mixed Composition Ankle Model''). We inject a pure composition of nuclei into the jet in order to study the impact of the injection composition on the UHECR and neutrino fluxes. While it is clear that a mixed injection composition will improve the UHECR fit, using a pure composition allows us to develop a deeper understanding of the behavior of UHECR nuclei in GRBs, and the impact of the injection composition on the predicted prompt and cosmogenic neutrino fluxes. We focus on the one zone emission model, which is considered in the prompt neutrino analyses, extended to heavier nuclei, and we perform systematic parameter space studies. While most of our results are derived for the internal shock model, we demonstrate how our results can be translated into other emissions scenarios; therefore the key parameters used in this study are emission radius $R$, gamma-ray luminosity $L_\gamma$, baryonic loading $\xi_i$, and injection isotope.

\section{Methods}
\label{sec:methods}

Our simulations are based on the NeuCosmA code, following the implementation for GRBs in \Refs~\cite{Hummer:2011ms,Baerwald:2011ee,Baerwald:2013pu,Baerwald:2014zga} for protons. We focus here on the extensions for nuclei; this description complements the short summary given in \Ref~\cite{Boncioli:2016lkt} for a specific case. 

\subsection{Treatment of the Nuclear Cascade}

\begin{figure*}[t!]
\includegraphics[width=0.75\textwidth,trim=2.5cm 7.5cm 5.0cm 1.0cm,clip]{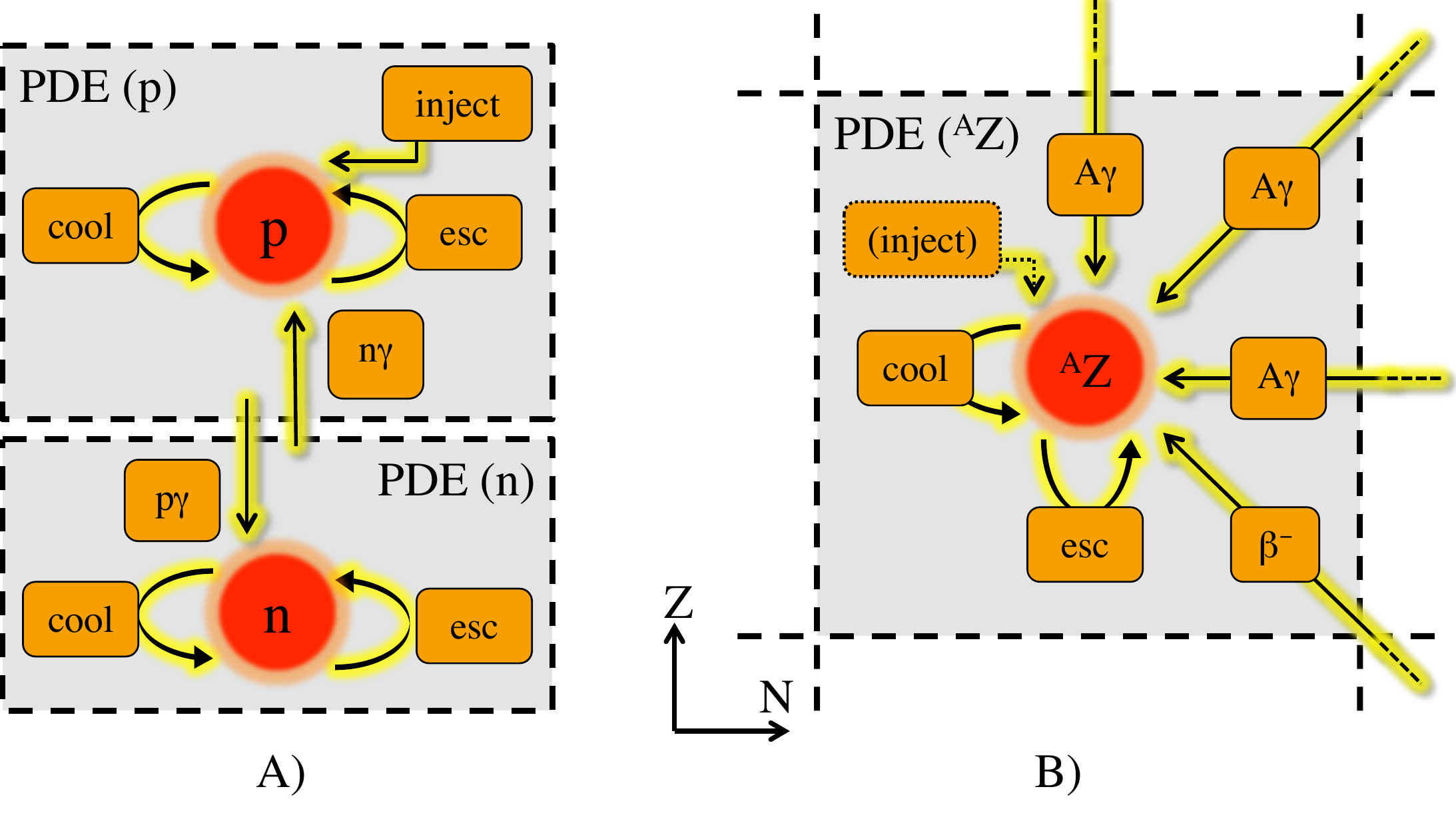}
\caption{Schematic illustration of the partial differential equations (PDEs) for A) an optically thick coupled proton-neutron system with proton injection (``inject'', from the acceleration zone), and B) an isotope in the nuclear cascade, possibly with injection (from the acceleration zone). Each dashed box corresponds to one species (or one PDE), to which the relevant cooling, escape and injection processes are ``attached''. Note that all disintegration, photo-meson, or decay channels destroying a species ($p$, $n$, or $^A$Z) show up as escape terms in the PDE of that species.}
\label{fig:pde}
\end{figure*}

Our method is fully deterministic, based upon solving a coupled system of partial differential equations (PDEs), which, for particle species $i$ (such as a nuclear isotope), reads:
\begin{equation}
\frac{\partial N'_i}{\partial t} = \frac{\partial}{\partial E'} \left( - b'(E') N'_i(E') \right) - \frac{N'_i(E')}{t'_{\text{esc}}} + \tilde Q'_{ji}(E) \, , \label{equ:master}
\end{equation}
where $b'(E) = E' {t'}^{-1}_{\text{loss}}$ (with the energy loss rate ${t'}^{-1}_{\text{loss}}$),  and ${t'}^{-1}_{\text{esc}}$ is the escape rate. The PDE system  is to be solved for the differential particle densities $N'_i \, [ \text{GeV}^{-1} \text{cm}^{-3} ]$ in the shock rest frame (SRF); note that primed quantities refer to the SRF. The coupled system arises because of the injection term 
\begin{equation}
 \tilde Q'_{ji}(E)= Q'_i(E) + Q'_{j \rightarrow i}(E) \, , \label{equ:inject}
\end{equation}
which allows for injection from an acceleration zone $Q'_i$ (typically a power law), as well as for injection from other species $j$ with the term  $Q'_{j \rightarrow i}$, such as from  photo-disintegration or $\beta^\pm$ decays. We will discuss the relevant interaction processes below. Fully-stripped ions are identified by $^A$Z, where $A$ is the mass number, $Z$ is the charge (or number of protons), and $N=A-Z$ is the number of neutrons.\footnote{We use the same symbol for the neutron number and differential number densities, as these cannot be mixed up and it is conventional notation.
}

A schematic illustration can be found in \figu{pde}, where each dashed (gray) box corresponds to one PDE. Note that we will deal with several hundred nuclear isotopes later, which means that automatic techniques for the PDE system setup will be used, and pictorial representations are useful to describe the models.  In the pictorial representation, the cooling, escape, and injection from the radiation zone $Q'_i$ act only on one species $i$, and are therefore ``attached'' to that -- meaning that the PDE system is set up with corresponding terms. The injection term $Q'_{j \rightarrow i}$ acts on species $i$ (and is attached to that PDE), but involves another species' density $N'_j$, illustrated by the arrows from a different dashed box corresponding to different species $j$; it couples the PDEs.

In example A) the computations in \Refs~\cite{Hummer:2011ms,Baerwald:2011ee,Baerwald:2013pu,Baerwald:2014zga}, which are effectively performed in the optically thin regime to photohadronic ($p\gamma$, $n\gamma$) interactions, are extended to optically thick sources. In this case, there are only two PDEs for the nuclear species (protons and neutrons).  Here, only protons are injected from an acceleration zone (the shock). Since the PDEs are coupled by the photohadronic interactions, protons will be converted into neutrons, and vice versa. In the optically thin regime, the interactions $p\gamma$ hardly take place, which means that the neutron species will not be populated. Typically cooling processes for protons are synchrotron cooling and adiabatic cooling, whereas the escape term is driven by the photohadronic interaction rate.\footnote{Since in fact a fraction of the interacting protons produces protons again, we treat the photohadronic energy losses discretely, \ie, we re-inject these protons at lower energy.} Neutrons, on the other hand, can typically escape from the region because they are not magnetically confined, or by photohadronic interactions.\footnote{The neutron decay lifetime is relatively long compared to the dynamical timescale relevant for GRBs.} Although this simple example is only meant for illustration, we note that the optically thick case to photohadronic interactions for proton injection is included in our work; see \App~\ref{app:optthick} for a more detailed discussion and a comparison to earlier works (accessible after a first pass through this section).

\begin{figure}[t!]
\includegraphics[width=\columnwidth]{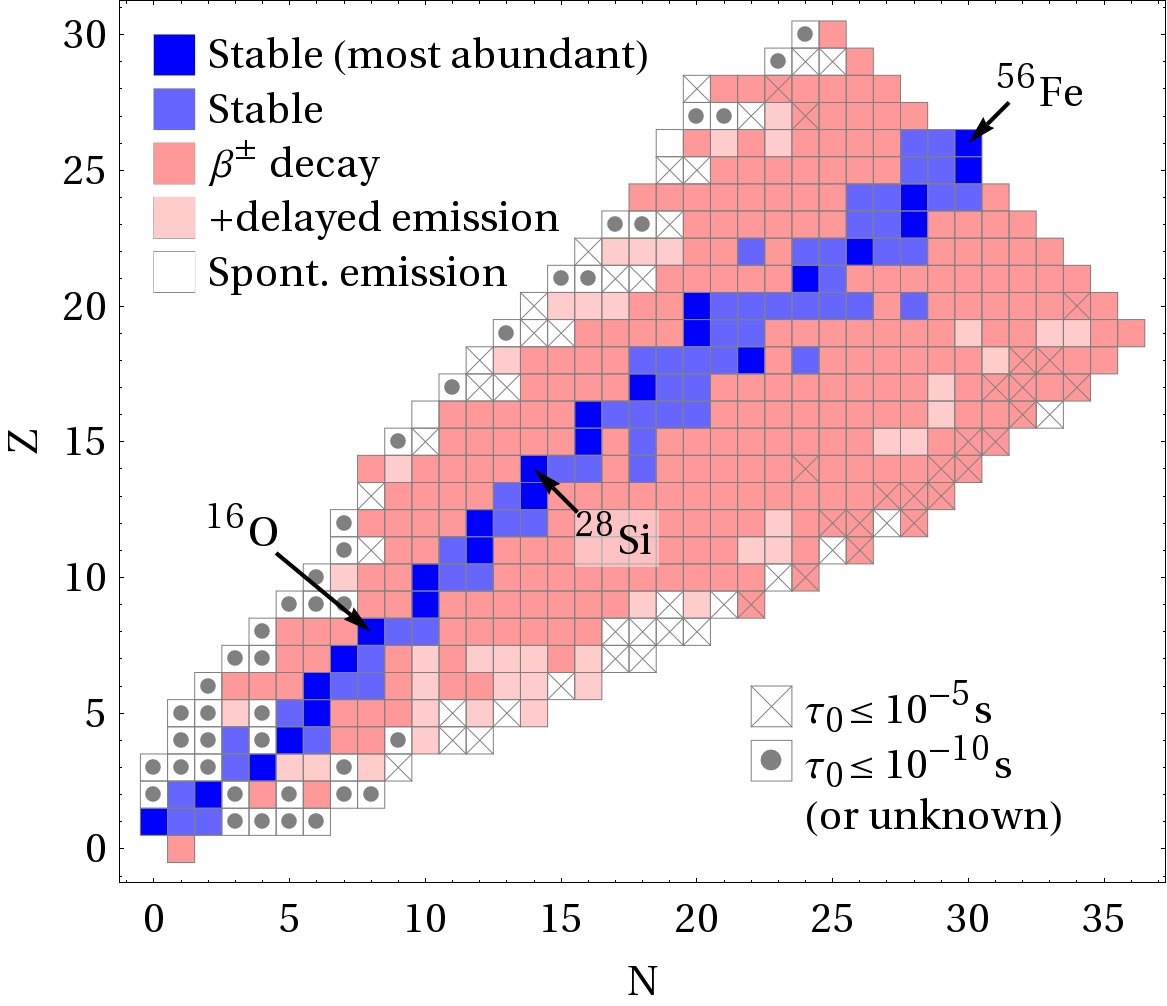}
\caption{Considered 481 isotopes in this work as a function of neutron number $N$ and proton number $Z$. Here it is assumed that the heaviest stable injection isotope is $^{56}$Fe. The color coding refers to the most abundant stable isotopes (for each element, \ie, equal $Z$), which are considered as injection isotopes (dark blue), stable isotopes (blue), $\beta^{\pm}$ emitters (dark red), possibly followed by the spontaneous emission of nucleons (light red), and spontaneous emitters of nucleons or $\alpha$ particles (white). Note that only the leading (largest branching ratio) processes are shown, and that isotopes with lifetimes longer than one month are regarded as stable in our scheme. Isotopes with rest frame lifetimes $\tau_0 \le 10^{-5} \, \mathrm{s}$ are marked by ``$\times$'' -- which are potentially interesting for neutrino production in GRBs if they are $\beta^{\pm}$ emitters. Isotopes with rest frame lifetimes $\tau_0 \le 10^{-10} \, \mathrm{s}$ (or unknown lifetimes) are marked by dots, which decay quickly enough even at the highest energies to be integrated out.}
\label{fig:isochart}
\end{figure}

Case B) in \figu{pde} refers to nuclei, where only one PDE for one species is illustrated. Similarly, the fully-stripped nuclear isotope will have cooling processes attached to it (synchrotron cooling, adiabatic cooling, and pair production cooling), as well as escape processes (photo-disintegration, photo-meson production, and decay, if applicable, which change the species). There may be injection from the acceleration zone, where we only inject a pure composition with one isotope at a time in this study for the sake of simplicity. The motivation for that is two-fold: first of all, it is clear that injecting an isotope cocktail will lead to better fit results. It is therefore interesting to see how far the simplest possible model can describe data. Second, we would like to develop an understanding of the impact of the injection composition, which is difficult if a cocktail of isotopes is injected from the beginning. As injection isotopes, we choose the most abundant stable isotope ($^A$Z) for each element ($Z$), whenever the results are shown as a function of $Z$ (dark blue isotopes in \figu{isochart}). The injection from different species can be photohadronic processes $A\gamma$ (photo-disintegration or photo-meson production, depending on the center-of-mass energy), beta decays, or spontaneous decays of proton- or neutron-rich isotopes, as illustrated in \figu{pde}. Note that escape processes, as we define them, do not only include escape from the region or escape over the dynamical timescale, but also conversion into a different species (such as by decay or interactions). 

\subsection{Hadronic Radiation Processes}

The injection rate $Q'_{j \rightarrow i}(E_i)$ in \equ{inject} of particles of species $i$ and energy $E_i$ from the interaction or decay of the parent particle $j$ can be approximated by
\begin{equation}
Q'_{j \rightarrow i}(E_i) = \int \frac{dx}{x} \, N'_j(E'_j) \, \Gamma'_j (E'_j) \, \frac{d n_{j \rightarrow i}}{dx} \left( x,\sqrt{s} \right) \, ,
\label{equ:prod}
\end{equation}
where $\Gamma'_j$ is the interaction rate, $x \equiv E'_i/E'_j$ is the fraction of primary energy taken by the secondary, and $\sqrt{s}$ is the center-of-mass energy (if applicable).\footnote{In fact, for photo-meson production one has to integrate over $dn_{j \rightarrow i}/dx$ within the photon energy integral, which is ``hidden'' in the interaction rate here; see \App~\ref{app:nuclear}.} The function 
\begin{equation}
 \frac{d n_{j \rightarrow i}}{dx} \left( x,\sqrt{s} \right) \simeq  M_{j \rightarrow i} (\sqrt{s}) \,  p_{j \rightarrow i} \left( x,\sqrt{s} \right) \, 
\label{equ:repro}
\end{equation}
 describes the distribution of secondaries of type $i$ per final state energy interval $dx$, the probability distribution is normalized $\int p_{j \rightarrow i}(x) dx = 1$, and the integral over that function gives the average number of secondaries $M_{j \rightarrow i}$ produced per interaction (multiplicity). The specific implementation of this injection function and the computation of the interaction rate depends on the nuclear process considered, for details see \App~\ref{app:nuclear}. In the main text, we focus on a qualitative description of the interactions. 

\subsubsection*{Beta Decays and Spontaneous Emission}

As a starting point, we choose all potentially relevant, known isotopes with $A<56$ and $^{56}$Fe, which is our heaviest (stable) injection isotope, taken from the database in \Ref~\cite{Mathematica160215}. These isotopes are shown as boxes in \figu{isochart} as a function of $N$ and $Z$, where the considered injection isotopes (most abundant stable elements) are shown in dark blue. Apart from the blue isotopes, which are stable (or have lifetimes longer than one month), all unstable isotope undergo various decay processes: $\beta$ decays (red), possibly followed by the delayed (later than about $10^{-14} \, \mathrm{s}$) spontaneous emission of one or two nucleons or an $\alpha$ particle (light red), or spontaneous emission of one or two nucleons or an $\alpha$ particle (white).\footnote{Note that our framework is already simplified by eliminating branchings $<5\%$. In the figure, we only show the leading branchings, as many isotopes have several decay channels -- which we take into account in the computations.}  

Typical lifetimes of $\beta^\pm$ emitters range from fractions of seconds to hours, and have to be Lorentz-boosted to be compared to the dynamical timescale $t'_{\mathrm{dyn}}$ of the prompt emission in the SRF (which will be order one second). Beta emitters with rest frame lifetimes $\tau_0 \lesssim 10^{-5} \, \mathrm{s}$ are potentially interesting for the secondary neutrino production in GRBs, which are the red boxes marked with ``$\times$'' in the figure.\footnote{These beta emitters decay faster than the dynamical timescale for $\gamma'_A \equiv E'_A/m_A  \lesssim 10^5$, which corresponds to neutrino energies $\lesssim 10^4 \, \mathrm{GeV}$ where the neutrino flux from beta decays may dominate.} One can however see from the figure that these interesting isotopes are relatively far off the main diagonal, which will be hardly populated. Neutrinos from $\beta^\pm$ decays therefore play a minor role from within the GRBs (but are included in the computation), while escaping neutrons and unstable isotopes may decay on the way to Earth. The neutrinos from neutron (and unstable nuclei) decays will only show up at very low energies for GRBs (see \eg\ \Refs~\cite{Baerwald:2010fk,Baerwald:2011ee}, where the component is explicitly shown). We include these neutrinos as part of our prompt neutrino flux.\footnote{For this component, we let the neutrons and unstable isotopes decay outside the source, since it can be shown that the neutrons and unstable isotopes will rather decay than interact at these low energies where the beta decay component dominates the neutrino flux. However, we do not show this component explicitly since its contribution will be over-estimated for high energies, where the nuclei may interact before they decay.}

The other interesting class of emitters are proton- or neutron-rich spontaneous emitters (white boxes in \figu{isochart}), which typically decay very quickly by the ejection of nucleons (without accompanying neutrinos). These spontaneous emitters can be integrated out (replaced by their daughters)  if the lifetime is short enough, \ie, they decay at the highest energies with rates faster than that of all the other radiation processes.  We estimate that isotopes with $\tau_0 \lesssim 10^{-10} \, \mathrm{s}$ can be integrated out (white boxes with dots), because then $\gamma'_A \, \tau_0 \ll t'_{\mathrm{dyn}}$ at the highest energies (where $\gamma'_A=E'_A/m_A \sim 10^{10}$).\footnote{Note that isotopes with unknown lifetimes are also assumed to decay faster than this threshold.}  Since isotopes far off the main diagonal will be hardly populated, this mainly effects the light unstable isotopes with $A \lesssim 6$.

\subsubsection*{Photo-disintegration and Photo-meson Production}

The disintegration of nuclei in interactions with photons can be qualitatively distinguished by the energy scale $\epsilon_r$ (photon energy in the nucleus' rest frame). For $8 \, \mathrm{MeV} \lesssim \epsilon_r \lesssim 150 \, \mathrm{MeV}$, the ``giant dipole resonance'' (GDR)~\cite{Goldhaber:1948zza} and other processes lead to an electromagnetic excitation of the primary nucleus with the emission of one or more nucleons (or light nuclei). We refer to this regime below the pion production threshold as ``photo-disintegration''. For $\epsilon_r \gtrsim 150 \, \mathrm{MeV}$, higher energy processes, such as baryonic resonances, dominate the disintegration, accompanied by meson production; we refer to this regime as ``photo-meson production'' regime.

Due to the power-law type of the projectile and photon spectra, the nuclear cascade will, to leading order, be determined by photo-disintegration above the threshold for GDR, what requires that target photons with the right energies are available as interaction partners. The photo-disintegration within the sources has been discussed in \Ref~\cite{Boncioli:2016lkt} from the perspective of the nuclear interaction model, where the impact of different model choices is demonstrated. For this work, our photo-disintegration model is based on cross-section information from TALYS 1.8 for nuclei with $A \geq 12$~\cite{Koning:2007} and for lighter nuclei on tabulated data, which is distributed with CRPropa2~\cite{Kampert:2012fi}; see \Ref~\cite{Boncioli:2016lkt} for details.

Our photo-meson interaction model is a superposition model based on the SOPHIA Monte Carlo generator~\cite{Mucke:1999yb}. Superposition means that the nuclei are treated as a bulk of independent nucleons with $E'_{p/n}=E'_A/A$. In each interaction, only one nucleon interacts and is ejected from the nucleus. The energy distributions of secondary pions and nucleons are computed with SOPHIA, whereas for the residual nucleus the energy per nucleon is conserved $E'_{A'}=(A-1)/A \, E'_A$. The probability that a neutron within the nucleus interacts is $N/A$ and $Z/A$ if a proton interacts, respectively. The inelastic cross section is assumed to scale like $\sigma_{A\gamma} = A \sigma_{p\gamma}$. Although this model is similar to what is being used as the state-of-the-art in the literature, see \eg\ \Refs~\cite{Anchordoqui:2007tn,Murase:2008mr,Kampert:2012fi}, it has clear deficits. First of all, the scaling of the  cross section may not be correct; for instance, Schlickeiser~\cite{schlickeiser2002} directly proposes the ``Glauber rule'' $A^{2/3}$. It is, however, well established that at energies above 150 MeV and several GeV in $\epsilon_r$, the cross section scales proportional to $A$ for nuclei heavier than $^4$He~\cite{MacCormick:1997ek}. At higher energies the photon interacts more hadron-like, where the $A^{2/3}$ behavior is more appropriate. We motivate our current assumption by the choice of the astrophysical source class, where lower interaction energies are more relevant.
Second, the probability for the ejection of a proton or a neutron is close to 50-50, what drives the residual nucleus further away from the main diagonal than one would expect from a spectator model~\cite{Rachen:1996ph}. 

In the main text of this work, we assume that the minimal photon energy is low enough for the high-energy cosmic ray nuclei to find interaction partners above the GDR threshold. It means that the photo-meson production will be of secondary importance. In the case, where the photon spectrum is cut off at low energy, what can occur due to synchrotron self-absorption~\cite{Wang:2007xj,Murase:2008mr}, the disintegration at the highest energies will be dominated by the photo-meson regime. We show an example demonstrating the quantitative impact in \App~\ref{app:ephmin} (which can be followed after reading \Sec~\ref{sec:classes}), where it is demonstrated that the neutrino flux is hardly affected, whereas the UHECRs can actually extend to higher energies in that case.  We will demonstrate in this work that the neutrino production in GRBs, depending on luminosity and collision radius, is typically dominated either by the photo-meson production off the injection nucleus, by the photo-meson production off the secondary nuclei  produced in the nuclear cascade, or by protons and neutrons produced in the nuclear cascade. The impact of the photo-meson model for this two cases is quantitatively different: while this photo-meson model produces reliable results for proton and neutron projectiles, we have to interpret the cases where the primary nucleus dominates with care due to the simplifications listed above. A more dedicated study of photo-meson production off nuclei will therefore be needed in the future.

The implementation of the photo-meson model follows a logic similar to \Ref~\cite{Hummer:2010vx}, where the idea was  to discretize one of the integrals in the double integration, needed to compute the secondary spectra, into so-called interaction types to gain linear (compared to quadratic) computational complexity. Our novel approach, described in \App~\ref{app:nuclear}, uses a new discretization scheme which extends the previous scheme to cope with interactions of various isotopes in the future. We apply this scheme to SOPHIA in the Appendix, and demonstrate that it can outperform \Ref~\cite{Hummer:2010vx} in terms of efficiency and precision.

\subsubsection*{Other Radiation Processes}

In addition to decays, photo-disintegration, and photo-meson production, a number of radiation processes are implemented as continuous energy loss (cooling) processes. 

At $\epsilon_r \lesssim 8 \, \mathrm{MeV}$, the $A \gamma$ interactions are determined by the quantum electrodynamics scale. As a consequence, charged isotopes will produce via the Bethe-Heitler process e$^+$e$^-$ pairs, which we include following \Ref~\cite{Blumenthal:1970nn}. However, for GRBs, this contribution is typically sub-dominant -- we therefore do not show it explicitly in our timescale plots.

All charged species are assumed to cool via synchrotron radiation and adiabatic cooling (roughly determined by $t'_{\mathrm{ad}} \simeq t'_{\mathrm{dyn}}$). Cosmic ray nuclei are assumed to be magnetically confined, but can directly escape within their Larmor radius from the shell boundaries (see discussion of cosmic ray escape below). This escape contribution will always have at least a sub-dominant impact on the source density. We add however an escape term for the neutrons with the free-streaming timescale.\footnote{This term is consistent with our assumptions for neutron escape and, in fact, necessary to reach the steady state within a few times the dynamical timescale: Since neutrons can only decay at the lowest energies, the steady state can be reached there at around the neutron rest frame lifetime in the absence of this term.}

\subsection{Automated Isotope Selection Scheme}

One of the key features of our new code is the automatic isotope selection scheme: it picks the isotopes from the 481 isotopes in \figu{isochart} which are relevant for the problem under investigation, depending on the requested precision. We first of all load the isotope data corresponding to \figu{isochart}, and flag all isotopes with lifetimes $\tau_0 \le 10^{-10} \, \mathrm{s}$ to be integrated out. As a second step, we load the photo-disintegration and photo-meson production models. If daughter nuclei are encountered which are to be integrated out, all paths are recursively followed to the next ``active'' isotopes to be treated. In this case, we replace the interaction by an effective one, directly producing the next active isotopes.\footnote{Consider, for example, disintegration of a nucleus $a$ with consecutive fast decays $a \rightarrow i_1 \rightarrow \hdots \rightarrow i_N \rightarrow b$, where the isotopes $i_1 \hdots i_N$ are to be integrated out because they interact quickly. One can show that the secondary injection of $b$ can be written in the form of \equ{prodsimple2} where effective $\chi_{a \rightarrow b} \equiv  \chi_{a \rightarrow i_1} \times \chi_{i_1 \rightarrow i_2} \times \hdots \chi_{i_N \rightarrow b}=A_{i_1}/A_a \times A_{i_2}/A_{i_1} \times \hdots \times A_b/A_{i_N}=A_b/A_a$ and $g_{a \rightarrow b} \equiv g_{a \rightarrow i_1} \times M_{i_1 \rightarrow i_2} \times \hdots M_{i_N \rightarrow b}$ are to be used. Thus, when one constructs the isotope chart, one can replace the daughter of that process recursively by all possible end states of fast decay chains with effective values of $\chi$ and $g$.} The implemented techniques are especially powerful and sophisticated for photo-meson production, where the interaction types ($\tilde x$-values, see \App~\ref{app:nuclear}) have to be re-defined automatically.  \figu{isochart} illustrates that the lifetimes of the $\beta^\pm$ emitters are longer than the chosen lifetime threshold for integrating out isotopes for GRBs (\ie, there are no dot-marked red boxes), which means that neutrino fluxes from integrated out $\beta^\pm$ emitters do not need to be taken into account. 

As a next step, we identify the relevant channels and set up the PDE system \equ{master} automatically. Starting from an injection isotope (such as $^{56}$Fe), we recursively follow all photo-disintegration and decay channels (including mixed paths) for different (close enough) center-of-mass energies, and flag the encountered isotopes as relevant for the computation.
We impose a threshold on the secondary multiplicity for these recursive paths, as daughters with very low multiplicities will be hardly filled.\footnote{Our software supports several different selection methods and activation schemes, but the results depend  on the chosen threshold values rather than the scheme.}  
Note that for a ``two-dimensional'' disintegration model (in $N$ and $Z$), such as TALYS, one should not pick a too large threshold, as many isotopes with low multiplicities will be filled off the main diagonal, which means that the nuclear cascade would otherwise cease.

\begin{figure}[t!]
\includegraphics[width=\columnwidth]{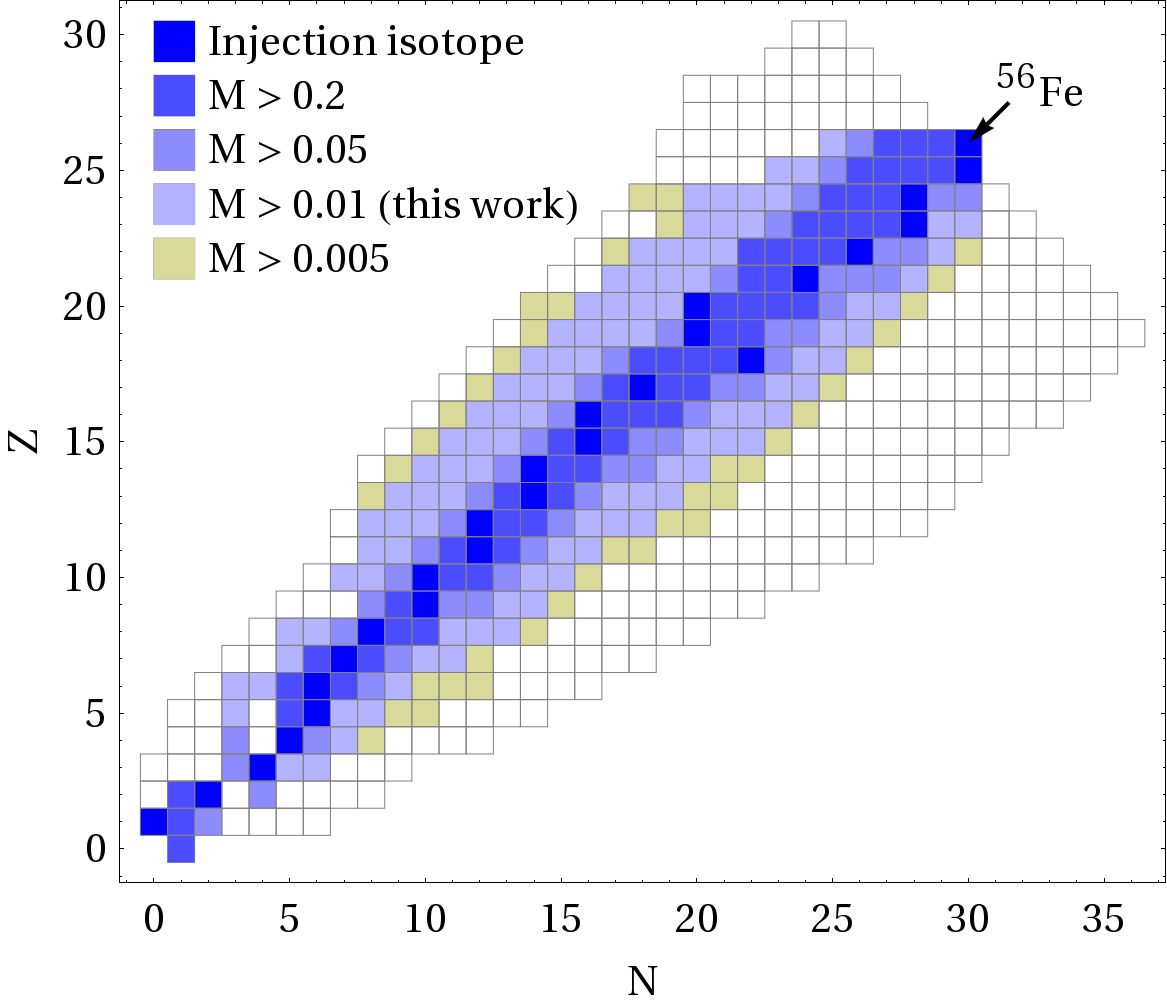}
\caption{Automatic isotope selection result as input for the PDE system setup for the injection of $^{56}$Fe, tracing recursively all possible disintegration, decay, and mixed paths. The isotopes are selected if their secondary multiplicity $M$ (identified with the pitch-angle averaged number of secondary nuclei $M \leftrightarrow g_{j \rightarrow i}(y)/f_j(y)$, see \App~\ref{app:nuclear}) exceeds a certain threshold, as indicated by the legend,  for any $8 \, \mathrm{MeV} \lesssim y \lesssim 125 \, \mathrm{MeV}$. The threshold $M>0.01$ has been used for the computations in this work. }
\label{fig:isoselect}
\end{figure}

\figu{isoselect} illustrates the selected isotopes for the injection of $^{56}$Fe for different threshold multiplicities. The thresholds $M>0.2$ and $M>0.05$ correspond to the most important isotope selections ``Astro, priority 1'' and ``Astro, priority 2'', respectively, in \Ref~\cite{Boncioli:2016lkt} (which were however computed for all possible injection elements there). In this work, we choose conservatively $M>0.01$, which leads to 233 selected isotopes with about 5000 attached photo-disintegration channels (from originally 41000 inclusive channels extracted from TALYS), 10 decay channels, and about 1300 photo-meson channels for the GRBs. We have checked that the results hardly change for smaller values of $M$.\footnote{In fact, for the code optimization, one would first of all choose a small threshold and then increase it as long as the results are not affected.} Of course, for different injection isotopes, \figu{isoselect} will look different -- it is therefore our procedure which defines the isotope chart, which can also be applied to different source classes.

\subsection{Energetics of the source}

 The photon spectrum in the source is assumed to follow observations. The spectrum for long-duration GRBs can be typically  described by a broken power law 
 \begin{equation}
 	N'_{\gamma}(\varepsilon') = C'_{\gamma} \cdot \left\{ \begin{array}{ll} \left( \frac{\varepsilon'}{\varepsilon'_{\gamma,\mathrm{br}}} \right)^{-1} & \varepsilon'_{\gamma,\text{min}} \leq \varepsilon' < \varepsilon'_{\gamma,\mathrm{br}} \\ \left( \frac{\varepsilon'}{\varepsilon'_{\gamma,\mathrm{br}}} \right)^{-2} & \varepsilon'_{\gamma,\mathrm{br}} \leq \varepsilon' < \varepsilon'_{\gamma,\text{max}} \\ 0 & \text{else} \end{array} \right. \quad , \label{equ:targetphoton}
 \end{equation}
 where $C'_{\gamma}$ is a normalization factor and $\varepsilon'_{\gamma,\mathrm{br}} \simeq 1 \, \mathrm{keV}$ (fixed in this work). Our minimal and maximal photon energies are chosen as small and large enough, respectively, such that there are no signifciant effects on the neutrino flux from the photon spectrum anymore. This especially implies that nuclei will always find interaction partners for disintegration at the GDR; see \App~\ref{app:ephmin} for the effect of the minimal photon energy cutoff.

In order to compute the photon density in the SRF, we define an ``isotropic volume'' of the interaction region
\begin{equation}
	V'_{\mathrm{iso}} = 4\pi \, R^2 \cdot \Delta d' 
	\label{equ:visoISM}
\end{equation}
with shell width $\Delta d'$ and the radius (distance from emitter) of the emission region $R$. 
Because of the intermittent nature of GRBs, the total fluence is typically assumed to be coming from $\Delta T/t_v$ such interaction regions, where $\Delta T$ is the duration of the prompt emission. We model the emission from one such region first, as this will allow for simpler interpretations in terms of multi-zone models such as \Refs~\cite{Bustamante:2014oka,Bustamante:2016wpu}.

The normalization of the photon spectrum in \equ{targetphoton} is obtained from
\begin{equation}
 u'_\gamma \equiv \int  \, \varepsilon' \, N'_{\gamma}(\varepsilon') \mathrm{d}\varepsilon' = \frac{L_{\gamma} \, \Delta d'/c}{\Gamma^2 \, V'_{\text{iso}}} = \frac{L_{\gamma}}{4 \pi c \Gamma^2 R^2} \, , \label{equ:photonorm}
\end{equation}
where $L_{\gamma}$ is the isotropic equivalent luminosity in gamma-rays and $\Gamma$ is the Lorentz boost factor. The integration limits are taken from the Fermi-GBM range from 8 keV to 40 MeV in the observer's  frame. Note that the factor $\Gamma^2$ comes from boosting the luminosity from the source (engine) frame to the SRF, and does not come from geometry estimators, \ie,  one can read from \equ{photonorm} how the target photon density scales with $\Gamma$ and $R$ without any geometry relationship for the radius. It scales with $R^{-2}$ -- which comes directly from the volume of the production region in \equ{visoISM} -- and therefore strongly with the radius. It is easy to show that the same dependence on $R$ enters the pion production efficiency (fraction of proton energy dumped into pion production in the optically thin case) $f_{p \gamma} \propto L_\gamma t_v/(R^2 \, \epsilon_{\gamma,\mathrm{br}} \, (1+z)^2)$ often used for analytical computations (assuming that $t_v$ is indicative for the shell width, see \equ{d}).\footnote{Note that this result is different from \Refs~\cite{He:2012tq} (\eq~(9)) and~\cite{Zhang:2012qy} (\eq~(6)), where the pion production efficiency scales $\propto 1/R$. Our formula is identical to \eq~(9) in \Ref~\cite{He:2012tq} if the model-dependent relationship \equ{rc} is applied to one power of $R$. See next subsection for the implications of different astrophysical models.}

If one defines a ``magnetic loading''  $\xi_B \simeq 1$ as the ratio between energy in magnetic field and gamma-rays, one can easily derive the magnetic field from the magnetic field energy density being equal to $\xi_B \cdot u'_\gamma$ in \equ{photonorm}.

\subsection{Implications of Different Astrophysical Models}
\label{sec:astromodel}
 
 For a review on the possible model assumptions on neutrino production models, see \eg\ \Refs~\cite{He:2012tq,Zhang:2012qy,Bustamante:2016wpu}. Here are some examples:
\begin{description}

 \item[Internal shock model (one-zone).] In this scenario, the shells of plasma are assumed to collide at a radius
\begin{equation}
R \simeq  2 \, \Gamma^2 \, \frac{c \, t_v}{1+\textit{z}} \, , \label{equ:rc}
\end{equation}
and the variability timescale is assumed to be indicative for the shell width 
\begin{equation}
 \Delta d' \simeq \Gamma \frac{c t_v}{1+z} \, .
\label{equ:d}
\end{equation}
All collisions are assumed to collide at same radius in the one-zone model. As a consequence, \equ{photonorm} strongly depends on $\Gamma$ and $t_v$, and the pion production efficiency becomes $f_{p \gamma} \propto L_\gamma/(\Gamma^4 t_v \epsilon_{\gamma,\mathrm{br}})$ as in \Ref~\cite{Guetta:2003wi}.

 \item[Internal shock model (multi-zone).] In the multi-zone collision models (\eg, \Refs~\cite{Kobayashi:1997jk,Daigne:1998xc}), shells of plasma are ejected from the central emitter, colliding at varying collision radii centered around a mean value. This case is similar to the one zone case, but the parameters of each collision $m$ can be very different: $\Gamma_m$ (Lorentz factor of merged shell), $R_{m}$ collision radius, $\Delta d'_m$ (width of merged shell, related to intermittent timescale of the central emitter), and $L_{\gamma,m}^{\mathrm{sh}}$ (radiated energy in gamma-rays). The observed variability timescale (from the lightcurve) roughly matches the intermittent timescale of the central emitter, and therefore the shell width -- as for the one-zone model -- but is a result of the model, rather than an input. Therefore, the key parameter determining the size of the interaction region is $R$, which covers a wider range than in the one-zone model, whereas the dependence on the other parameters is milder.

 \item[Photospheric models.] Below the photosphere, Thomson scattering hinders gamma-rays to escape. In photospheric prompt emission models (\eg, \Refs~\cite{Rees:2004gt,Giannios:2007yj}), a fraction of energy in gamma-rays is therefore released once the radiation zone becomes optically thin to Thomson scattering, where the photon spectrum is thermal and may be different from our assumed spectrum. The production region is in this case described by the photospheric radius $R_{\mathrm{ph}}$, which is typically smaller than \equ{rc} and the densities are therefore higher. Since the shell width hardly changes in the coasting phase of the GRB, \equ{d} should still be a good description. 
 \item[Magnetic reconnection models.] Magnetic reconnection models (\eg, \Refs~\cite{Lyutikov:2003bz,Zhang:2010jt}) often describe pulses and the fast time variability in the light curves at the same time. There are therefore two variability timescales, a fast one $t_v^f$ (indicative for the shell width in \equ{d}) and a slow one  $t_v^s$  (indicative for the collision radius in \equ{rc}). If the slow scale is about a factor 10-100 larger than the fast scale, $R$ will be about this factor larger and the particle densities correspondingly smaller. The pion production efficiency can be written as $f_{p \gamma} \propto L_\gamma t^f_v/(\Gamma^4 (t^s_v)^2 \epsilon_{\gamma,\mathrm{br}})$, which means that it is suppressed by the factor $(t_v^s/t_v^f)^2$ compared to the standard internal shock model.
\end{description}
Our strategy is to formulate the physics of the nuclear cascade as independent of the model as possible. Our results will therefore be shown as a function of $L_\gamma$ and $R$ for one collision only, such that one can read off the physical implications as a function of the most relevant parameters. Unless stated otherwise, we use $z=2$, $\Gamma=300$ and $t_v=0.01 \, \mathrm{s}$ for the other parameters with a milder dependence. In the main text, most results will apply to the internal shock scenarios, \ie, \equ{rc} holds. We show however in \Sec~\ref{sec:model} how our results can be (at least qualitatively) translated to other astrophysical models.

\subsection{Injection of Nuclei and \\ Maximal Primary Energy}
\label{sec:injection}

Nuclei of species $i$ are assumed to be injected with a cut-off power law with a spectral index $k \simeq 2$ expected from Fermi shock acceleration from an acceleration zone: 
 \begin{equation}
 	Q'_i(E_i') = C'_i \cdot \left\{ \begin{array}{ll} \left( \frac{E_i'}{\mathrm{GeV}} \right)^{-k} \cdot e^{ - \left( \frac{E_i'}{E'_{i,\mathrm{max}}} \right)^P } &  E_i' \; \ge \;  E'_{i,\text{min}}  \\ 0 & \text{else} \end{array} \right. . \label{equ:targetA}
 \end{equation}
We use $P=2$ for the cutoff function, unless noted otherwise. Note that there is basically no impact of the cutoff function on the neutrino spectra, whereas a fit to cosmic ray data will be sensitive to it.
The maximal energy $E'_{i,\text{max}}$ is determined automatically by balancing the acceleration rate $t'^{-1}_{\mathrm{acc}}=\eta \, c/R'_L$ (where $R'_L$ is the Larmor radius) with the sum of synchrotron loss, adiabatic cooling, photo-disintegration, and photo-meson production rates, where we choose $\eta=1$ for the acceleration efficiency.

In order to define a ``nuclear loading'' $\xi_i$ of species $i$, we take into account the optically thick case to $A \gamma$ interactions (photo-disintegration, photo-meson processes) and normalize the injection luminosity to the $\gamma$-ray luminosity. If the photons can escape freely, which is (at least at around the break energy, where $\tau_{\gamma \gamma} \ll 1$) typically a valid assumption beyond the photosphere (see \eg\ \Ref~\cite{Bustamante:2016wpu}), one finds
\begin{equation}
\int_0^{10 \, E'_{i,\mathrm{max}}} \, E'_i \, Q'_i(E'_i) \, \mathrm{d}E'_i =  \xi_i \cdot u'_{\gamma}  \cdot \frac{c}{\Delta d'}
\label{equ:Anorm}
\end{equation}
to determine $C'_i$ in \equ{targetA}, with the nuclear loading $\xi_i$, and $u'_{\gamma}$ from \equ{photonorm}. As indicated earlier, we only inject one isotope at a time in this work (pure composition), where we use $\xi_i = 10$ as frequently assumed for protons, unless noted otherwise (such as in the cosmic ray fits), and inject the most abundant stable isotope for each element ($Z$). With this assumption, we can inject the same luminosity for different compositions, and check how the neutrino flux depends on composition. 

\subsection{Neutrino Production and Cosmic Ray Escape}

In addition to the nuclei, we add $\pi^+$, $\pi^-$, and $K^+$ mesons to the system. These are injected from photo-meson production off protons, neutrons, and all nuclear isotopes. Adiabatic cooling, synchrotron cooling and escape (through decays) are taken explicitly into account. We also include four muon species for left- and right-handed $\mu^+$ and $\mu^-$, since the helicity-dependence of the muon decays is taken into account~\cite{Lipari:2007su}. For the neutrinos, we have four species at the source ($\nu_e$, $\bar\nu_e$, $\nu_\mu$, $\bar\nu_\mu$) which receive injection from the pions and muons according to the usual decay chains, from kaon ($\nu_\mu$, leading mode only), and from the $\beta^\pm$ decays of neutrons and unstable isotopes ($\bar\nu_e$, $\nu_e$) from inside and outside the source. Flavor mixing between source and detection is taken into account with the mixing angles $\theta_{12}=33.48^\circ$, $\theta_{23}=42.3^\circ$, and $\theta_{13}=8.50^\circ$, and the CP phase $\deltacp=306^\circ$~\cite{Gonzalez-Garcia2014}.

The PDE system is, after its setup and the pre-computation of interaction rates, which do not change as a function of time in this study, evolved until (after a few times $t_{\mathrm{dyn}}$) the steady state is reached. For the integration, we re-parameterize the PDE system in terms of $E'^2 N'$ and use a Crank-Nicolson solver.

As far as the escape of cosmic rays is concerned, we follow \Ref~\cite{Baerwald:2013pu}, where the ``direct'' escape from a GRB shell has been discussed in greater detail.\footnote{These assumptions for particle escape are similar to \Ref~\cite{Globus:2014fka}, where this escape mechanism is called ``high-pass filter''.} It is conservatively assumed that charged particles can only escape from within the Larmor radius of the edges of the shells. 
  This means that a fraction $f_{\mathrm{esc}}=\min(R'_L(E'),\Delta d')/\Delta d' \le 1$ escapes over the dynamical timescale. Note that $f_{\mathrm{esc}}=1$ if the Larmor radius reaches (or exceeds) the shell width, which corresponds to the free-streaming limit. It can be shown that $f_{\mathrm{esc}}$ is invariant even if the shell expands~\cite{Baerwald:2013pu}. The escape fraction can be translated into an effective escape rate $t'^{-1}_{\mathrm{esc,dir}} = \min(R'_L(E'),\Delta d')/c \,  t'^{-2}_{\mathrm{dyn}}$ (for $ct'_{\mathrm{dyn}} \simeq \Delta d'$). It is interesting to compare that to the escape from the whole shell: in this case $t'^{-1}_{\mathrm{esc,shell}} = D' \,  t'^{-2}_{\mathrm{dyn}}$, where $D'$ is the (spatial) diffusion coefficient. This assumption yields the same result if $D' \propto R'_L$ (Bohm diffusion). For neutrons, the escape rate is given by the free-streaming rate; see \App~\ref{app:optthick} for a comparison to the model in \Ref~\cite{Baerwald:2013pu}.

As a consequence, protons and nuclei can escape with a rate $\propto R'_L \propto E'$, which means that they can escape freely when/if the Larmor radius reaches the shell width. It can be shown that this condition is satisified at the highest energy for $\eta=1$ if the maximal primary energy is limited by the dynamical timescale. 
This situation can be typically found in sources where the radiation densities and the maximal primary energy are low. If, on the other hand, the radiation densities are high such that the maximal energy is limited by the photohadronic interactions, the escape at the highest energy will be suppressed. 

\subsection{Propagation of cosmic rays}

Once the accelerated particles are able to escape from the source, they propagate through the extragalactic space, encountering photon fields, such as the Cosmic Microwave Background (CMB) and the Extragalactic Background Light (EBL, from infrared to ultraviolet), with which they interact. Similarly to what happens in the source,  the energy scale of the process is given by the photon energy in the nucleus rest frame. The dominant processes for nuclei are photo-disintegration on CMB at the highest energies and on EBL at intermediate energies, while the photo-meson production is shifted towards $A$ times the threshold for protons. At the lowest energies, nuclei lose energy adiabatically because of the expansion of the Universe. For protons, the intermediate energy range is dominated by energy losses by electron-positron pair production on the CMB. Interactions of cosmic ray with the EBL is more relevant for nuclei than for protons. However, it provides the majority of low-energy cosmogenic neutrinos and it is, therefore, included in our calculations.

The propagation of cosmic rays ejected from the GRBs is computed using the simulation code {\it SimProp} \cite{Aloisio:2012wj}; the isotopes produced during interactions in the source depend on the adopted photo-disintegration model. In the present paper we use the {\it SimProp} code with the PSB model for photo-disintegration processes \cite{Puget:1976nz,Stecker:1998ib}, that contains one representative isotope for each nuclear mass. The nuclei ejected from the source are then grouped and the corresponding fluxes are summed, assigning the sum to a mass chosen as representative for each group. The ejected particles are propagated through the extragalactic space: for this study we use the Gilmore Extragalactic Background Light (EBL) \cite{Gilmore:2011ks}, that is one of the models available in {\it SimProp}. We assume a homogeneous distribution of identical sources up to $z=6$. The events are simulated with a uniform distribution of $\log_{10}(E_{\mathrm{inj}}/\mathrm{eV})$ and a uniform distribution of sources in the cosmologically co-moving frame. Each event is then weighted with a source evolution equal to the star formation rate, as given for example in \cite{Yuksel:2008cu}. An order four tensor is computed, containing the number of nuclei arriving at Earth with a given mass number and energy, that were ejected with a certain mass number and energy from the sources. The matrix is multiplied with vectors containing the spectrum of the representative mass ejected from the source, in order to obtain the spectrum at Earth. Neutrinos ejected from the sources and those produced during the propagation are treated similarly. The former ones are computed with NeuCosmA following the procedures described in previous chapters. A bi-dimensional propagation matrix is used to apply weights to these fluxes according to the choice of the source evolution. The latter ones are computed with {\it SimProp} and their flux at Earth is computed using equal weights as for the spectra of cosmic rays ejected from the sources.
The all-particle cosmic-ray spectrum at Earth is normalized to the energy spectrum measured by the Pierre Auger Observatory \cite{Valino:2015}. The baryonic loading is computed as in \cite{Baerwald:2014zga}.

\section{Nuclear Cascade Source Classes}
\label{sec:classes}

In order to describe the nuclear cascade in the sources and the characteristics of the ejected nuclear isotopes from a GRB shell (interaction region),  we introduce three qualitatively different cases in this section: ``Empty Cascade'', meaning that the nuclear cascade hardly develops, ``Populated Cascade'', meaning that the nuclear cascade develops, and ``Optically Thick Case'', meaning that the source is optically thick to $A\gamma$ interactions for nucleons and all nuclei. While there are continuous transitions among these cases, we discuss three prototypes which exhibit the behavior characteristic for its class. In order to keep the discussion as simple as possible, we inject a pure composition of $^{56}$Fe only, which is the heaviest commonly discussed injection isotope. We will show continuous parameter space scans in the following sections.

\subsection{Empty Cascade}
\label{sec:emptycascade}

\begin{figure*}[ht!]
\includegraphics[width=0.49\textwidth]{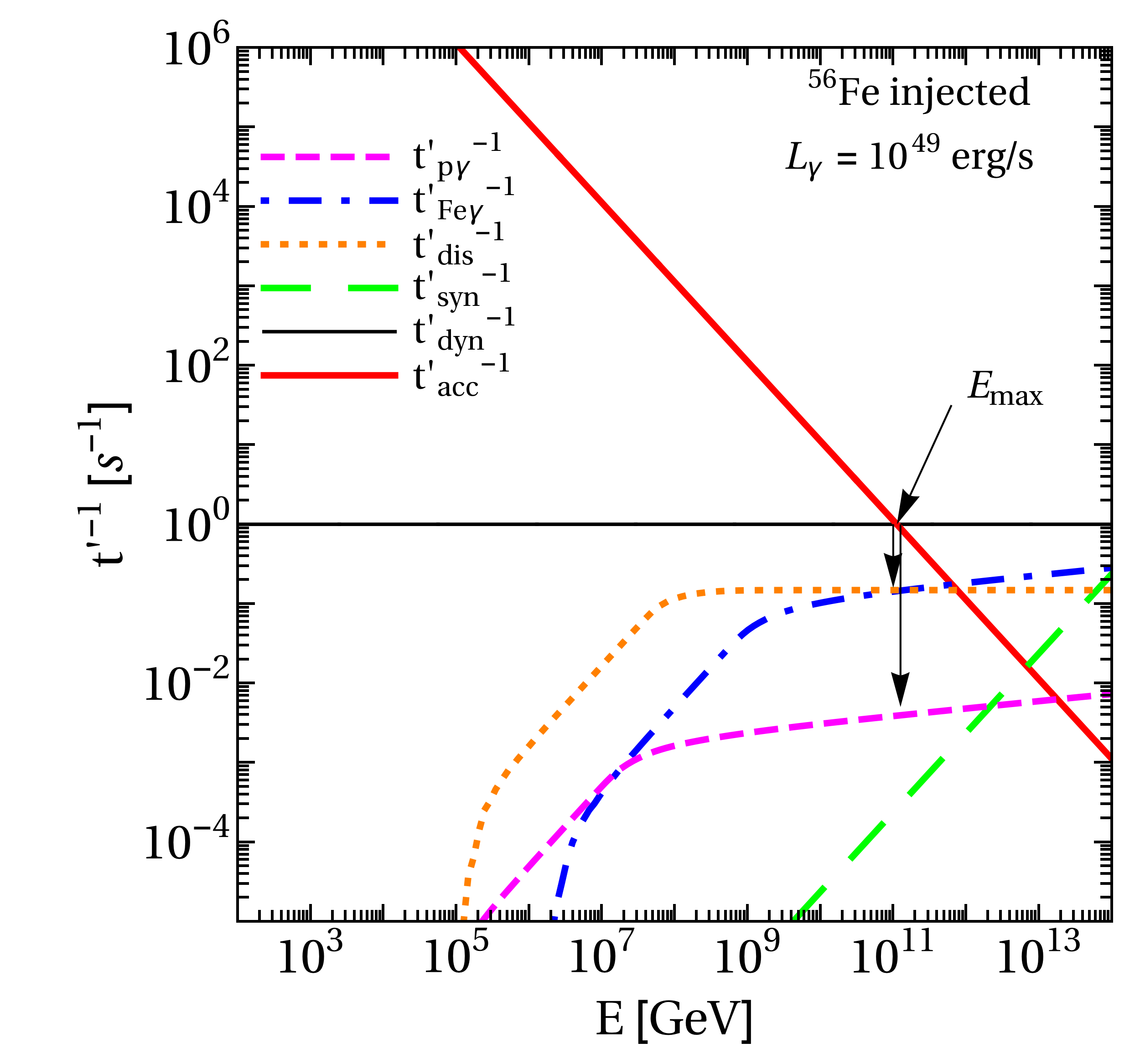}
\includegraphics[width=0.49\textwidth]{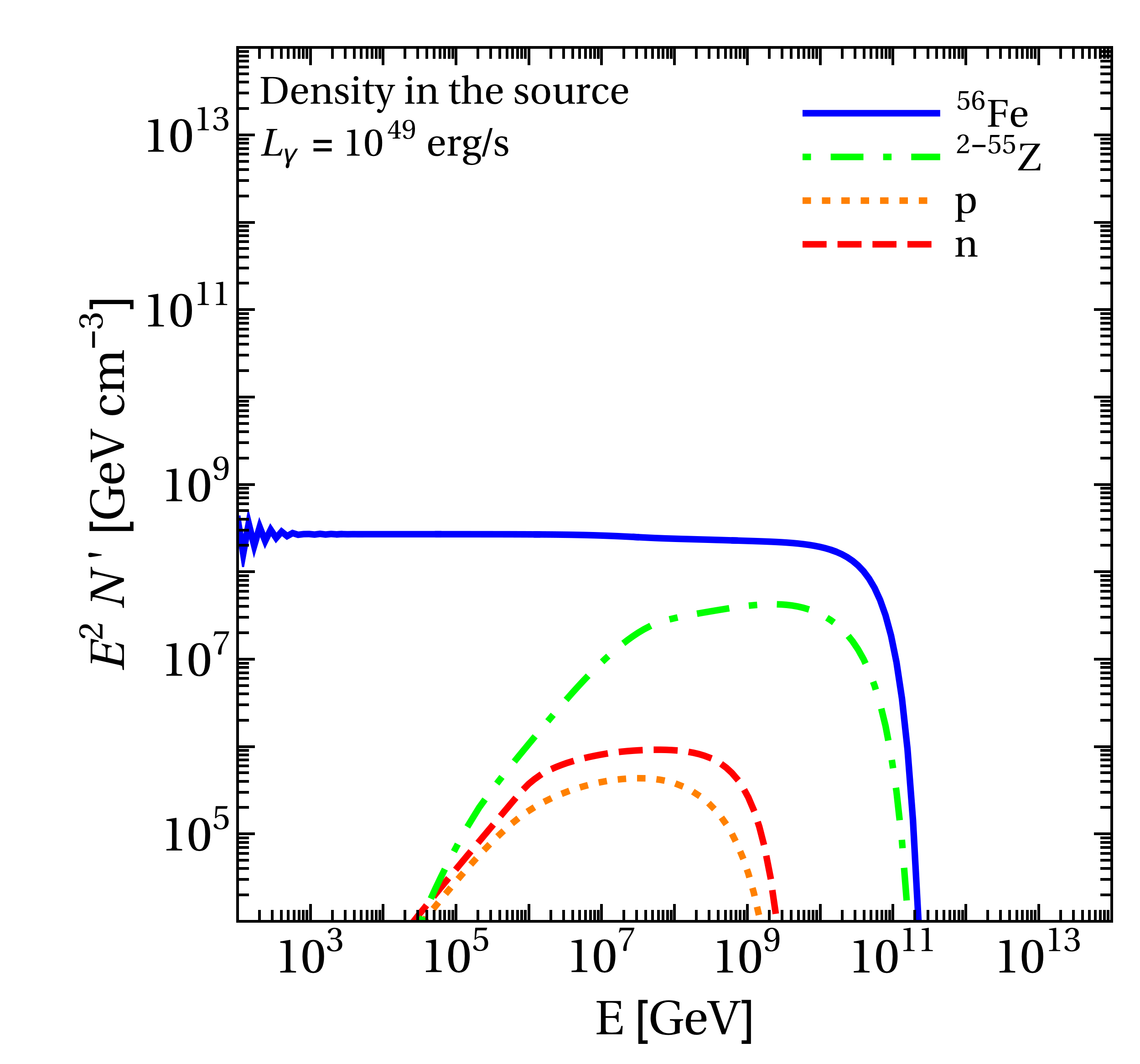}
\includegraphics[width=0.47\textwidth]{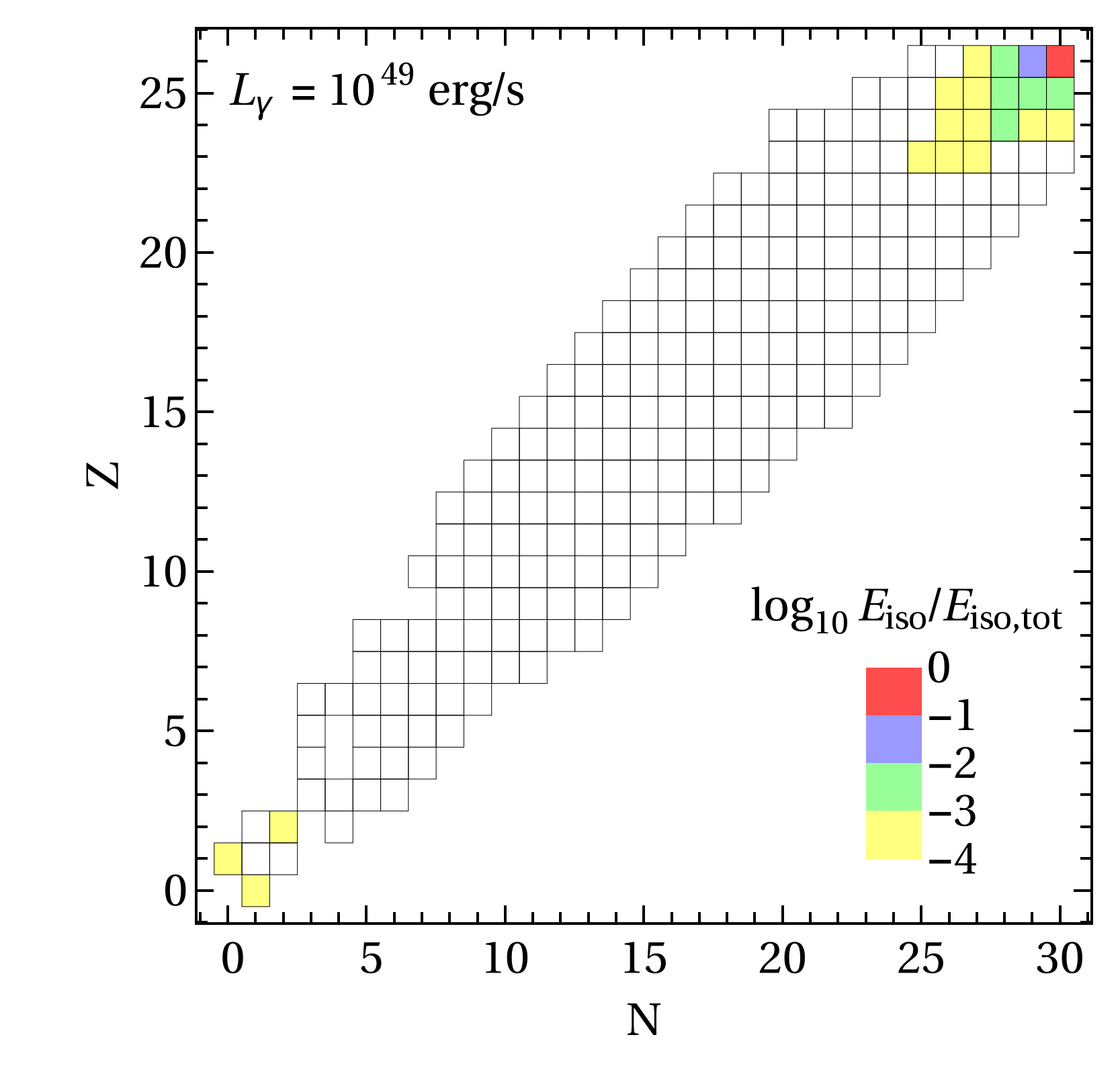}
\includegraphics[width=0.49\textwidth]{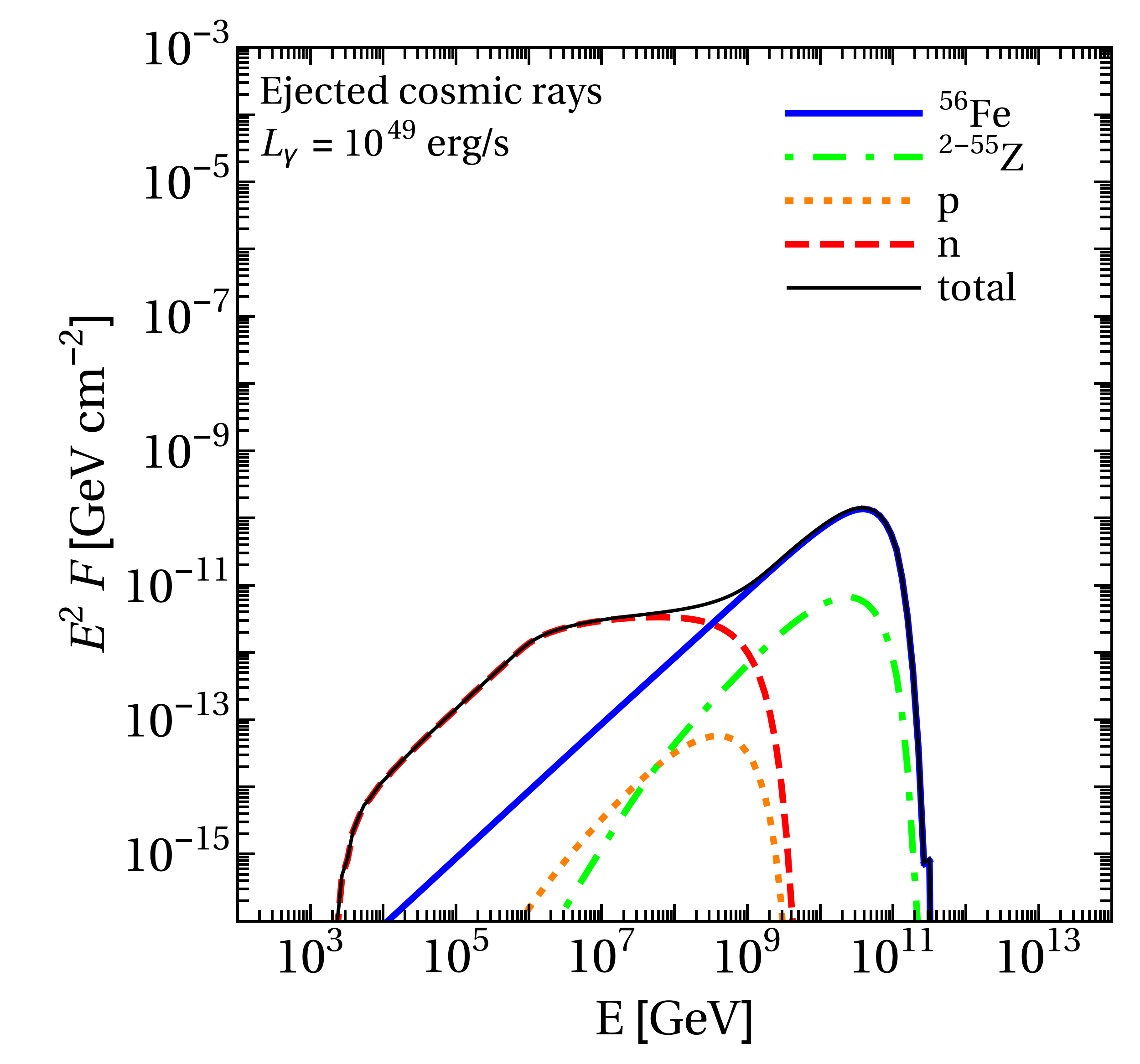}
\caption{Example for the ``Empty Cascade'' source class (isotropic luminosity $L_\gamma = 10^{49}$ erg s$^{-1}$) for injection of pure $^{56}$Fe. Interaction rates (top left), particle densities in the source (top right), nuclear cascade (bottom left) and ejected cosmic ray fluence per shell (bottom right, without CMB and EBL interactions) as a function of the energy in the observer's frame. The different curves in the right panels correspond to the different isotopes:  primary $^{56}$Fe in blue, secondaries produced in nuclear cascade according to legend, where the result from secondary nuclei other than nucleons are summed over.
The other GRB parameters are chosen to be $R  \simeq 10^{8.3} \, \mathrm{km}$, $\Gamma = 300$, $\xi_{\mathrm{Fe}} = 10$, $\varepsilon'_{\gamma,\mathrm{br}} = 1$~keV and $z = 2$.
}
\label{fig:proto49}
\end{figure*}

We define the Empty Cascade  as the case which is optically thin to $A \gamma$ interactions 
of the injection isotope (and, consequently, all lighter isotopes including nucleons). We have checked that this definition is almost equivalent to asking for the neutrino flux being dominated by photo-meson production off the injection isotopes. The luminosity of this source class is relatively low (or the production radius is large), as low enough radiation densities in the source are a necessary requirement for that.

The rates of the dominant processes (apart from Bethe-Heitler pair production) for primary nuclei and photo-meson production off protons $t'^{-1}_{p \gamma}$ are shown in the upper left panel of \figu{proto49} for one example. By comparing $t'^{-1}_{\mathrm{Fe} \, \gamma}$ and $t'^{-1}_{\mathrm{dis}}$ with $t'^{-1}_{\mathrm{dyn}} \simeq t'^{-1}_{\mathrm{ad}} \equiv c/\Delta d '$ (see arrows), one can clearly see that the source is optically thin to $A \gamma$ interactions at the highest energy (dominated by the dynamical timescale or adiabatic losses). 

The particle spectra in the source at the end of the time evolution are shown in the upper right panel of \figu{proto49}. The primary ($E^{-2}$) injection spectra of $^{56}$Fe is indeed hardly modified and extends to the mentioned maximal energy. Secondary nuclei and nucleons are suppressed, and their maximal energies follow basically the conserved Lorentz factor. The spectra are somewhat harder than $E^{-2}$ because of the high-energy behavior of the $A \gamma$ cross section (which is not peaked such as a resonance there). 

The nuclear cascade (integrated energy of isotopes relative to total injection energy) is shown in the lower left panel of \figu{proto49}. One can clearly see that apart from $^{56}$Fe a few closeby isotopes are populated, while most of the nuclear cascade is,  relative to the injection luminosity, empty. Nucleons and light nuclei (especially $^4$He) are produced as disintegration products, but their occupation is relatively small compared to the injection isotope. 

\figu{proto49}, lower right panel, illustrates the ejected cosmic ray spectra for the assumptions stated earlier. Note that this type of figure shows the fluence at Earth including adiabatic losses, but no interactions with the CMB and EBL (which deform the spectrum).  Here charged cosmic rays escape via direct escape, while neutrons are not magnetically confined, see \Ref~\cite{Baerwald:2013pu} for a more detailed discussion. This leads to a characteristically different shape of the ejected spectra: relatively hard ejection spectra for the charged cosmic rays, and softer spectra for the neutrons, which will eventually decay into protons on their way to Earth. There is evidence that such hard ejection spectra are actually  the required input for the cosmic ray propagation of UHECR nuclei to describe Auger data~\cite{Aab:2016zth}, we will come back to this issue later.

A characteristic of the Empty Cascade is that the cosmic ray ejection spectrum is dominated by the hard spectrum of the injected primaries. Nevertheless, the contribution of the neutrons is, especially at low energies, more substantial than one may think from the upper right panel in the source. The reason is that the neutrons basically escape unmodified, while the nuclei are somewhat depleted even at the highest energies by the escape mechanism (see \figu{optthickness} for the magnitude of the effect).

\subsection{Populated Cascade}
\label{sec:partialcascade}

\begin{figure*}[ht!]
\includegraphics[width=0.49\textwidth]{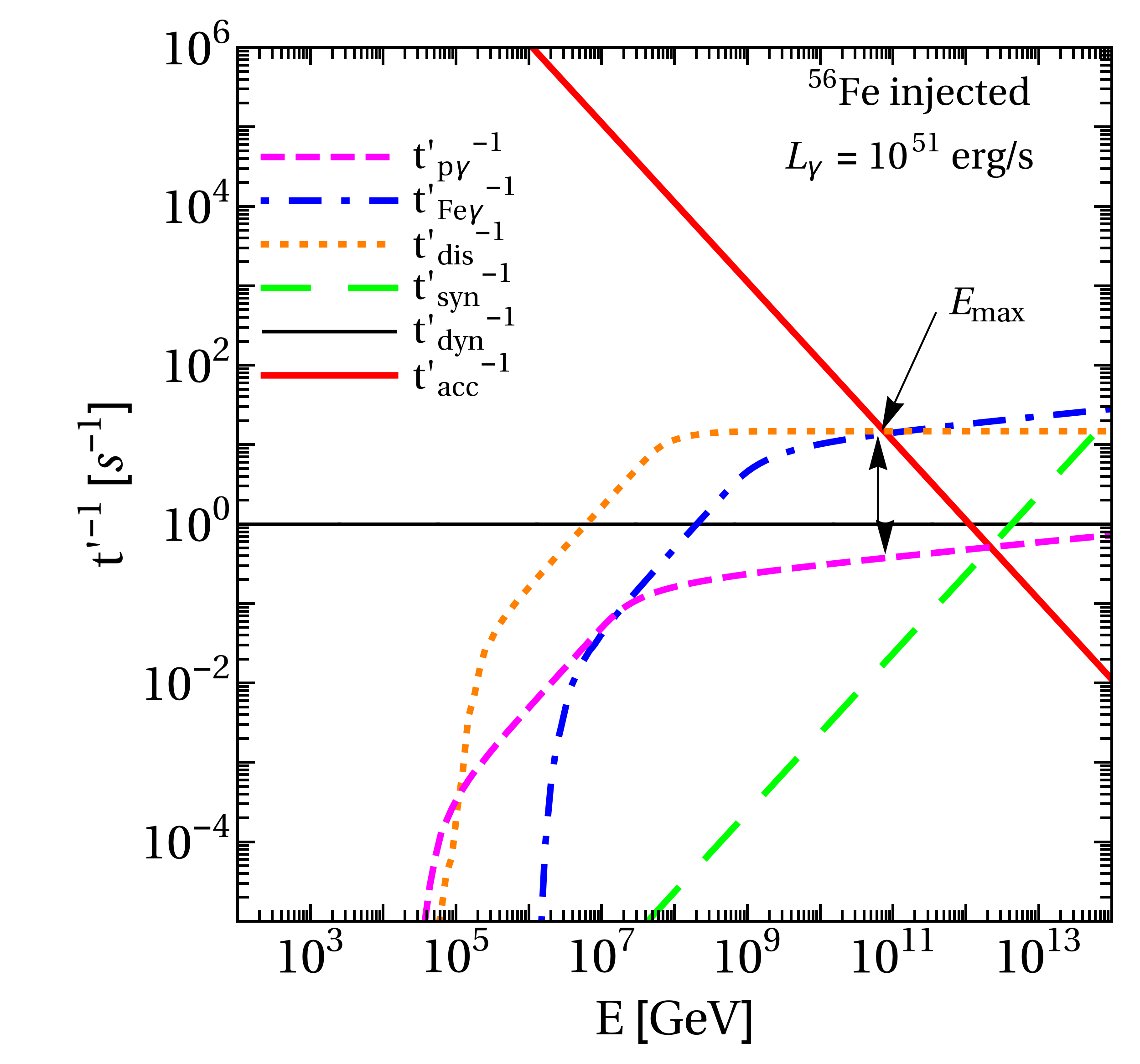}
\includegraphics[width=0.49\textwidth]{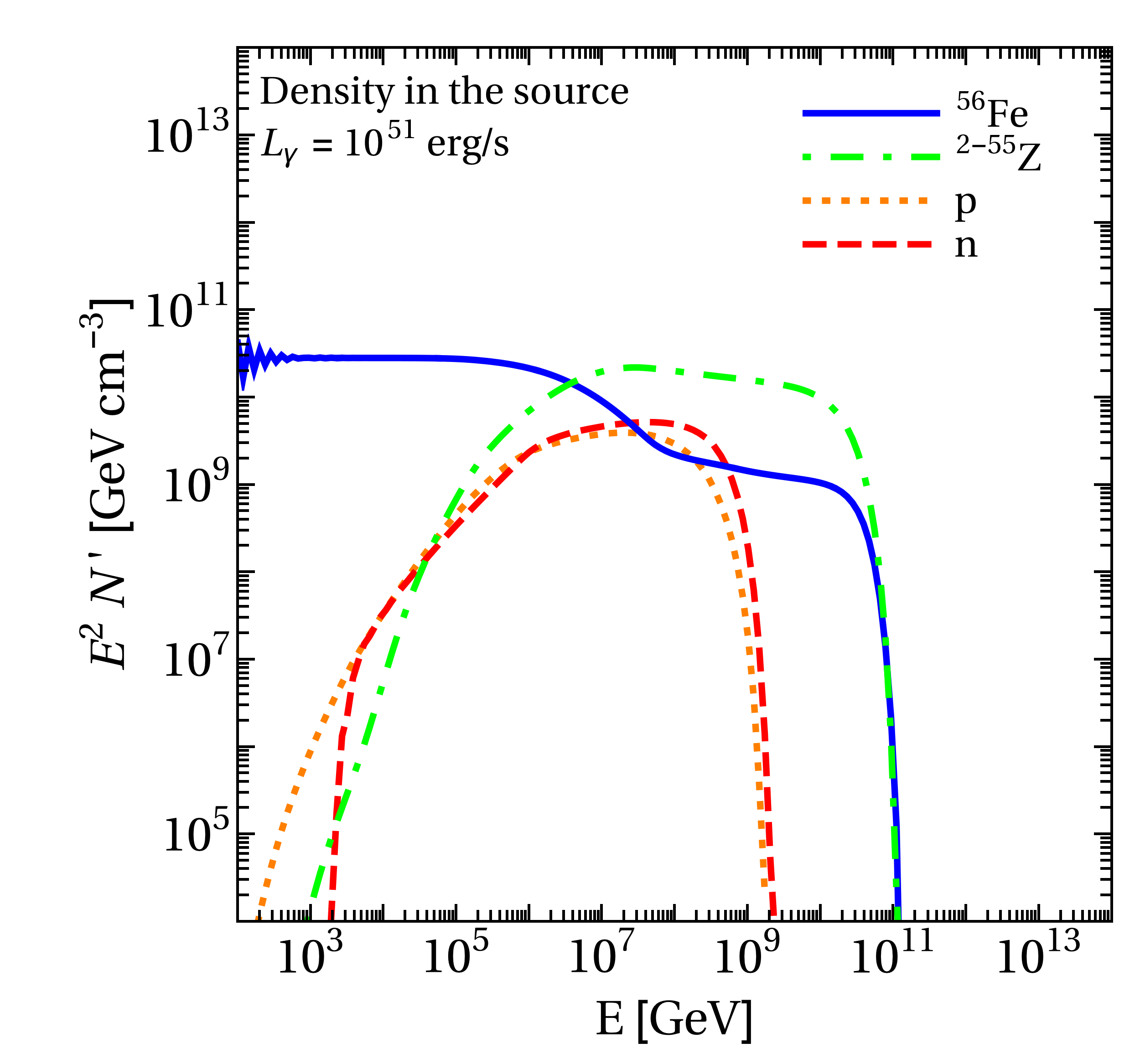}
\includegraphics[width=0.47\textwidth]{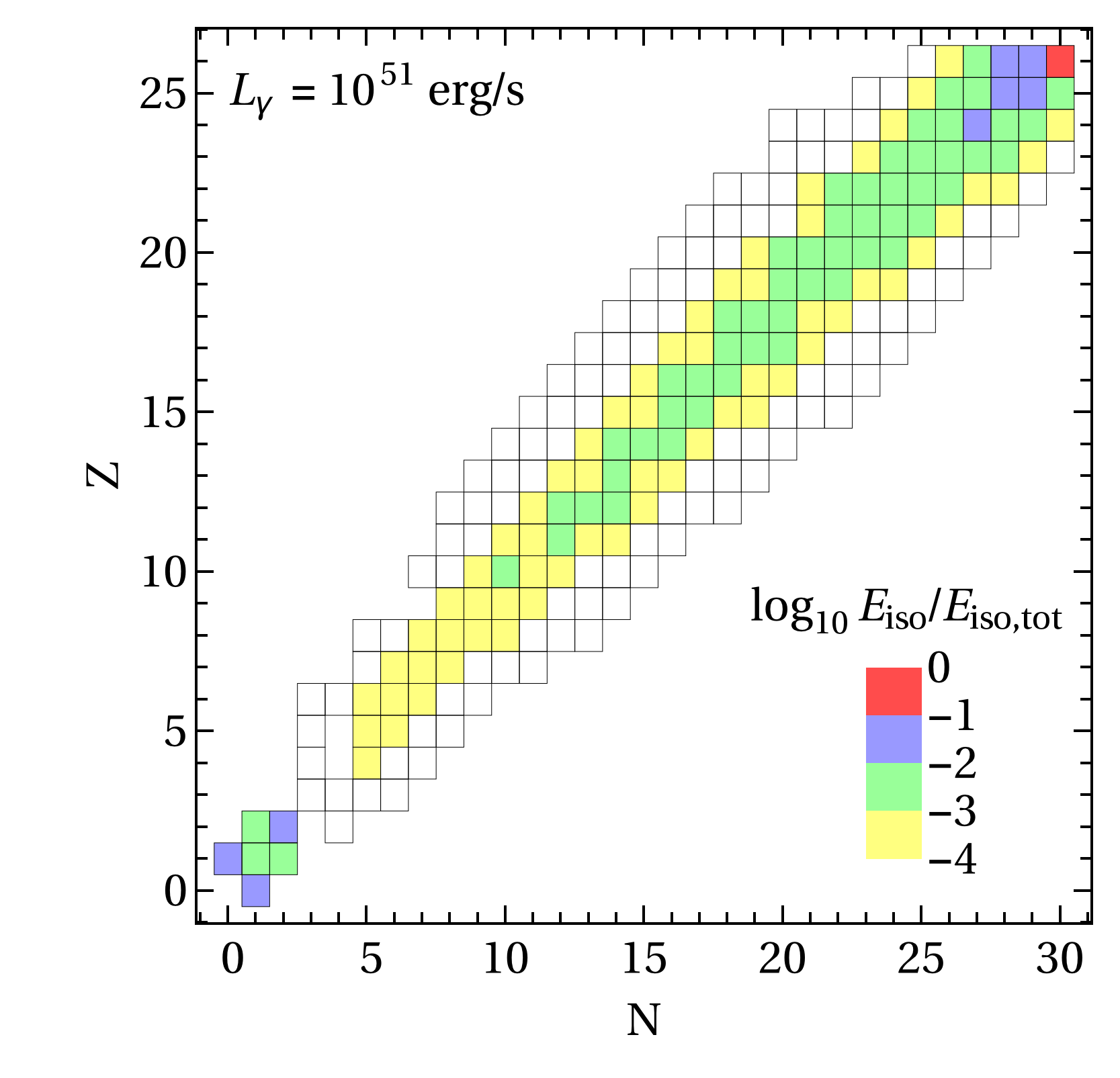}
\includegraphics[width=0.49\textwidth]{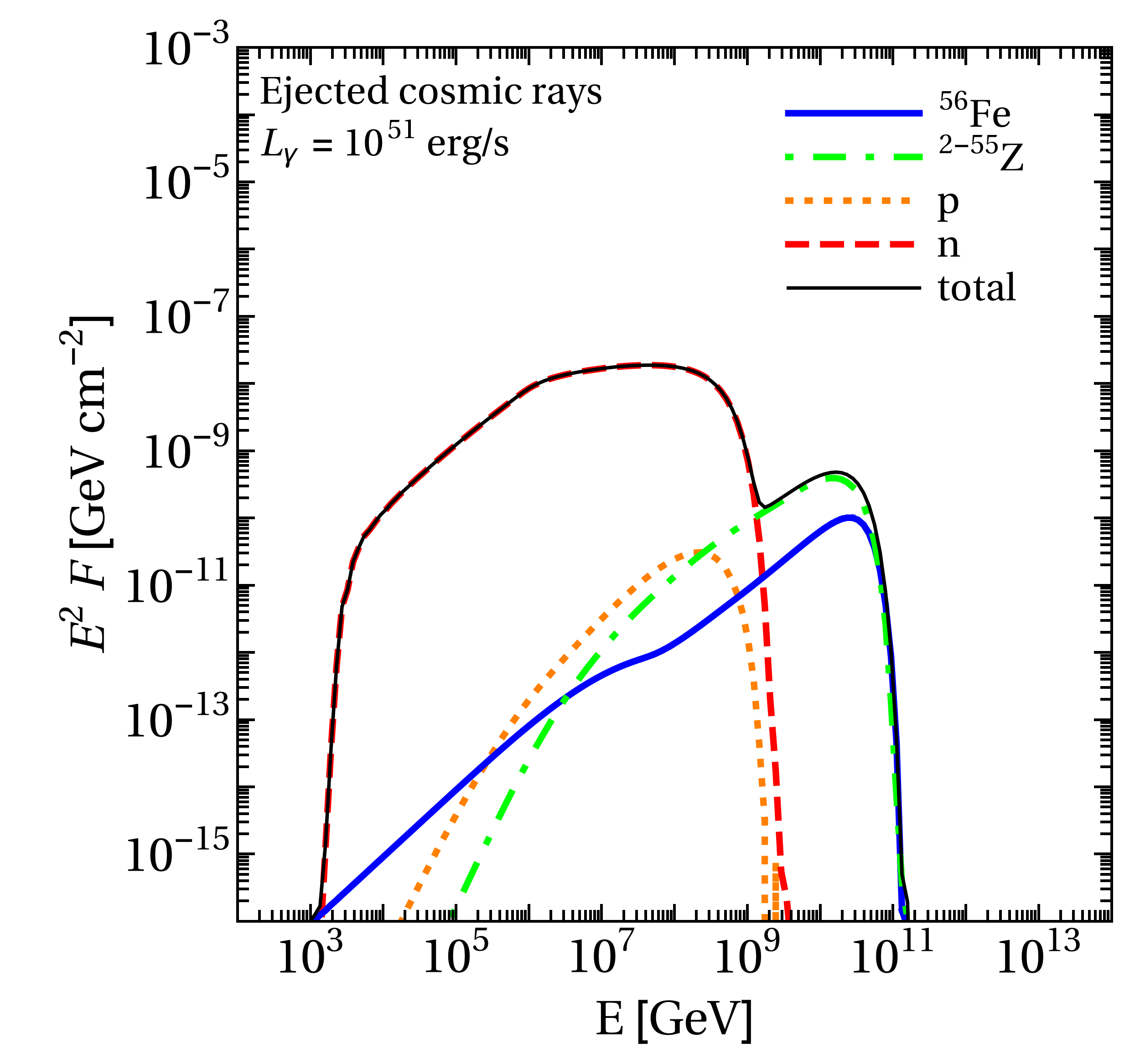}
\caption{Example for the ``Populated Cascade'' source class (isotropic luminosity $L_\gamma = 10^{51}$ erg s$^{-1}$) for injection of pure $^{56}$Fe. See caption of  \Fig~\ref{fig:proto49} for details.}
\label{fig:proto51}
\end{figure*}

We define the Populated Cascade as the case, which is optically thick to $A \gamma$ interactions of the injection isotope that will disintegrate and populate the cascade. At the same time, the source environment is optically thin to $p \gamma$ interactions, so that the proton and neutron fluxes will be hardly affected by these interactions; see arrows. 
We have checked that this definition is almost equivalent to asking for the neutrino flux being dominated by photo-meson production off the secondary isotopes lighter than the primary, but heavier than the nucleons. The luminosity of this source class is intermediate. The source class corresponds to the typical assumption of the optically thin case for protons, extended to nuclei.

\figu{proto51} shows one example for this source class in the same format as \figu{proto49}. From the upper left panel, we can indeed see that the source is optically thick to Fe$\gamma$ interactions, while it is optically thin to $p\gamma$ interactions. The maximal primary energy is given by $A \gamma$ interactions. The nuclear cascade (lower left panel) is, relative to the injection energy, well populated, where light nucleons are occupied similar to the near--$^{56}$Fe isotopes. 

From the densities in the source, upper right panel, one can see a clear depletion of the injection $E^{-2}$ spectrum beyond the disintegration threshold, while the secondary isotopes are populated with a density comparable to the original primary density. The proton and neutron densities are larger than in the Empty Cascade, but not yet comparable to the injection density. 

Nevertheless, the ejected cosmic ray spectrum, see lower right panel, is already dominated by the neutrons. The reason is that the maximal energy is dominated by $A \gamma$ interactions and therefore the Larmor radius can reach only about 1/30 of the size of the region at the maximal energy (see upper left panel: compare acceleration and dynamical rates at the maximal primary energy). Consequently, the ejected secondary spectra are suppressed by a similar factor at the maximal energy because cosmic rays from the innermost regions of the shell take too long to escape. The effective (all energy) ejection spectrum including neutrons and nuclei will be relatively soft because of the neutron escape component, with a potential dip.

Comparing the Populated and Empty Cascades, we notice that the isotropic luminosity, which determines the photon density in the source, controls the relative height between the neutron and nuclei escape components. Since we use the isotropic luminosity as fit parameter later, the right luminosity to UHECR data will be automatically determined later.

\subsection{Optically Thick Case}
\label{sec:fullcascade}

\begin{figure*}[ht!]
\includegraphics[width=0.49\textwidth]{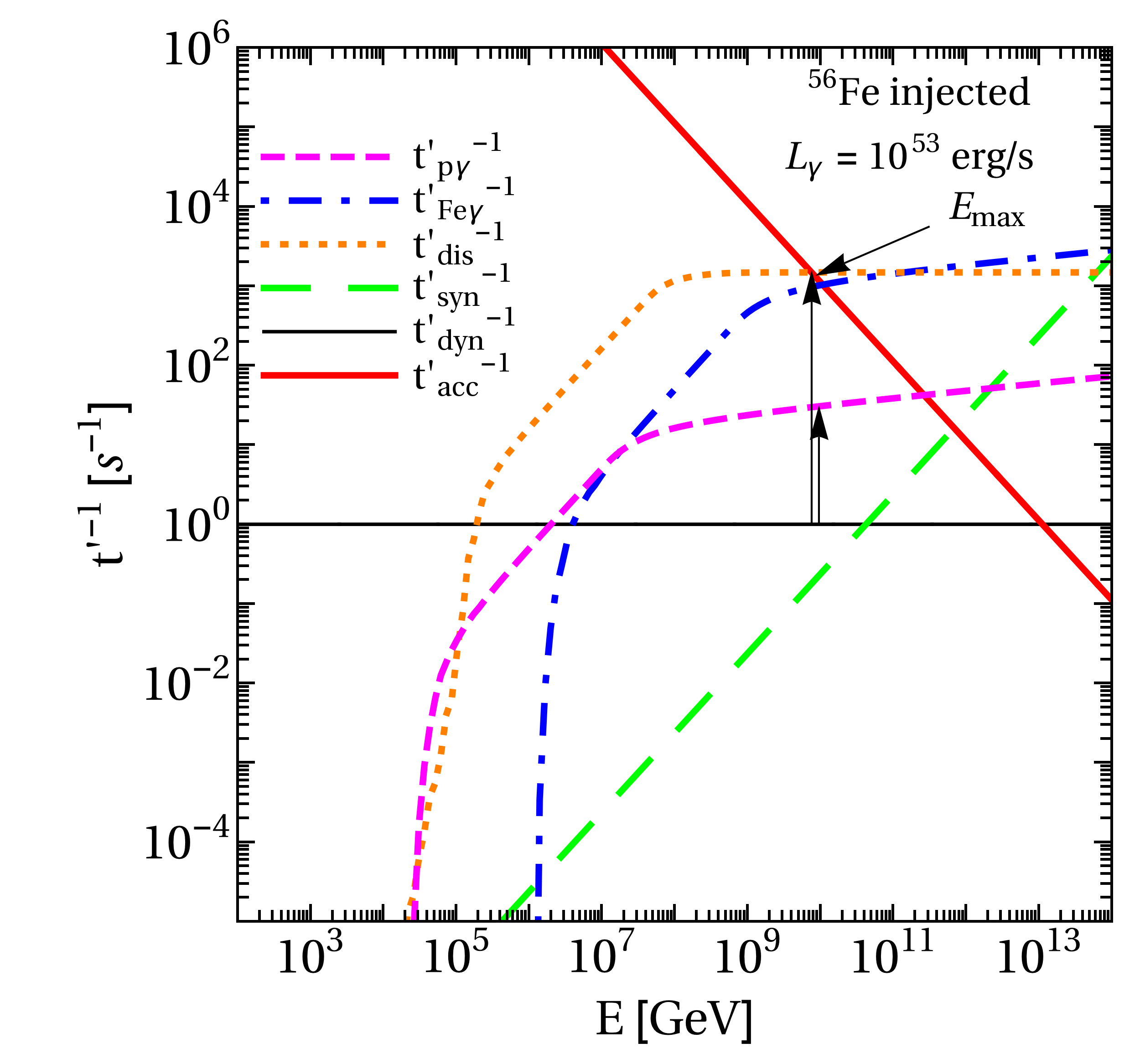}
\includegraphics[width=0.49\textwidth]{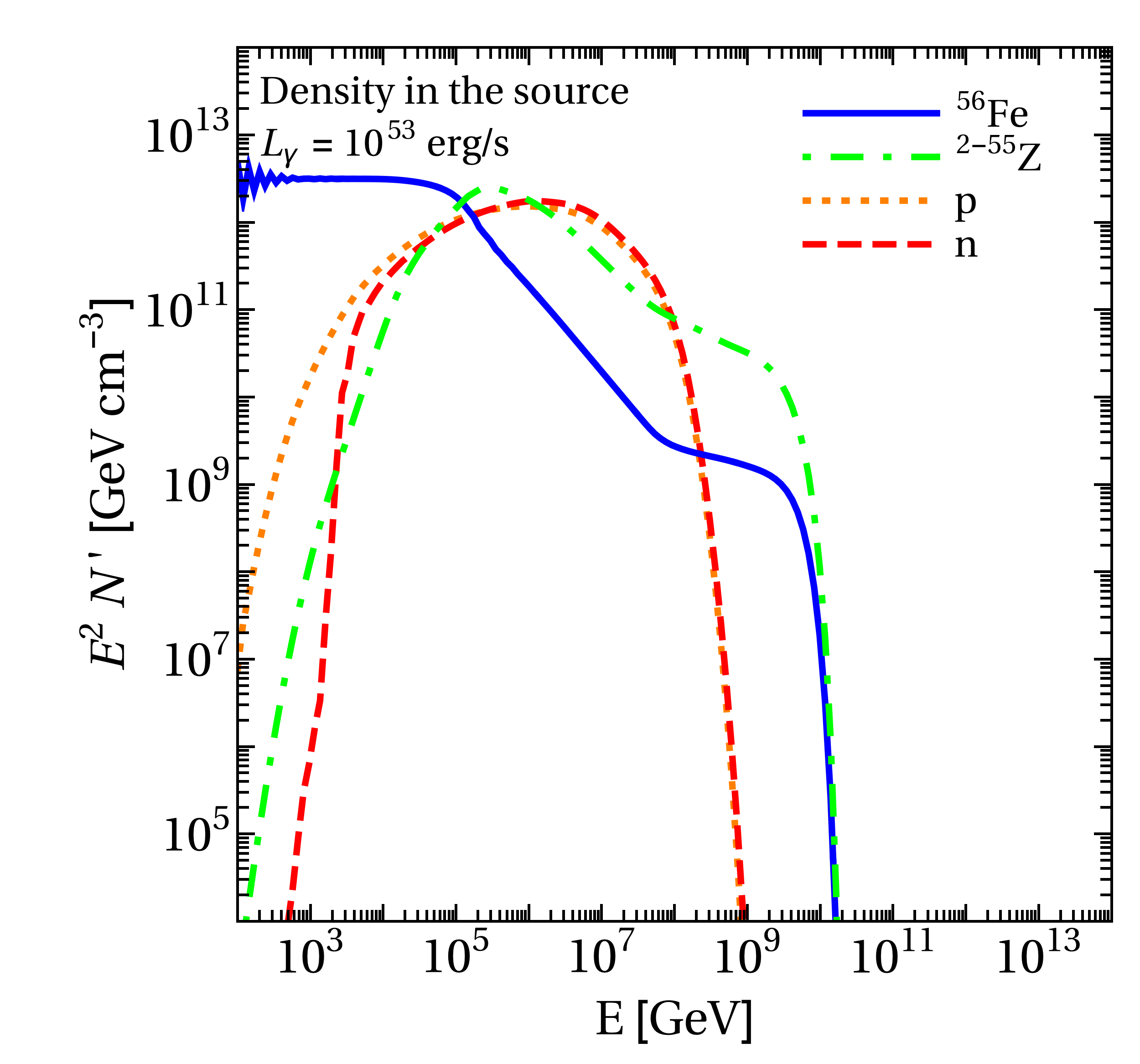}
\includegraphics[width=0.47\textwidth]{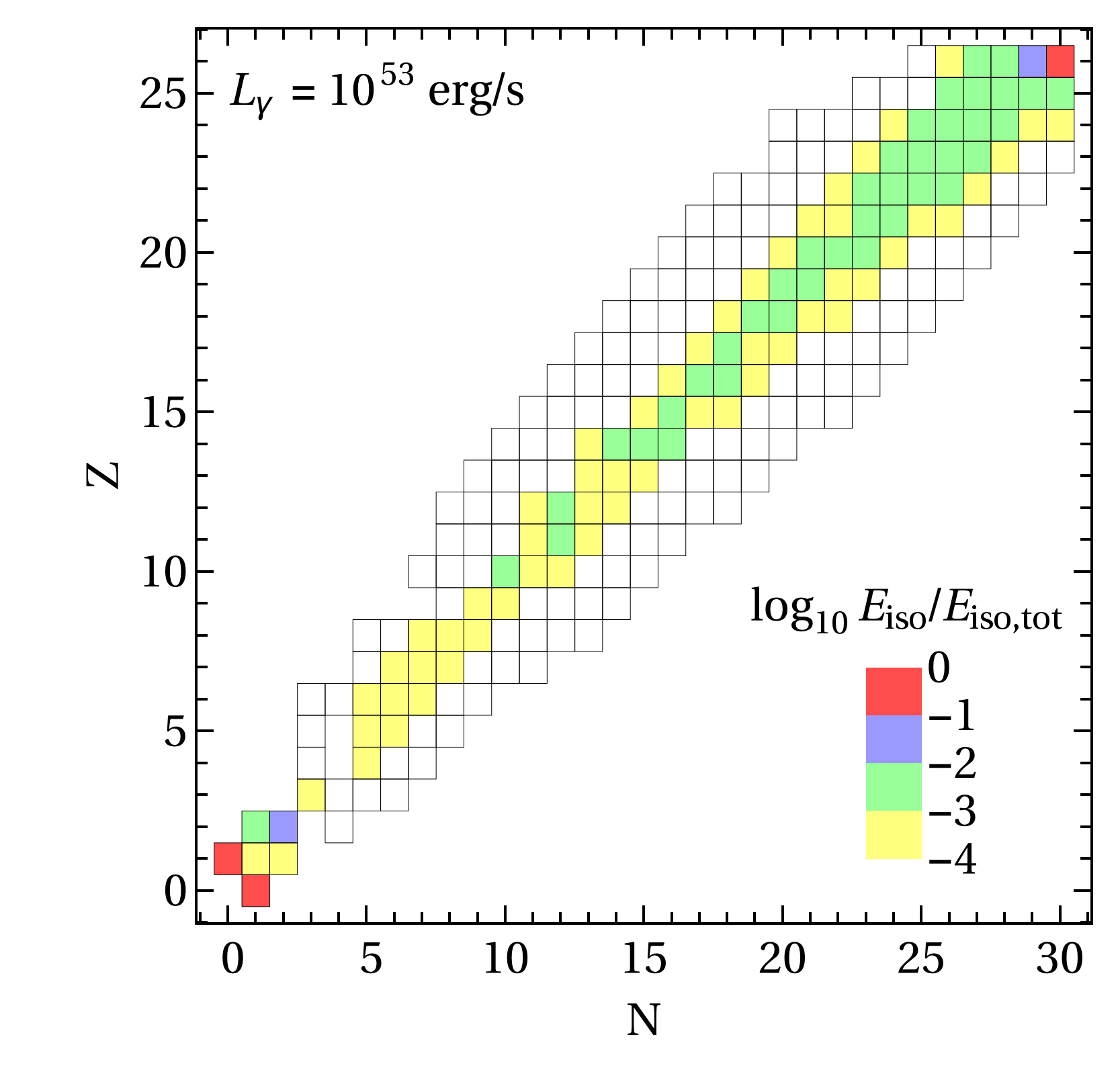}
\includegraphics[width=0.49\textwidth]{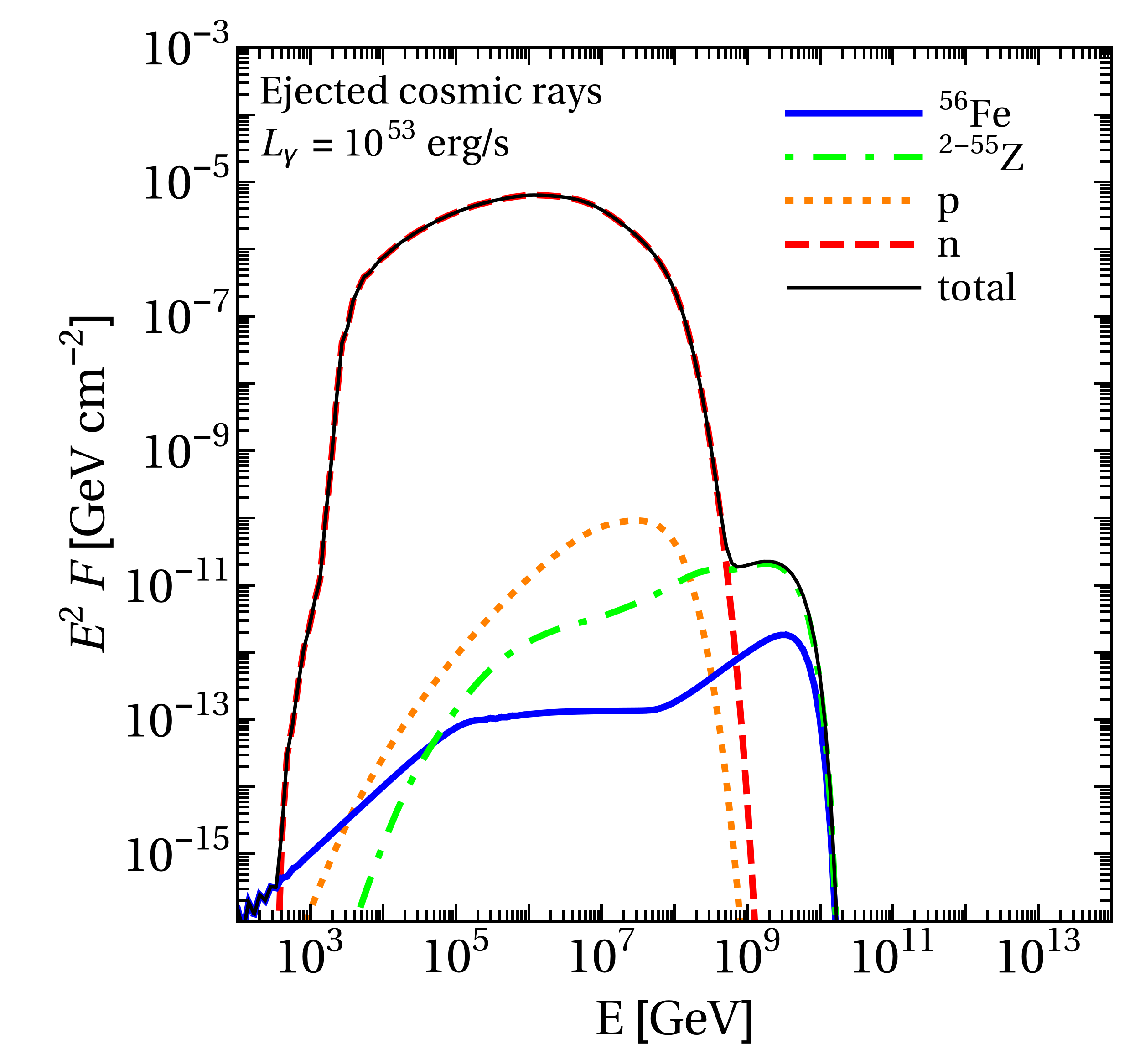}
\caption{Example for the ``Optically Thick Case'' (to nucleons and nuclei) source class (isotropic luminosity $L_\gamma = 10^{53}$ erg s$^{-1}$) for injection of pure $^{56}$Fe. See caption of  \Fig~\ref{fig:proto49} for details.}
\label{fig:proto53}
\end{figure*}

We define the Optically Thick Case as the one which is optically thick to $A \gamma$ interactions of nucleons and nuclei.
This can be clearly seen in the upper left panel of \figu{proto53}, which shows one example with an extremely high luminosity (arrows). It turns out that the neutrino production in this case will be dominated by the protons and neutrons produced in the disintegration chain.

It is first of all instructive to compare the nuclear cascade in the lower left panel of \figu{proto53} with the corresponding one in \figu{proto51}. The cascade appears to be less populated off the main diagonal because the intensity of the cascade is actually reduced at intermediate isotopes, and most energy is dumped into nucleons produced in the cascade -- which are populated similarly to the primaries. The maximal primary energy is determined by photo-disintegration (see upper left panel). 

The primary and secondary isotope densities are strongly suppressed beyond the photo-disintegration threshold (see upper right panel), while the protons and neutrons are populated at a level comparable to the injected primary flux. Compared to the Populated Cascade, the nucleons peak at the photo-meson threshold (break in $t'^{-1}_{p \gamma}$ in upper left panel), because they cascade down in energy by multiple interactions. 

The ejected charged cosmic rays (lower right panel) dominate at the highest energies but are strongly suppressed because of the low densities in the source and because the Larmor radius at the highest energy is extremely small compared to the size of the region. In fact, it is difficult to reach the UHECR energy range because of the $A \gamma$ interactions limiting the maximal primary energy.

\section{Impact of Astrophysical Parameters and Models}
\label{sec:model}

\begin{figure*}[t]
\includegraphics[width=0.49\textwidth]{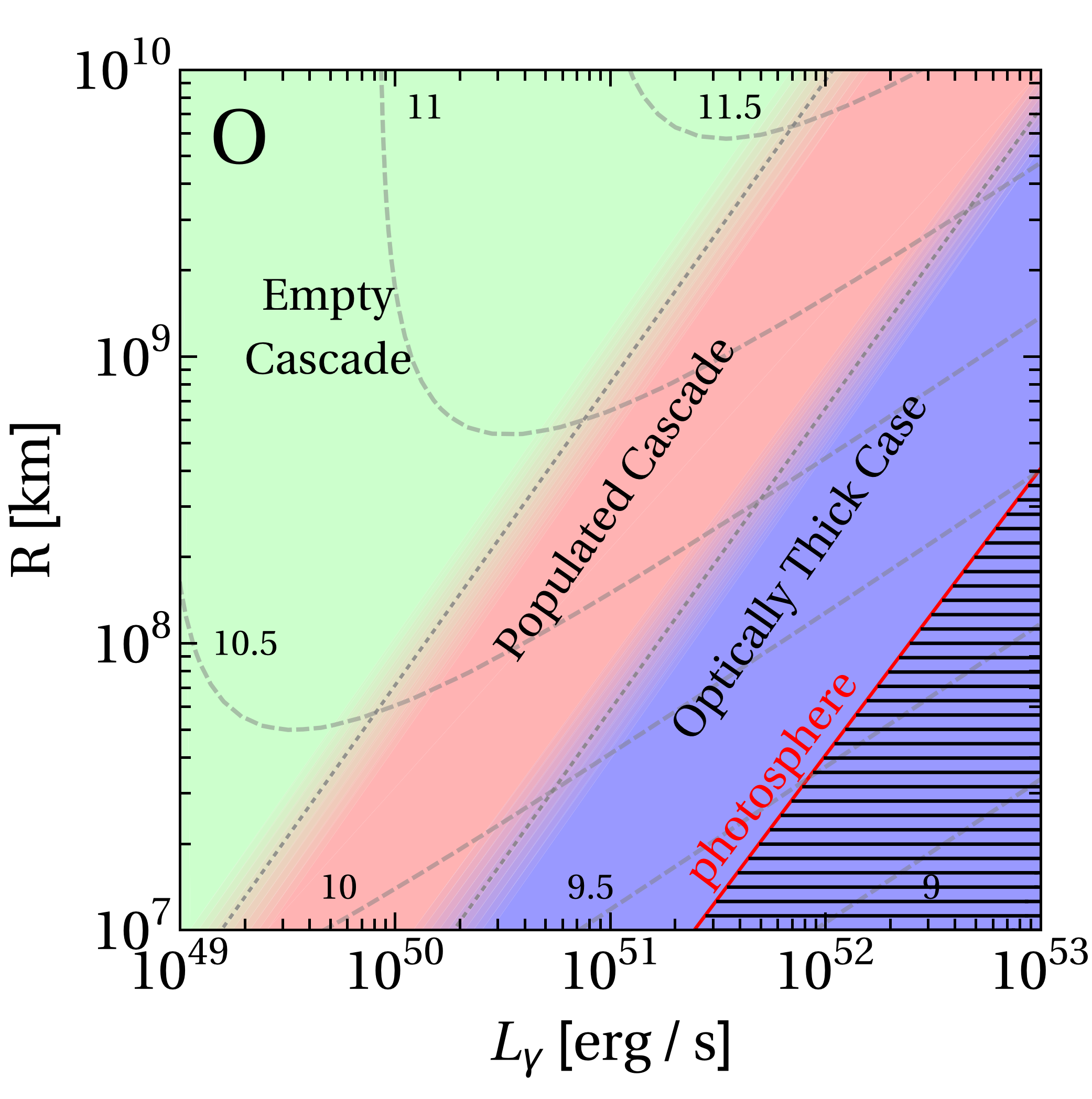}
\includegraphics[width=0.49\textwidth]{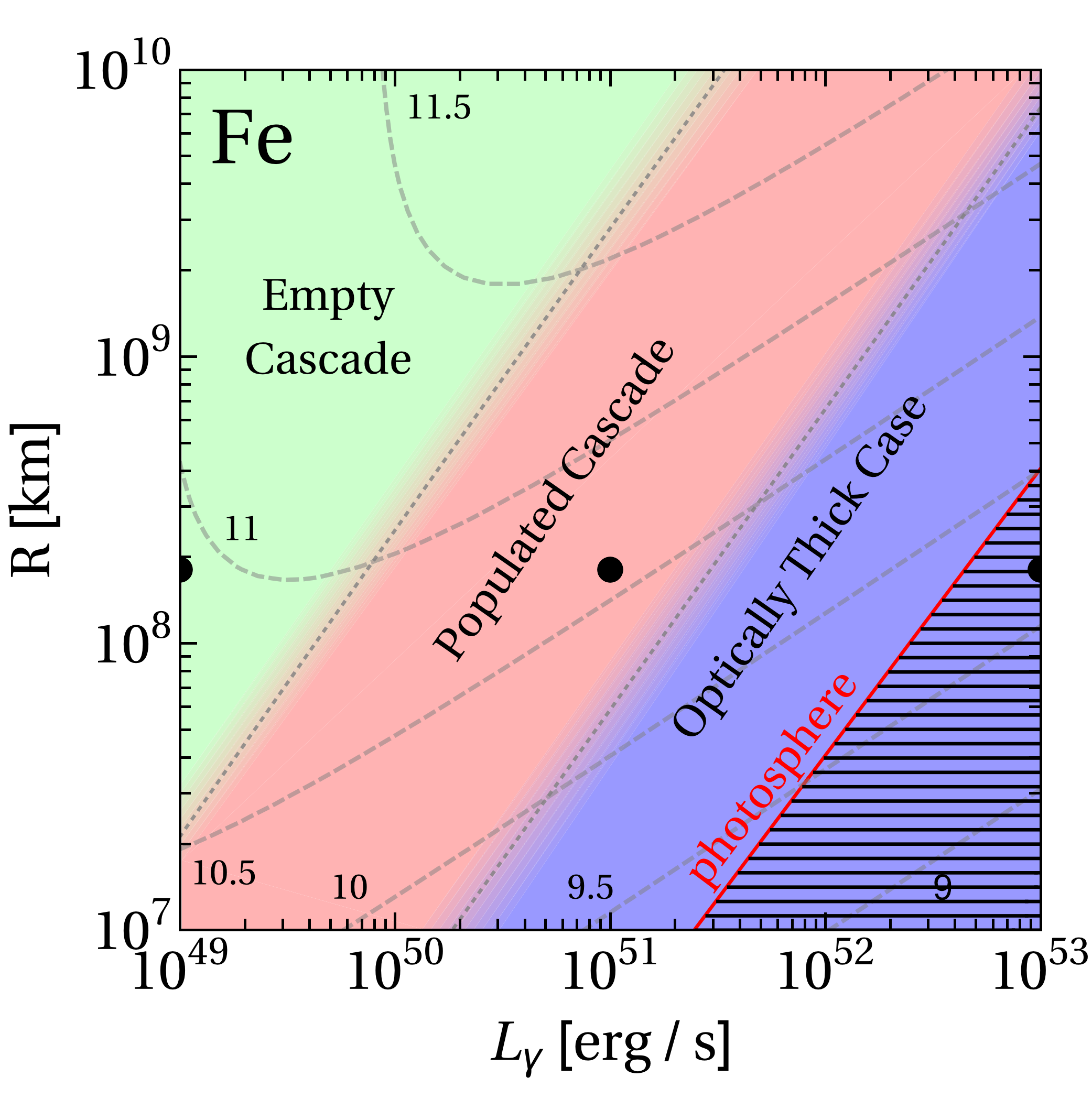}
\caption{Nuclear cascade class as a function of luminosity $L_\gamma$ and collision radius $R$ in the internal shock scenario. The injection composition is pure  \isotope[16]{O} (left left) or \isotope[56]{Fe} (right panel). The other parameters are the same as in \Fig~\ref{fig:proto49}. Black dots indicate the position of the examples shown in \Sec~\ref{sec:classes}. The gray dashed contours display  $\mathrm{log}_{10}(E_{i,\mathrm{max}} [\mathrm{GeV}])$  in the observer's frame. Below the photosphere (red solid line), gamma rays cannot leave the source anymore due to electron-positron pair production. Note that in this figure we fix $\Gamma=300$, and scale $t_v \propto R$ according to \equ{rc}. 
}
\label{fig:nuscan}
\end{figure*}

In the previous section, we discussed the results for three prototypes from the Empty Cascade, Populated Cascade and Optically Thick Case regimes. We follow the previous classification using the interaction rates at the maximal energy, and show in \figu{nuscan} these three regimes as a function of $L$ and $R$ for the injection of pure oxygen (left panel) and iron (right panel). Note that the transition among the regions in terms of the development of the nuclear cascade is continuous, while our classification scheme will produce well-defined different regimes. 
In \figu{nuscan} we also show the sub-photospheric region, \ie, the region from which photons cannot escape because of Thomson scattering.\footnote{Assuming that the plasma is electron-(fully stripped) isotope-dominated and that the electron density is similar to the proton density (from charge conservation, electron-positron pair production assumed to be sub-dominant), the photospheric radius -- where the optical thickness to Thomson scattering is one -- is given by $R_{\mathrm{ph}} \simeq (1/2 \, \xi_i \, L_\gamma \, t_v \, \sigma_{\mathrm{Th}}/(4 \, \pi \, \kappa \, (1+z) \Gamma c^2 m_p) )^{0.5}$. Here $\kappa \simeq 0.25$ (see \eg\ last column, \Tab~2 in \Ref~\cite{Bustamante:2016wpu} for simulation results) is the re-conversion efficiency from the kinetic into the radiated (nuclei dominated) energy. Note that the factor $1/2$ comes from the assumption of heavier nuclei (which carry about as many protons as neutrons), which descreases the photospheric radius by $1/\sqrt{2}$.
}

Since high luminosities or small collision radii mean high photon densities, it is clear that the optically thick case and even the photosphere are reached in the lower right corners of the panels, whereas the cascade is hardly populated in the upper left corner. Our chosen prototypes (circles in right panel) correspond to extreme examples for the Empty Cascade and Optically Thick Case, whereas the prototype for the Populated Cascade lies well within the corresponding region. 
The maximal primary energy (gray-dashed contours) is typically given by $A \gamma$ interactions (lower right) or adiabatic cooling (upper left).\footnote{Since the interaction rates $t'^{-1}_{A \gamma} \propto u'_\gamma \propto L_\gamma/R^2$ (\cf, \equ{photonorm}) and $t'^{-1}_{\mathrm{acc}} \propto B'/E' \propto \sqrt{u'_\gamma}/E' \propto \sqrt{L_\gamma}/(E' R)$, one finds the maximal primary energy to be constant along the curves $R \propto \sqrt{L_\gamma}$ in the $A \gamma$ dominated region. If, however, adiabatic losses dominate the maximal energy, the dependence on $R$ cancels and the maximal energy hardly depends on $R$ because of the implied internal shock relation \equ{rc} leading to $t'^{-1}_{\mathrm{ad}}=2 \Gamma/R$.} Note that the adiabatic cooling limited case corresponds to a rigidity-dependent maximal energy, and that it roughly coincides with the Empty Cascade case.

The transition between the Empty and Populated Cascades depends on the injection composition (compare left and right panels) because the $A \gamma$ interaction rates increase with mass number -- which determine this classification. Correspondingly, the transition between adiabatic and interaction dominated maximal energy shifts (see gray-dashed contours). Since the transition between Populated Cascade and Optically Thick Case depends on the optical thickness to proton and neutron interactions, it is not affected by the injection composition. Therefore, the lower the injection mass is, the smaller the Populated Cascade region will be, because the $A \gamma$ interaction rate approaches more and more the $p \gamma$ rate.

\begin{figure*}[tp]
\includegraphics[width=0.6\textwidth]{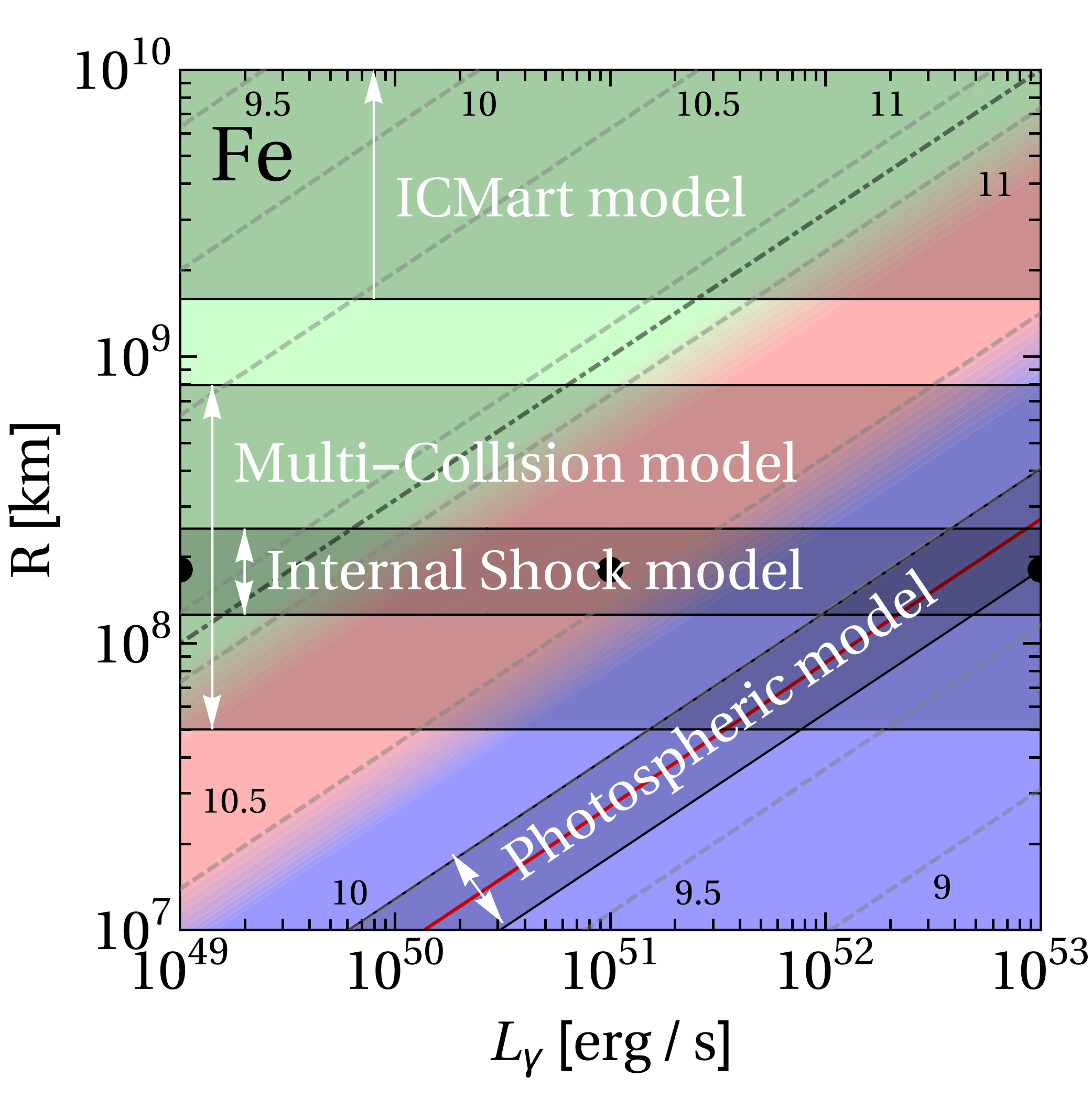}
\caption{Impact of different astrophysical model assumptions (see marked shadings) for fixed $t_v=0.01 \, \mathrm{s}$. Similar to \Fig~\ref{fig:nuscan} for $^{56}$Fe, but the  internal shock model relationship \equ{rc} among collision radius, $t_v$, and $\Gamma$ is not applied, and only holds in the region marked ``Internal Shock model'' -- which implies that for the prototypes (circles) the same results as in \Sec~\ref{sec:classes} are obtained. Here the idea is that $R$ is the main control parameter for the target density in the shells, whereas $t_v$ determines the shell thickness and $\Gamma$ the energy shift only, \ie, the dependence on these parameters is milder (the pion production efficiency scales $\propto t_v$ similar to $L_\gamma$, but typically does not vary in such a large range; therefore $L_\gamma$ is chosen as parameter here); for details, see \Sec~\ref{sec:astromodel}. The regions applying to different astrophysical production models (taken from \cite{Zhang:2012qy}, for multi-collision model see \Sec~\ref{sec:astromodel}) are sketched in the figure for the chosen value of $t_v$  (fast time variability in light curve). 
Note that the maximal energies are shown as gray dashed curves as in \figu{nuscan}, but not the transition curves among the cascade regions. Instead, the dashed-dotted line separates the $A\gamma$ dominated (lower right corner) and  adiabatic loss dominated (upper left corner) maximal primary energy regions. 
}
\label{fig:modeldep}
\end{figure*}

In order to address the astrophysical model-dependence of the nuclear cascade, \figu{modeldep} shows the same result as \figu{nuscan} for $^{56}$Fe with $t_v=0.01 \, \mathrm{s}$ fixed (instead of scaling it with the  internal shock model relationship \equ{rc} among collision radius, $t_v$, and $\Gamma$). Note that the pion production efficiency $f_{p \gamma} \propto L_\gamma t_v/R^2$  (see \Sec~\ref{sec:methods}), which means that the dependence on $\Gamma$ (which shifts the energies) is very mild, and $R$ is the dominant control parameter -- which we have chosen on the vertical axis compared to earlier papers~\cite{Baerwald:2014zga}.

The maximal proton energy behaves similarly to \figu{nuscan} in the lower right corner, whereas in the adiabatic loss dominated upper left corner, the scaling is the same as in the lower right corner.\footnote{Since $t'^{-1}_{\mathrm{acc}} \propto B'/E' \propto \sqrt{u'_\gamma}/E' \propto \sqrt{L_\gamma}/(E' R)$ is proportional to the (constant) adiabatic cooling rate, one has $R \propto \sqrt{L_\gamma}$ for fixed maximal energy.} This observation is interesting because the cosmic ray fit is sensitive to the maximal energy: since maximal energy and $A\gamma$ interaction rates scale in the same way in the parameter space shown, we do not expect significant changes of the results if one moves parallel to the lines (for the fixed chosen value of $t_v$). Consequently, our protoype behavior of the nuclear cascade (compare to the dots in the figure) does not change very much along these lines. 

The figure also displays typical regions expected for different prompt emission models, as outlined in the figure caption and discussed in \Sec~\ref{sec:methods}. For example, larger collision radii are expected in the ICMart model compared to the internal shock model because the pulse timescale is indicative for the collision radius. While \equ{rc} holds for the internal shock scenario, the chosen $\Gamma$ and $t_v$ lead to a relatively narrow region for the internal shock model, while the multi-collision model has a large intrinsic spread of the collision radii around this nominal value. Photospheric models typically predict a prompt emission dominated by the photosphere. One can read off from \figu{modeldep} what happens for a certain choice of $L$ and $R$, such as for a specific collision in the multi-zone model. Moving along the ``diagonals'', one can then easily find a prototype which displays the qualitative behavior. For example, the result of our $L=10^{49} \, \mathrm{erg \, s^{-1}}$, $R=10^{8.3} \, \mathrm{km}$ internal shock model prototype will correspond to a $L=10^{51} \, \mathrm{erg \, s^{-1}}$, $R=10^{9.3} \, \mathrm{km}$ ICMart collision in terms of nuclear cascade development, maximal primary energy, and neutrino production efficiency. However, note that the ICMart collision has higher $\gamma$-ray and proton luminosities, which means that in the cosmic ray fit a correspondingly lower baryonic loading will be required.

\section{Prompt Neutrino production}
\label{sec:neutrinos}

\begin{figure*}[ht!]
\includegraphics[width=0.49\textwidth]{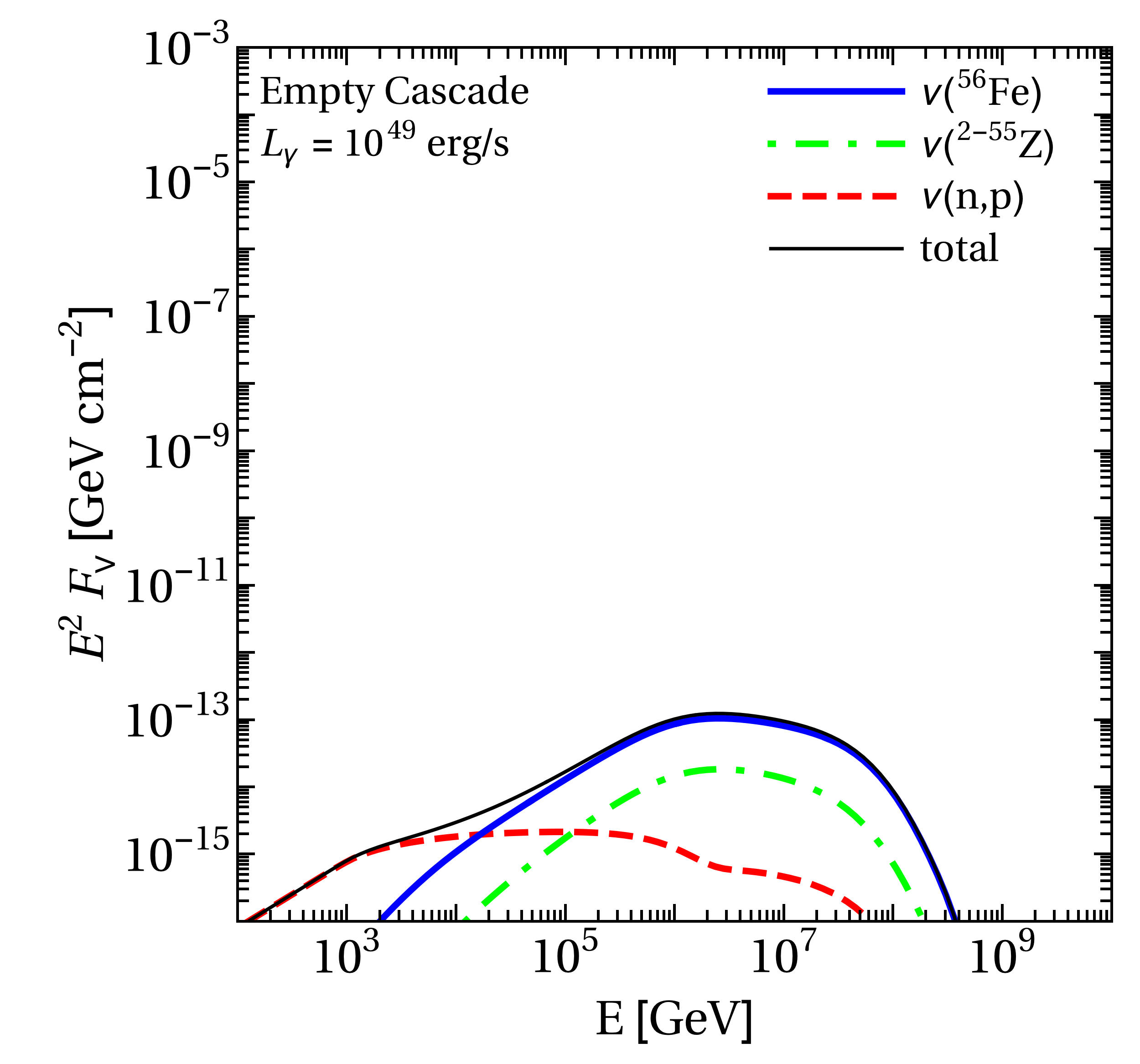}
\includegraphics[width=0.49\textwidth]{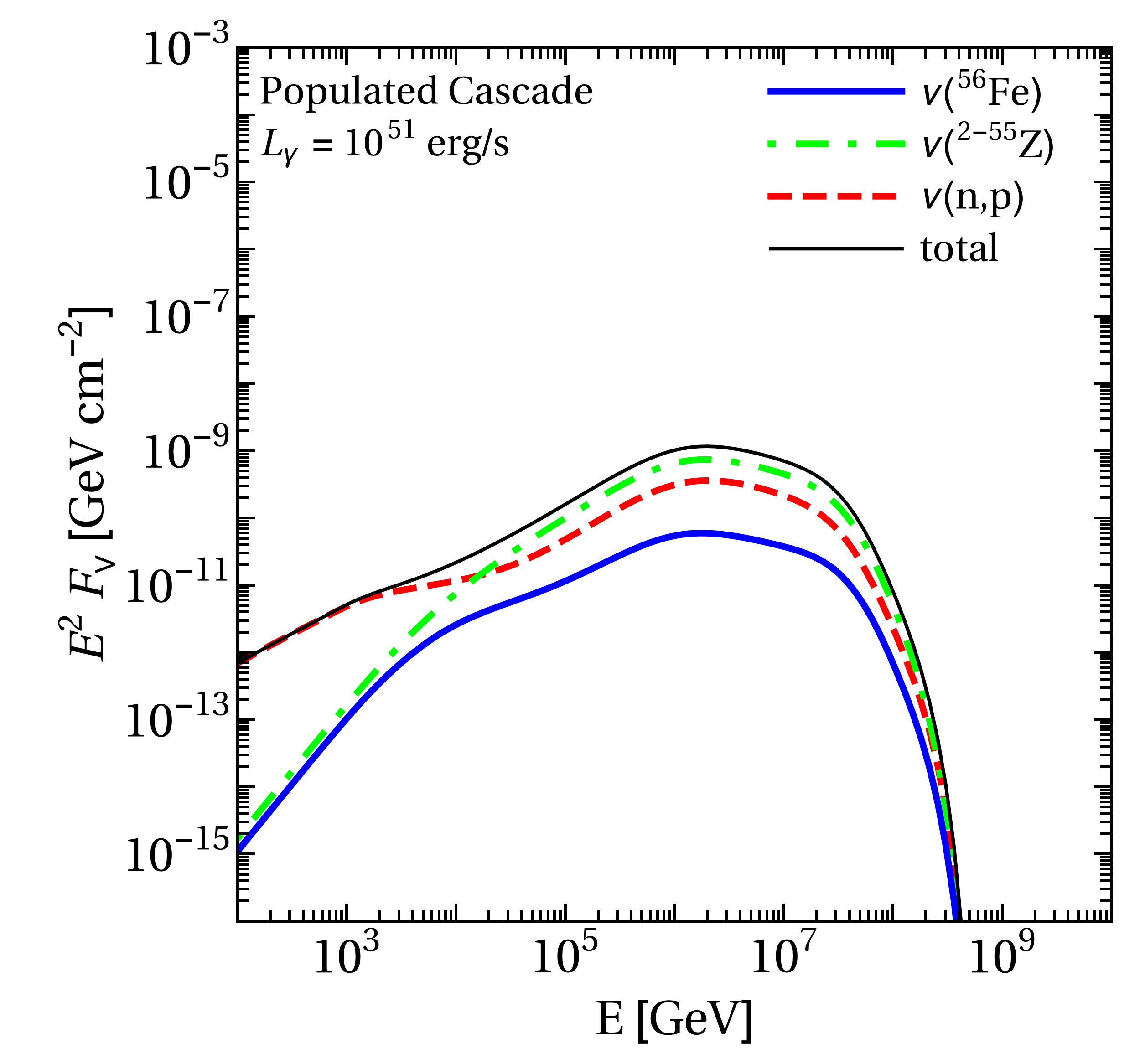}
\begin{minipage}[c]{0.5\textwidth}
\includegraphics[width=0.98\textwidth]{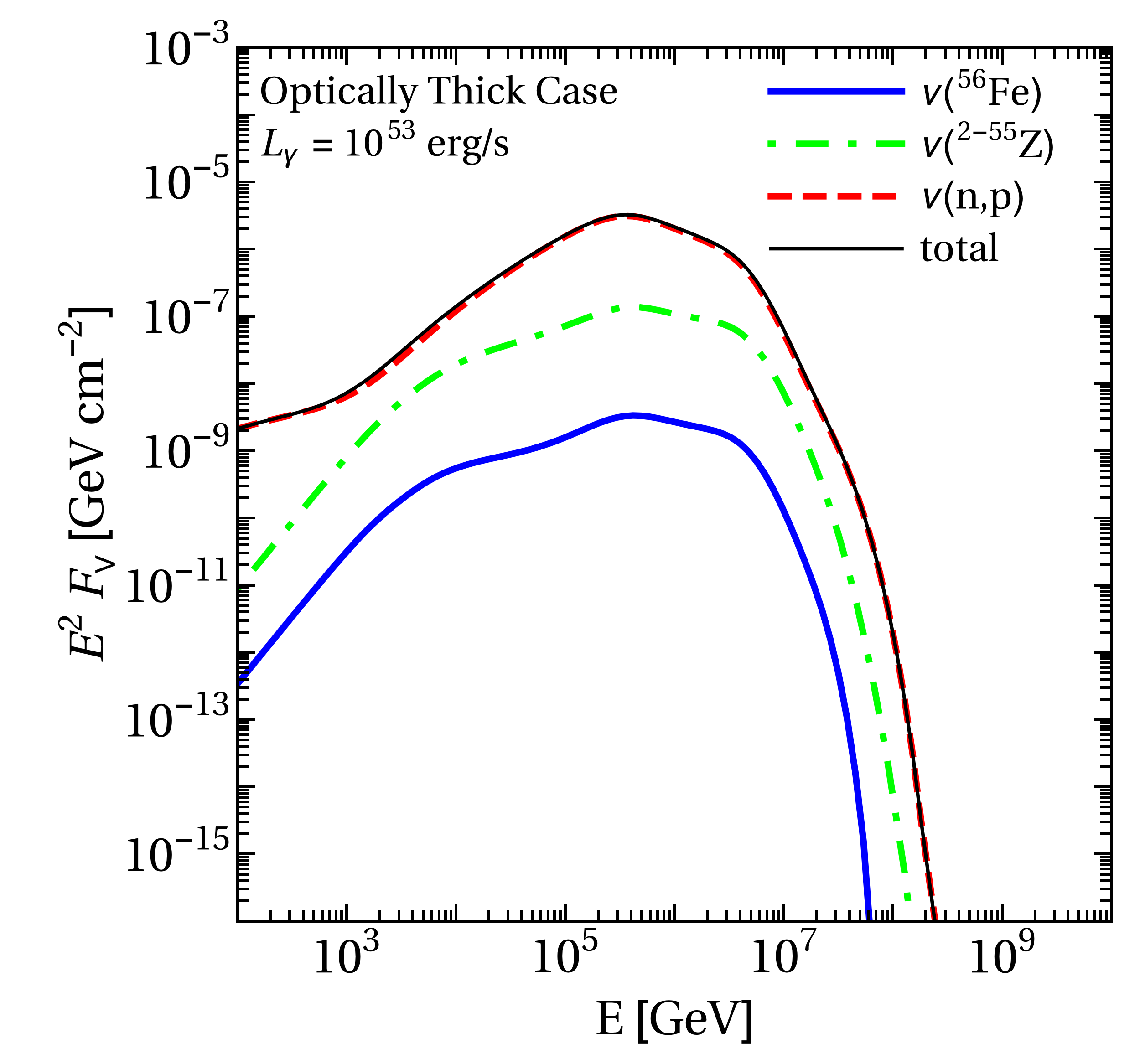}
\end{minipage}\hfill
\begin{minipage}[c]{0.45\textwidth}
\vspace{-3cm}
\caption{Contribution of primary and secondary isotopes to neutrino production for the injection of pure $^{56}$Fe. The panels show the total all-flavor neutrino fluence per shell of a GRB as a function of the energy in the observer's frame. Different luminosities (panels) correspond to the examples for the prototypes in \Sec~\ref{sec:classes}. The different curves show the contribution of the photo-meson production off the primary ($^{56}$Fe) and secondary isotopes produced in the photo-disintegration, where the proton/neutron contribution is separated.
}
\label{fig:nuflux}
\end{minipage}
\end{figure*}

Let us now study the neutrino production for the prototypes in \Sec~\ref{sec:classes}. \figu{nuflux} shows the all-flavor neutrino fluence per shell for the injection of pure $^{56}$Fe, split up by the contribution from the injected primary, all secondaries, and the interactions of nucleons produced in the cascade.
As indicated earlier, we have checked that the classification of the three regimes (Empty Cascade, Populated Cascade, Optically Thick Case) roughly corresponds to a classification according to the dominant contribution of the neutrino fluence: For the Empty Cascade (see upper left panel), the injected primary dominates the neutrino production; for the Populated Cascade, the contribution from all secondary isotopes produced in the nuclear cascade  dominates; and for the Optically Thick Case, the nucleons produced by nuclear disintegration dominate. The contribution from neutron decays is visible as a bump at low energies in all cases, but it never dominates the neutrino fluence. Note that the neutrino fluence grows quadratically with luminosity since photon and baryon density each scale with luminosity.

This separation of the contributions to the neutrino fluence is not only for conceptual reasons, but has more profound implications. As indicated in \Sec~\ref{sec:methods}, the photo-meson production off nucleons is relatively well known and has been studied in detail in the past with the SOPHIA interaction model. However, the photo-meson production off nuclei relies typically on a superposition model, which (in our case) scales the cross section $\propto A$. Therefore we {\em know} that the neutrino fluence prediction off nucleons (red curves in \figu{nuflux}) is more or less robust, whereas the other contributions carry large uncertainties to be quantified in the future. As a consequence, the neutrino prediction for the Populated Cascade (where the nucleon contribution is sub-dominant but large) and Optically Thick cases (where the nucleon contribution dominates) is more reliable than in the Empty Cascade case -- where however the neutrino flux is low anyways.

\begin{figure*}[ht!]
\includegraphics[width=0.49\textwidth]{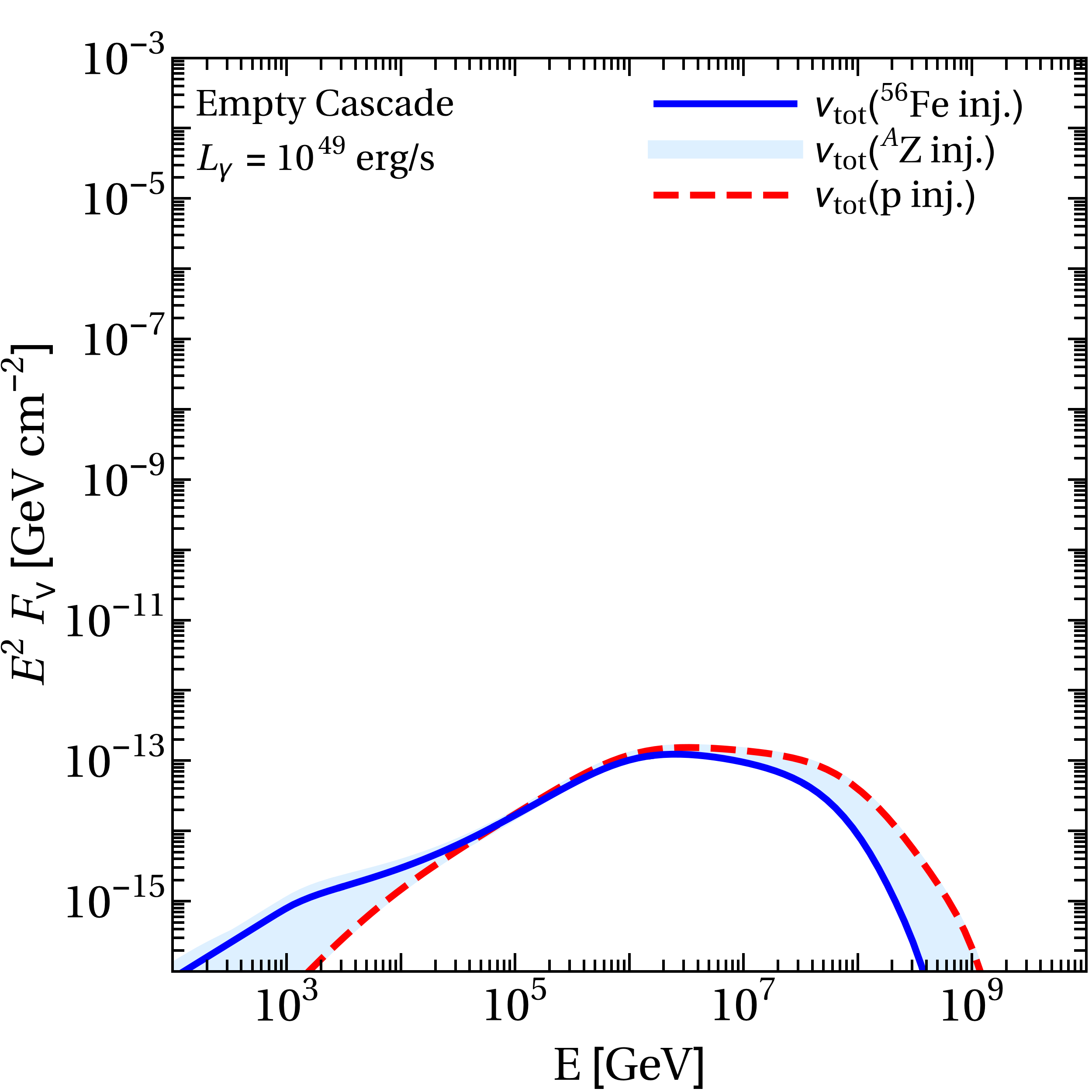}
\includegraphics[width=0.49\textwidth]{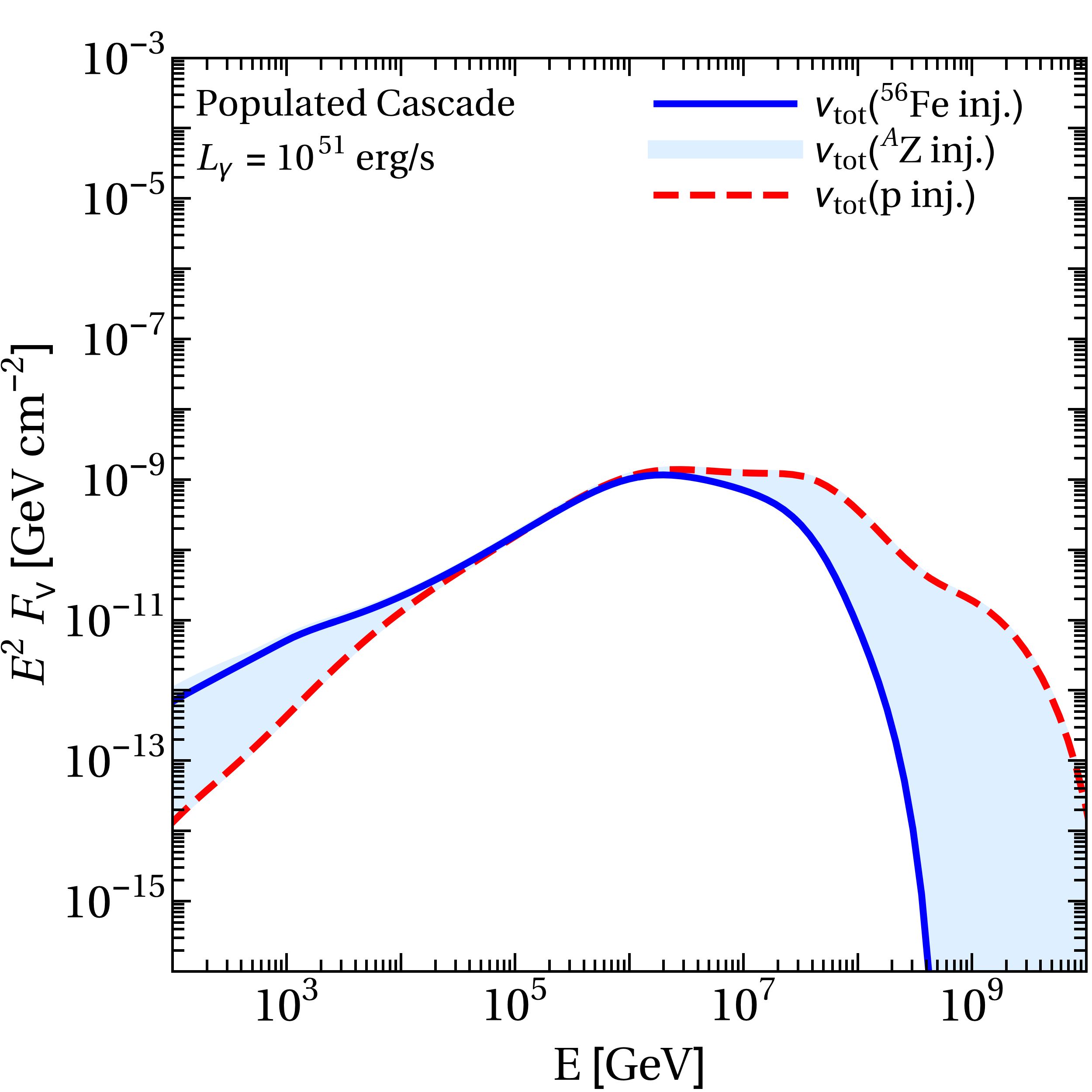}
\begin{minipage}[c]{0.5\textwidth}
\includegraphics[width=0.98\textwidth]{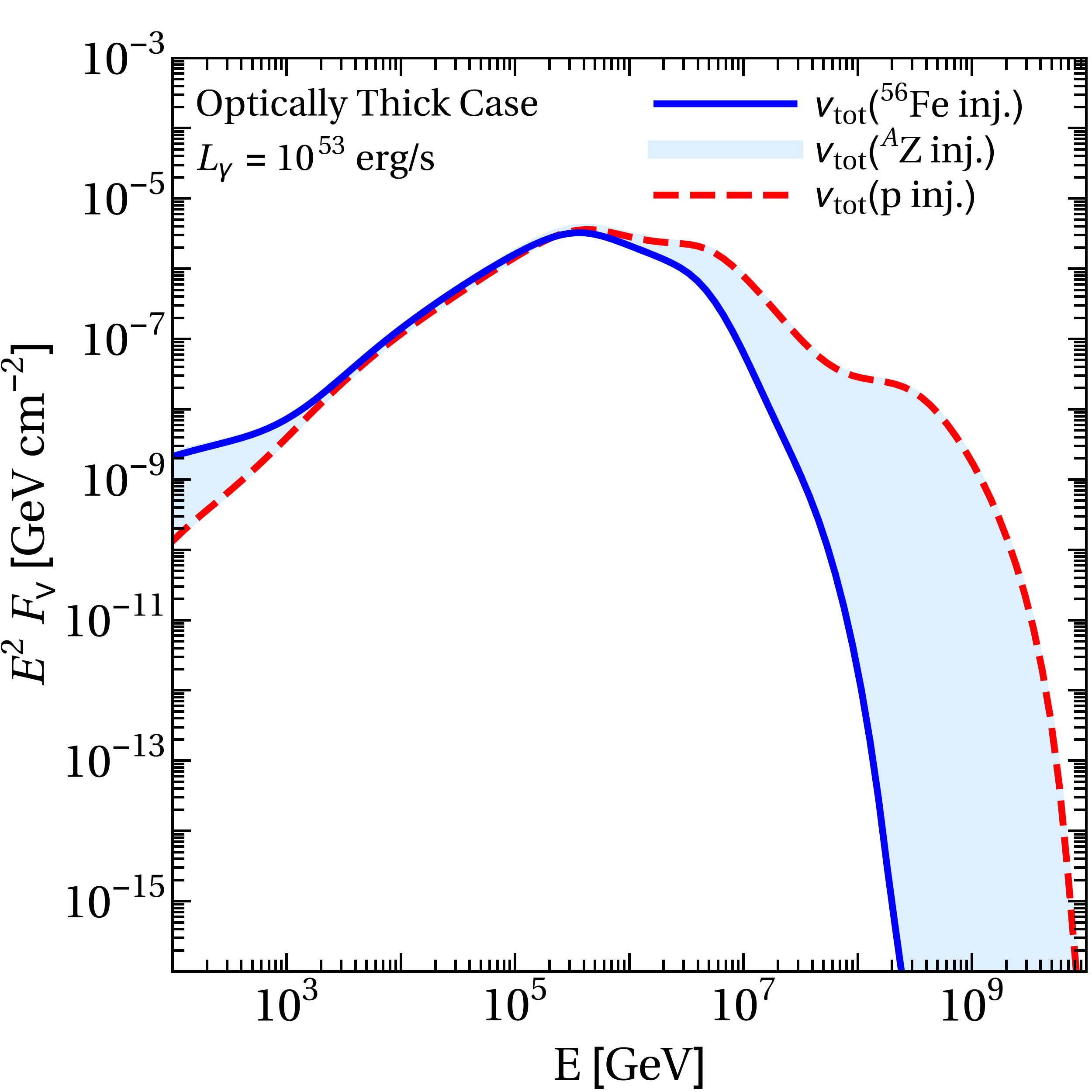}
\end{minipage}\hfill
\begin{minipage}[c]{0.45\textwidth}
\vspace{-3cm}
\caption{Impact of injection composition: Total all-flavor neutrino fluence per shell of a GRB as a function of the energy in the observer's frame. Different luminosities (panels) correspond to the examples for the prototypes in \Sec~\ref{sec:classes}. The different  curves correspond to the injection of pure isotopes: Protons, $^{56}$Fe, or any other isotope in between (see legend).
}
\label{fig:nucomp}
\end{minipage}
\end{figure*}

The impact of the injection isotope (for pure composition injection) is shown in \figu{nucomp}. That is, we inject the same luminosity of different isotopes, and we show the resulting neutrino fluence for pure proton, pure $^{56}$Fe, and pure other (intermediate injection isotope, shaded region) in the figure for the different prototypes.

As one of our most important results, the neutrino fluence at the peak hardly depends on the injection composition. 
This result is a consequence of several factors: Lorentz factor conservation in the disintegration, which splits up the nucleus into lighter nuclei or nucleons, the $E^{-2}$ injection spectrum, which conserves the energy per decade, the photo-meson interaction rate being almost flat beyond the threshold\footnote{One can, for instance, treat the nucleus as a superposition of nucleons and re-write the secondary production in terms of the nucleons, see \eg\ \Ref~\cite{Joshi:2013aua} for nucleus-nucleus interactions. Then one can see that the secondary production is roughly the same as before if the primary flux $\times$ interaction rate roughly scale $\propto E^{-2}$ and the photo-meson cross section $\sigma_{A \gamma} \simeq A \sigma_{p \gamma}$. Because of the Lorentz factor conservation in disintegration (there is effectively hardly any energy lost), disintegration does not affect this result.}, and strong magnetic field effects on the secondary muons, pions, and kaons, which determine the maximal energy cutoff.

Regarding the shape of the spectrum, the high-energy cutoff is somewhat higher for lighter injection isotopes because the maximal primary energy does not follow the Peters cycle \cite{Peters:1961} (rigidity-dependent maximal energy) for the Populated Cascade and Optically Thick cases (as we will discuss in \Sec~\ref{sec:composition}). This means that the maximal energy per nucleon, which affects the maximal neutrino energy, actually decreases. If this effect is stronger than the cutoff from magnetic field effects on the secondaries, it becomes visible in the neutrino fluence. Furthermore, the neutrino fluence from nuclei exhibits a stronger low energy component which mostly comes from neutron decays produced in the nuclear cascade.

\section{Description of cosmic-ray data}
\label{sec:cosmicrays}

In order to connect with cosmic-ray data, we extrapolate now from the previously studied single collision zone to a population of alike GRBs with a fixed duration of $\Delta T=10 \, \mathrm{s}$, which implies that the emission from each GRB comes from $\Delta T/t_v$ such collisions.
We perform a fit of UHECR observations from the Pierre Auger Observatory, combining the modeling of interactions in the source (computed as described in the previous sections) with the propagation of cosmic rays in the extragalactic space. For the propagation the {\it SimProp} code \cite{Aloisio:2012wj} has been used with Gilmore EBL \cite{Gilmore:2011ks} and PSB cross section model (as defined in \cite{Batista:2015mea}). We also consider the the extensive air shower in the Earth's atmosphere and EPOS-LHC \cite{Pierog:2013ria} is assumed for UHECR-air interactions.
We find that a good description of the data is obtained by distributing the GRBs as sources of cosmic rays following the star formation rate \cite{Yuksel:2008cu}, assuming a pure $^{28}$Si at the injection. The primary spectrum is described in \equ{targetA} and we use here $k=1.8$ and $P=2$. We fix the following parameters: source evolution, spectral index and cutoff shape at injection, and the nuclear species at the injection. In the present work, we keep this procedure as simple as possible, in order to show the power of the method, while leaving a more detailed analysis for a future work. The fit is performed above $10^{18}$ eV (Mixed Composition Dip Model) and above $10^{19}$ eV (Mixed Composition Ankle Model) by using the combined spectrum \cite{Valino:2015} and the shower depth ($X_{\mathrm{max}}$) distributions \cite{Aab:2014kda}, which contain information about the mass of the nucleus interacting with the atmosphere, with a similar procedure as used in \cite{diMatteo:2015,Aab:2016zth}.
A scan over $(R,L_{\gamma})$ is performed and for each pair the normalization to the experimental flux is found. For each point of the parameter space the number of expected prompt and cosmogenic neutrino events is calculated following \cite{Baerwald:2014zga}. For the prompt neutrino flux, the exposure for muon neutrinos is calculated by summing the exposure relative to the IceCube analyses of \cite{Aartsen:2015knd} (3 yr) and the one of \cite{Aartsen:2017wea} (3 yr) for a total of 1014 GRBs occurred in the Northern Hemisphere and the one of \cite{Aartsen:2017wea} (5 yr) for 664 GRBs occurred in the Southern Hemisphere, and comparing the total number of burts of the combined sample with the assumed 667 bursts per year as in \cite{Baerwald:2014zga}. For the cosmogenic neutrino flux, the exposure for muon neutrinos is taken from \cite{Aartsen:2016ngq}. The exclusion regions (90\% C.L.) of the prompt and cosmogenic neutrino analyses are calculated assuming both analyses as background free.

\subsection{Mixed Composition Dip Model}

\begin{figure*}[tp]
\includegraphics[width=0.6\textwidth]{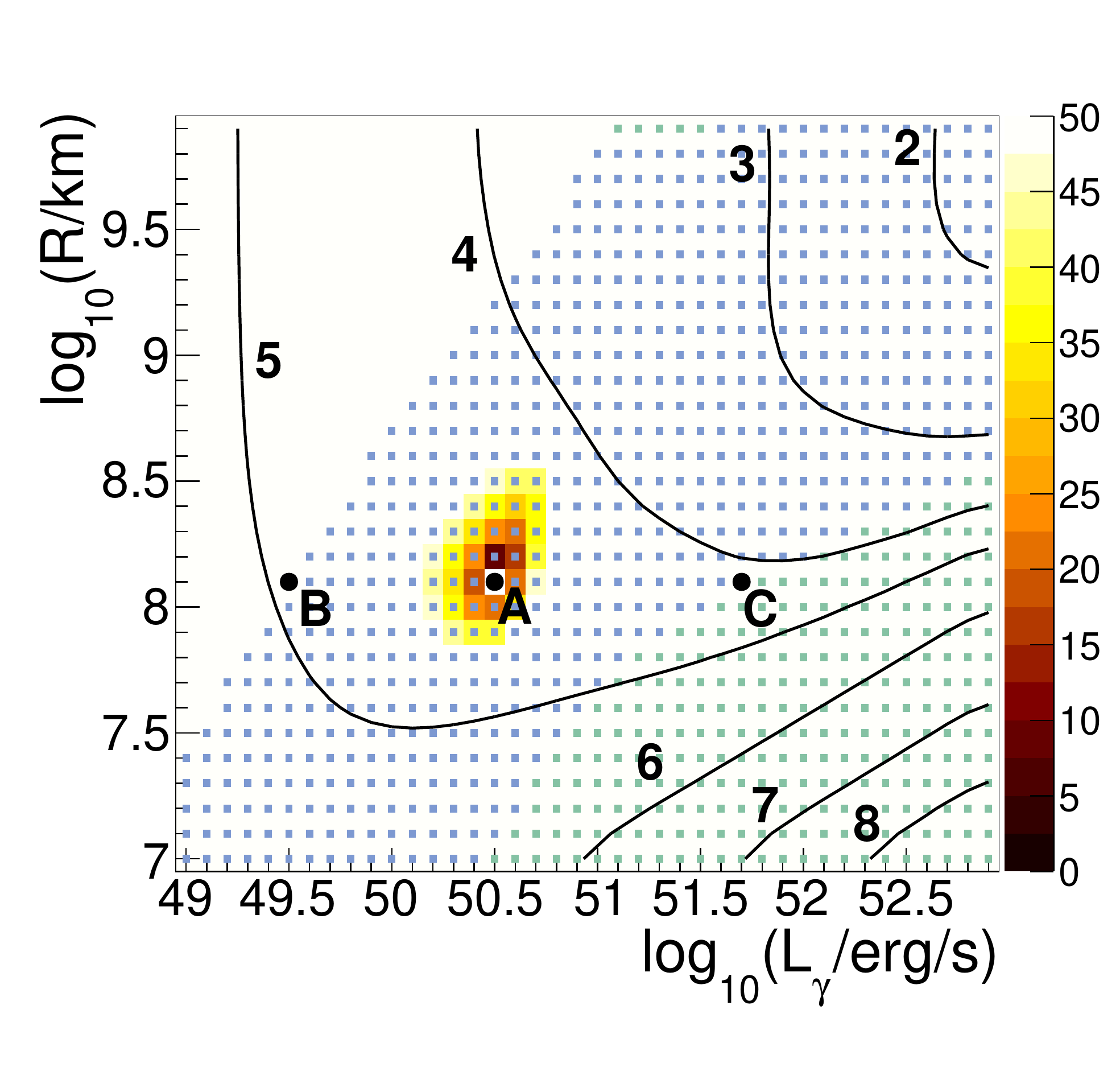}
\caption{Parameter space scan for Mixed Composition Dip Model in the internal shock scenario. Here $\sqrt{\chi^2-\chi^2_{\mathrm{min}}}$ is shown as a function of $R$ and $L_{\gamma}$ for the fit to cosmic ray data of the Pierre Auger Observatory \cite{Valino:2015,Aab:2014kda} above $10^{18}$ eV;  pure silicon injection with $k=1.8$ is assumed. The sources are distributed in the range from $z=0$ to 6 following the star formation rate (SFR). The blue squares mark the current (90\% C.L.) IceCube-excluded region from the GRB stacking analysis from Northern and Southern sky muon tracks \cite{Aartsen:2015knd,Aartsen:2017wea}, while the green ones mark the current (90\% C.L.) IceCube-excluded region from the cosmogenic neutrino analysis \cite{Aartsen:2016ngq}, applied to $\nu_\mu+\bar{\nu}_\mu$. The contrours show the nuclear loading ($\log_{10} \xi$).}\label{fig:bestSi18}
\end{figure*}

\begin{figure*}[tp]
\includegraphics[width=0.6\textwidth]{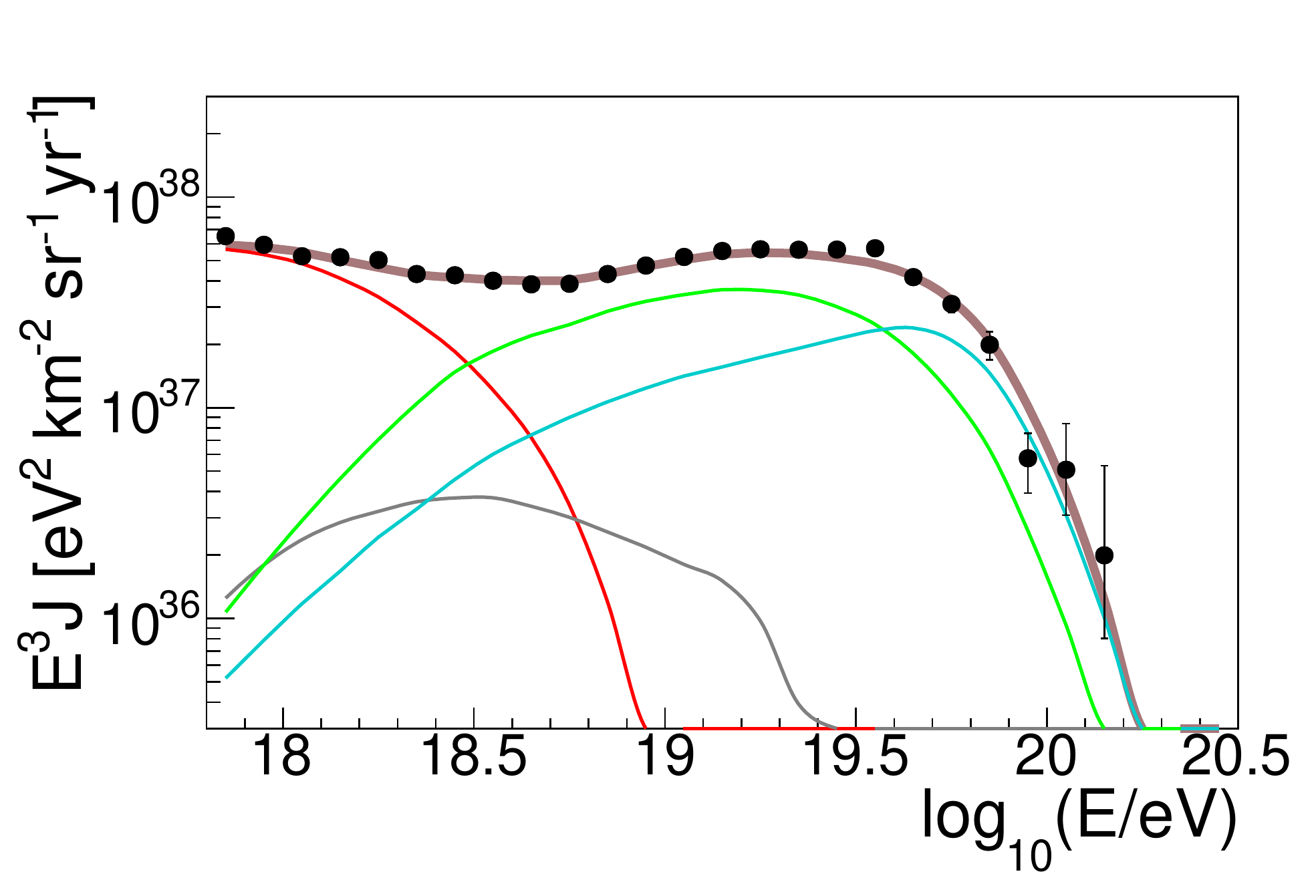}\\
\includegraphics[width=1\textwidth]{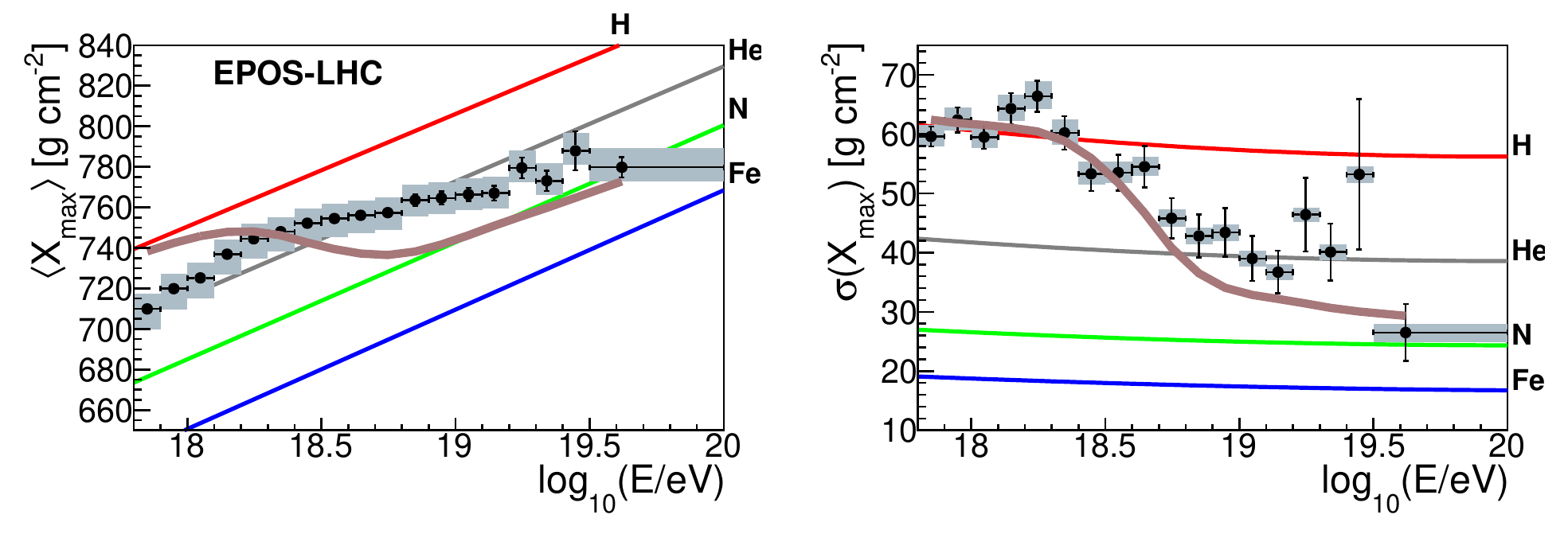}
\caption{Cosmic-ray observables obtained with the best-fit parameters in the Mixed Composition Dip Model, corresponding to point A for $(\log_{10}(R/\mathrm{km}),\log_{10}(L_{\gamma}/(\mathrm{erg/s})))=(8.1,50.5)$ in \figu{bestSi18}. Top: simulated energy spectrum of UHECR (multiplied by $E^3$) compared to data from \cite{Valino:2015}. Spectra at Earth are grouped according to the mass number as follows: $A = 1$ (red), $2 \leq A \leq 4$ (grey), $5 \leq A \leq 22$ (green), $23 \leq A \leq 28$ (cyan), total (brown). Bottom: average and standard deviation of the $X_{\mathrm{max}}$ distribution as predicted (assuming EPOS-LHC \cite{Pierog:2013ria} for UHECR-air interactions) for the model versus pure (${}^{1}$H (red), ${}^{4}$He (grey), ${}^{14}$N (green) and ${}^{56}$Fe (blue)) compositions, compared to data from \cite{Porcelli:2015}.}\label{fig:crSi18}
\end{figure*}

We first fit the spectrum and composition above $10^{18}$ eV, including the ankle region. The result of the fit is shown in \figu{bestSi18}, where the $\sqrt{\chi^2-\chi^2_{\mathrm{min}}}$ as a function of $R$ and $L_{\gamma}$ is given. The blue region indicates the current IceCube exclusion region from the prompt GRB analysis \cite{Aartsen:2017wea}, and the green marks refer to the cosmogenic neutrinos limits \cite{Aartsen:2016ngq}, both at the 90\% C.L. The best-fit spectrum at Earth and the composition observables are shown in \figu{crSi18}. The ankle feature is well reproduced without the need of additional components, since the interactions in the source naturally produce a light component at the lowest energies and a heavy one at the highest energies, allowing for the reproduction of the composition trend. However, we emphasize that the current procedure is not optimized to find the best description of the data, for which at least a mixed composition at the injection and a shift in the energy scale would be required.  
The contours of the baryonic loading in \figu{bestSi18} follow the behavior of the maximal energy, confirming \cite{Baerwald:2014zga}. The baryonic loading reaches extreme values in the region of the photosphere. 
The baryonic loading required at the best fit is $3\times10^4$. The best fit clearly lies in the Populated Cascade region, where the prompt neutrino flux is dominated by photo-meson production off the secondary isotopes in the cascade. This region is excluded by the GRB stacking analysis. In the Optically Thick region, the amount of nucleons emitted at the source is maximal due to the development of the cascade; as a consequence, this region is also excluded by the fit because of the expected cosmogenic neutrino flux. Since a number of parameters are fixed in our calculation, this observation should not be perceived too general. Moreover, the goodness of the fit including the ankle is low due to the very small statistical errors at the lowest energies and to the fact that in the current work the systematic uncertainties are not included. 

\begin{figure*}[tp]
\includegraphics[width=0.6\textwidth]{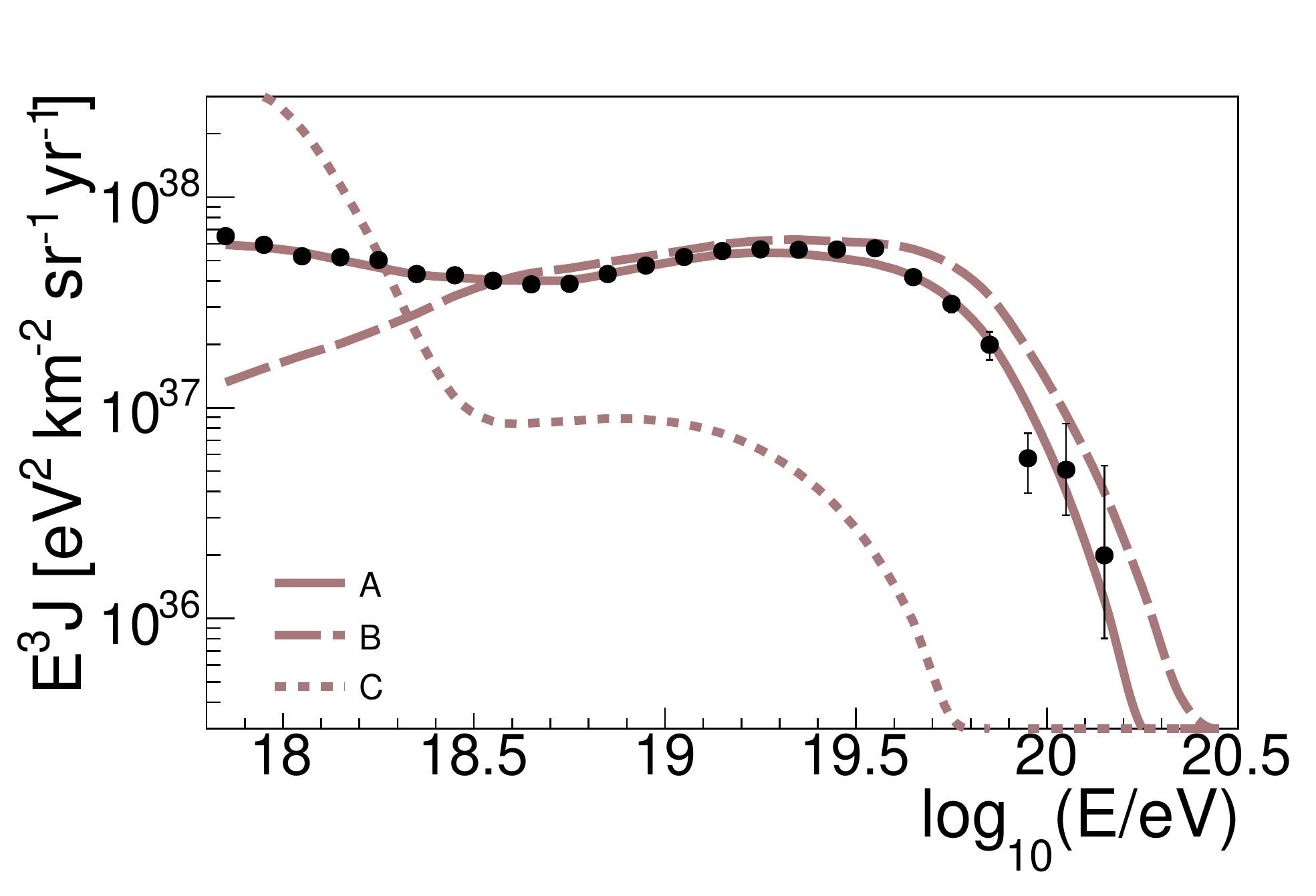}\\
\includegraphics[width=1\textwidth]{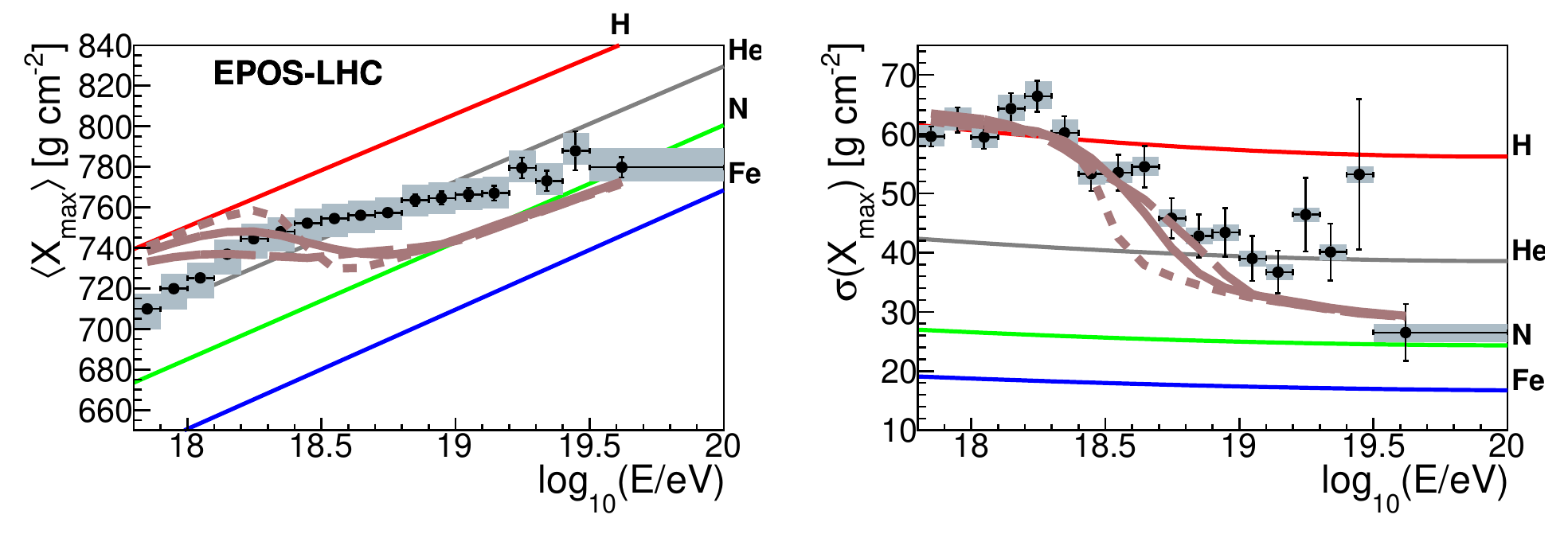}
\caption{Cosmic-ray spectra and composition observables  obtained for selected points in the parameter space in \figu{bestSi18} of the Mixed Composition Dip Model (similar to \figu{crSi18}). Here the points are: A: $(\log_{10}(R/\mathrm{km}),\log_{10}(L_{\gamma}/(\mathrm{erg/s})))=(8.1,50.5)$, B: $(8.1,49.5)$, C: $(8.1,51.7)$.}\label{fig:bench_crSi18}
\end{figure*}

\begin{figure*}[tp]
\includegraphics[width=0.5\textwidth]{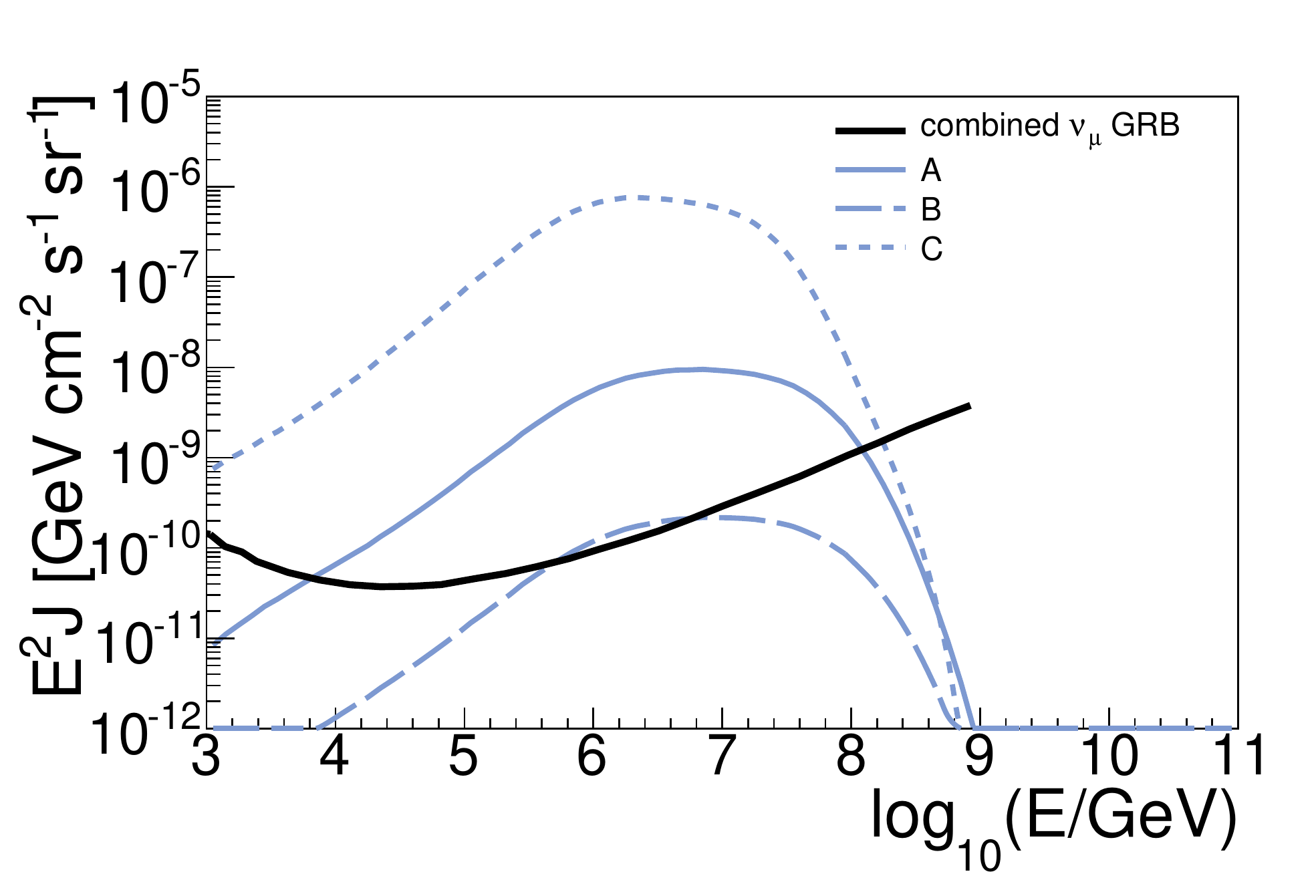}\hspace{-0.5cm}
\includegraphics[width=0.5\textwidth]{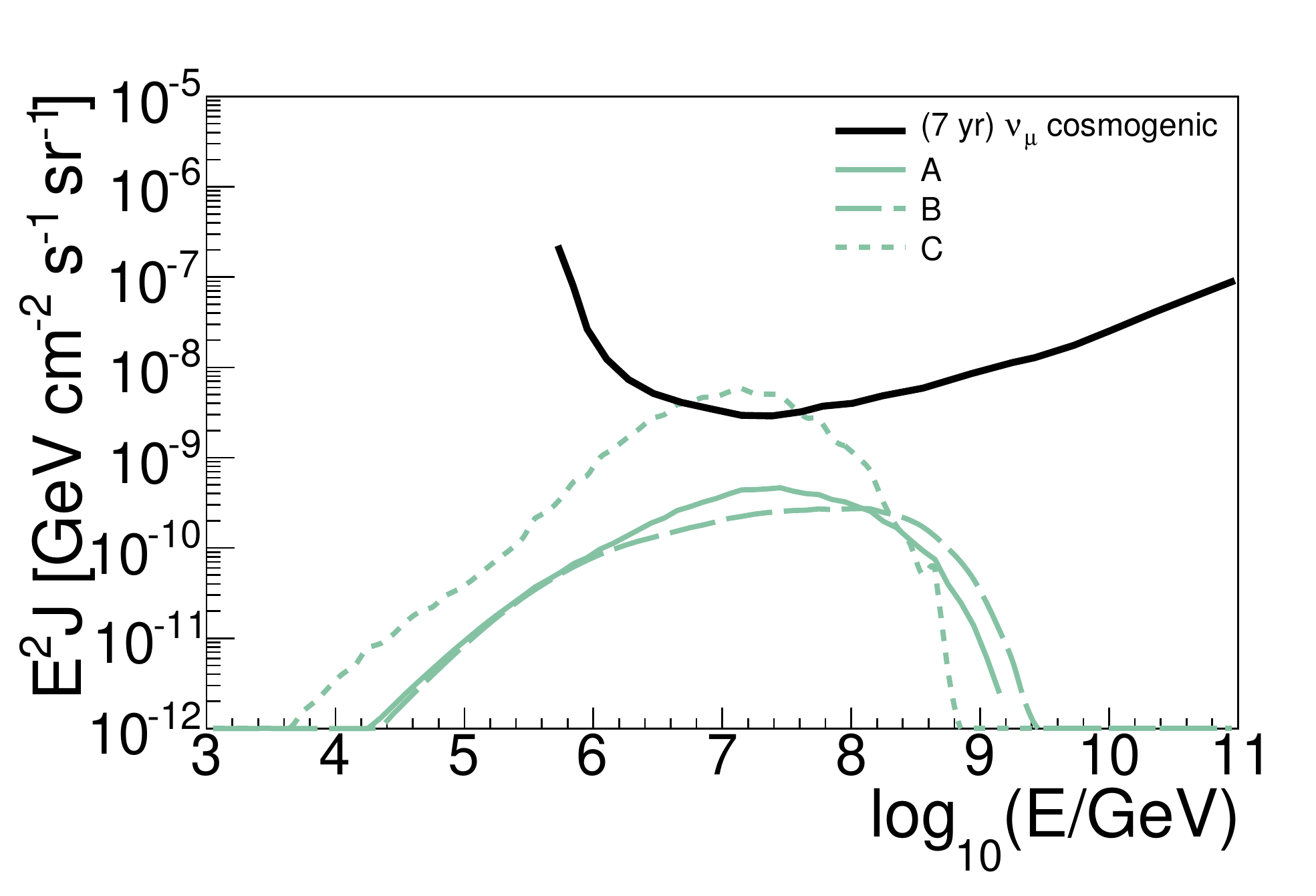}
\caption{Prompt and cosmogenic (muon flavor) neutrino spectra obtained for selected points in the parameter space in \figu{bestSi18} for the Mixed Composition Dip Model. Here we compare to  the differential limits obtained from \cite{Aartsen:2017wea} (considering the Northern exposure) and \cite{Aartsen:2016ngq}. The different points are: A: $(\log_{10}(R/\mathrm{km}),\log_{10}(L_{\gamma}/(\mathrm{erg/s})))=(8.1,50.5)$, B: $(8.1,49.5)$, C: $(8.1,51.7)$. Here the differential limits are defined as in \cite{Baerwald:2014zga}, such that following the differential limit curve for one decade in energy will yield one event.}\label{fig:bench_neuSi18}
\end{figure*}

In \figu{bench_crSi18}, the cosmic-ray spectra and composition observables are shown for selected points. Point A corresponds to the best fit, while points B and C correspond to cases with the same collision radius as the best-fit case, with a different choice for the luminosity. The B case is at the border between the Empty and Populated Cascade. This can be seen also in the lower energy part of the energy spectrum in \figu{bench_crSi18}, where the amount of propagated nucleons cannot account for the flux. The opposite case is represented by point C, where high photo-disintegration in the source results in a large contribution from nucleons at lower energies, which further increases during propagation in the extragalactic space. The composition observables (lower panels of \figu{bench_crSi18}) react on increasing luminosity at the source with a sharper transition from light to heavy masses as energy rises.  
In \figu{bench_neuSi18} the corresponding prompt and cosmogenic neutrino fluxes are shown. The best-fit case (A) is excluded by the prompt neutrino flux (left panel), but it is still allowed within the cosmogenic neutrino limits (right panel). The prompt flux clearly shows the discussed enhancement as a function of the luminosity of the sources. The cosmogenic flux reaches the IceCube limit only when approaching the Optically Thick Case (C), because of the amount of nucleons injected in the extragalactic space.
 
\subsection{Mixed Composition Ankle Model}

\begin{figure*}[tp]
\includegraphics[width=0.6\textwidth]{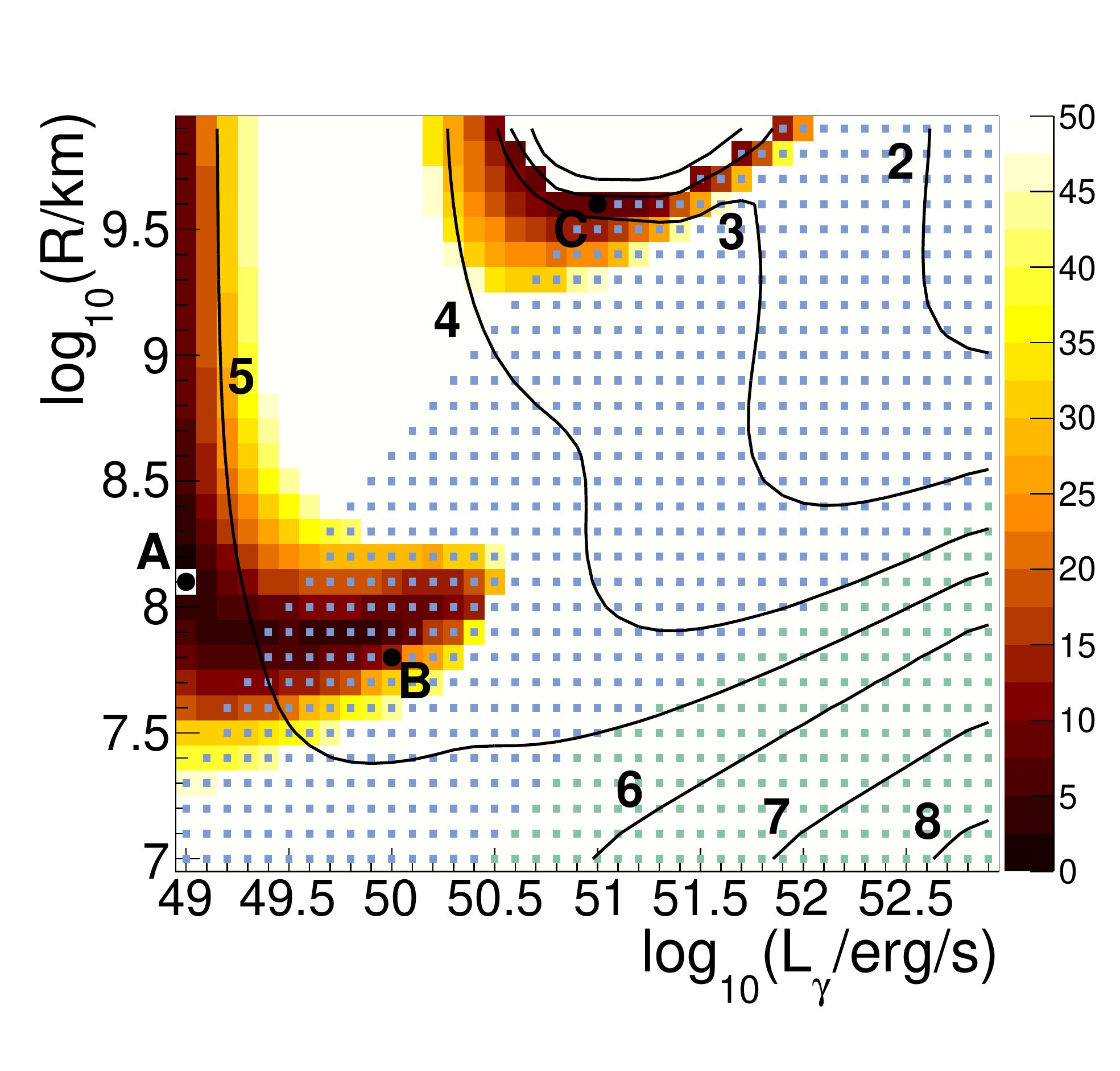}\\
\caption{Parameter space scan for Mixed Composition Ankle Model in the internal shock scenario. Here $\sqrt{\chi^2-\chi^2_{\mathrm{min}}}$ is shown as a function of $R$ and $L_{\gamma}$ for the fit to cosmic-ray data of the Pierre Auger Observatory \cite{Valino:2015,Aab:2014kda} above $10^{19}$ ev, including a penalty for the overshooting of the flux at lower energies. See caption of \figu{bestSi18} for details.}\label{fig:bestSi19}
\end{figure*}

\begin{figure*}[tp]
\includegraphics[width=0.6\textwidth]{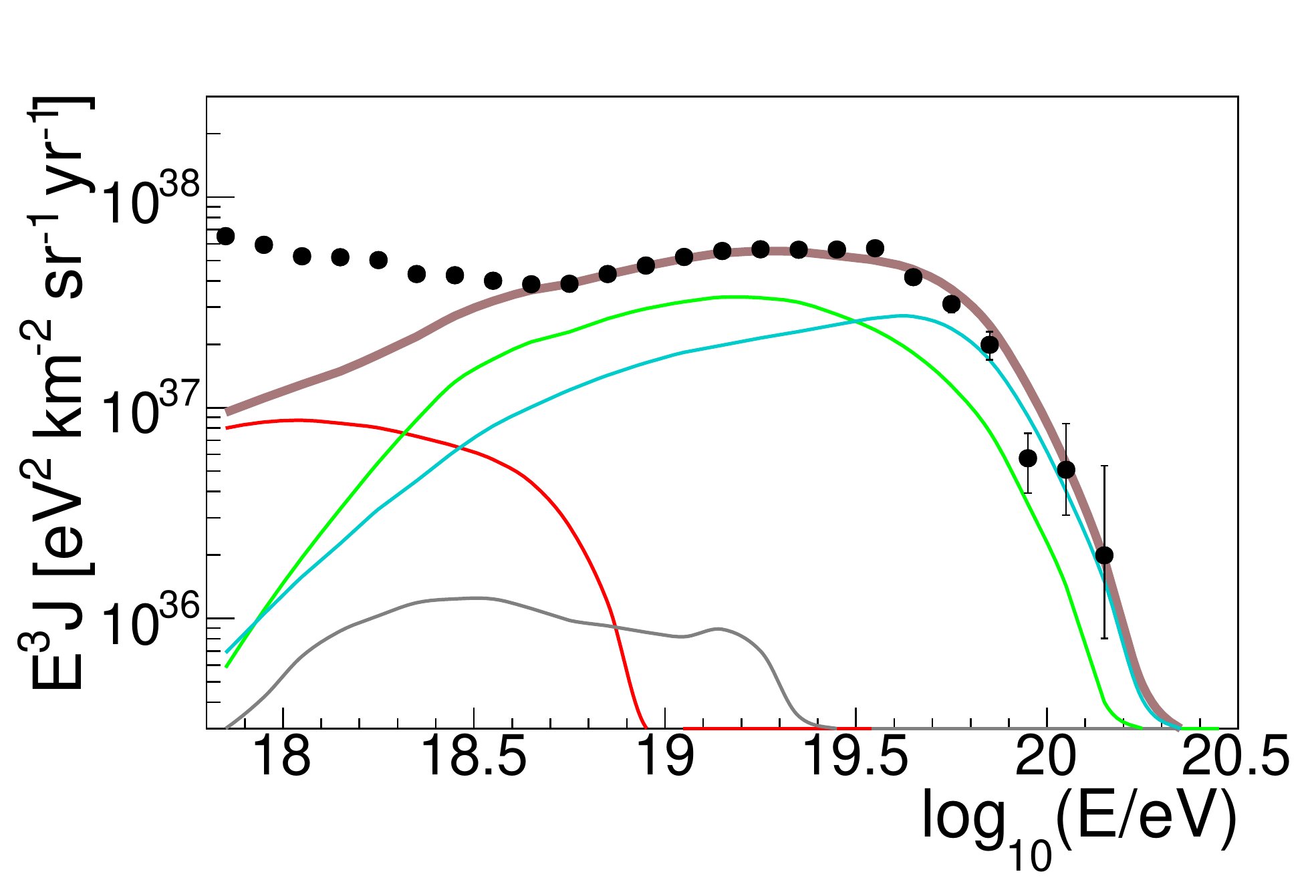}\\
\includegraphics[width=1\textwidth]{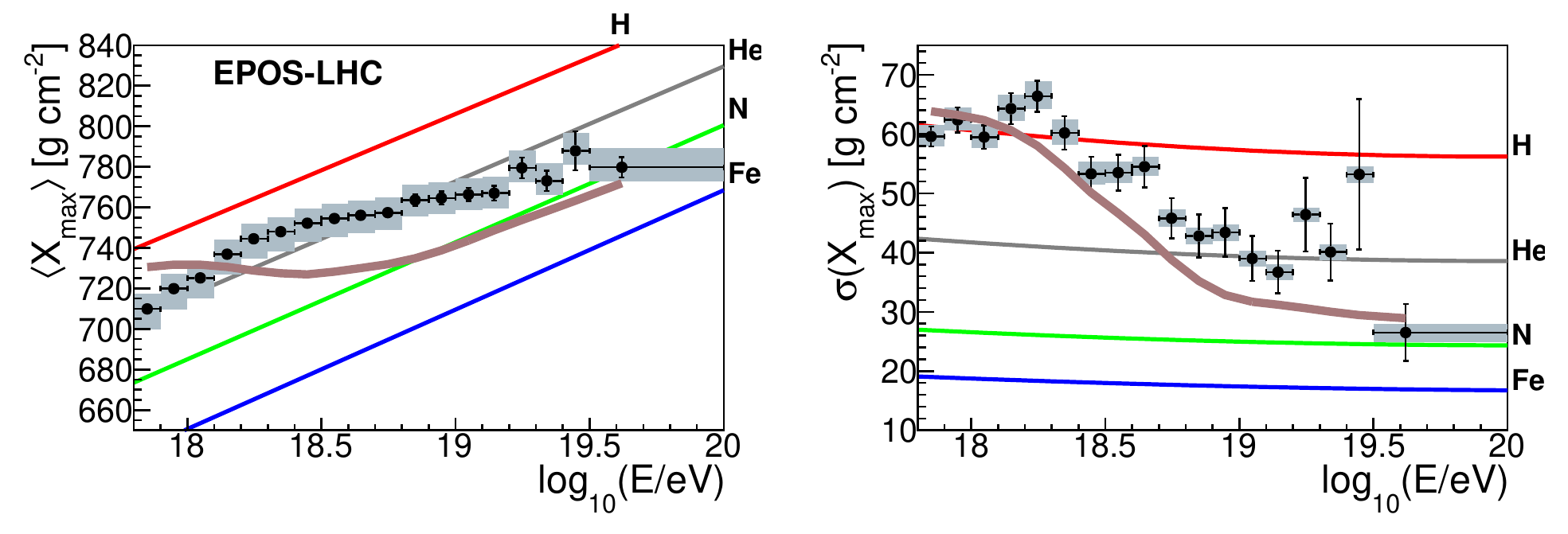}
\caption{Cosmic-ray observables obtained with the best-fit parameters in the Mixed Composition Ankle Model, corresponding to point A $(\log_{10}(R/\mathrm{km}),\log_{10}(L_{\gamma}/(\mathrm{erg/s})))=(8.1,49)$ in \figu{bestSi19}. See caption of \figu{crSi18} for details.}\label{fig:crSi19}
\end{figure*}

The fit is performed above $10^{19}$ eV, with a penalty for overshooting the flux at lower energies. The goodness of fit is here enhanced with respect to the Mixed Composition Dip Model due to the absence of the data points at the lowest energies. The best fit is found at low source luminosity and intermediate collision radius (point A in \figu{bestSi19}). These parameters are not excluded by the existent neutrino limits. This case corresponds to the Empty Cascade: the protons in the energy spectrum at Earth (\figu{crSi19}, upper panel) come, therefore, mainly from propagation in the extragalactic space. 
The obtained GRB parameters require either very low $\gamma$-ray luminosities (points~A and B), which are significantly lower than expected from $\gamma$-ray observations (see \Ref~\cite{Atteia:2017dcj} for an overview), or large collision radii (point~C), which may be indicative for magnetic reconnection models.
The transition of the mass composition from light to heavy is less sharp than in the best fit for the Mixed Composition Dip Model (compare the lower panels of \figu{crSi18} and \figu{crSi19}).

\begin{figure*}[tp]
\includegraphics[width=0.6\textwidth]{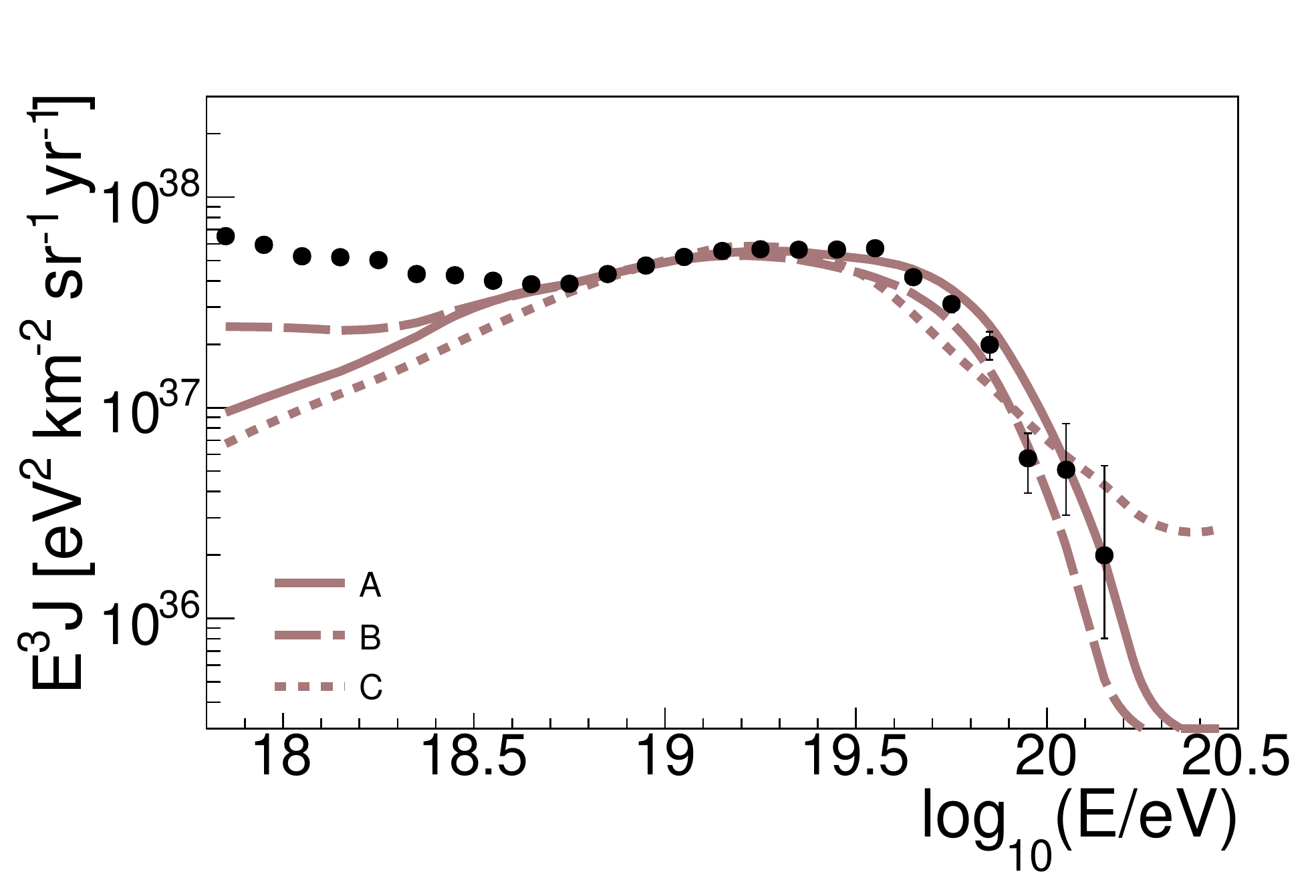}
\includegraphics[width=1\textwidth]{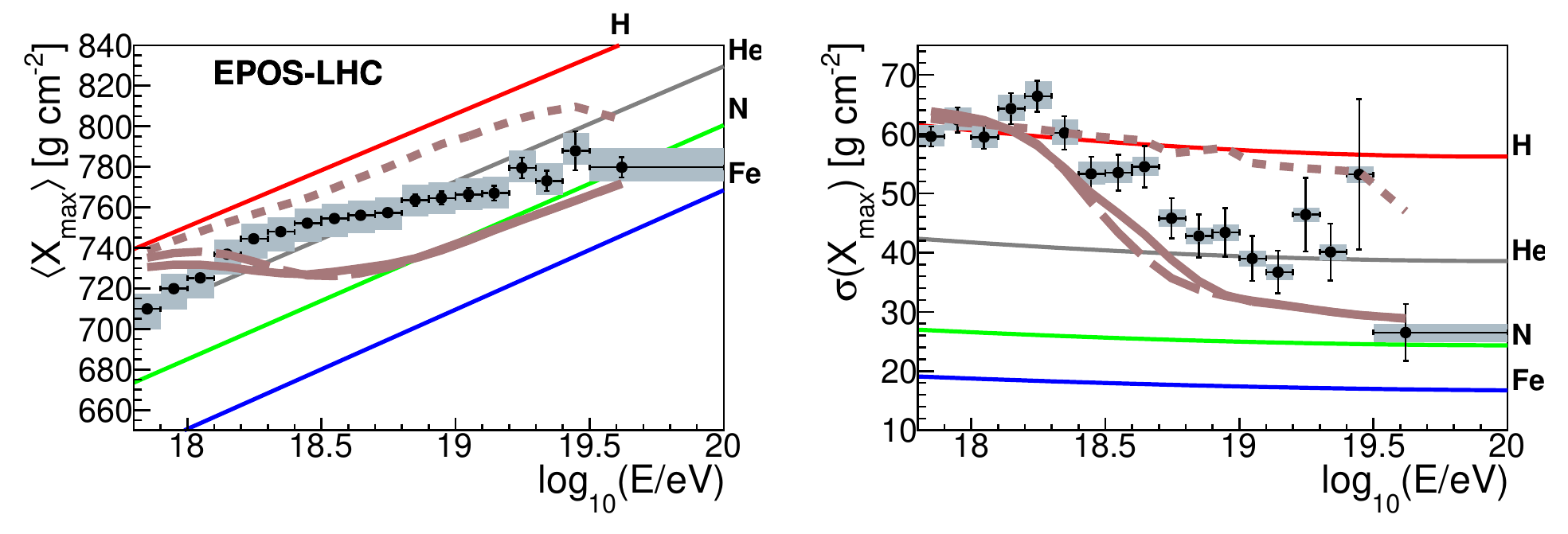}
\caption{Cosmic-ray spectra  and composition observables  obtained for selected points in the parameter space in \figu{bestSi19} of the Mixed Composition Dip Model (similar to \figu{crSi19}). Here the points are:
 A: $(\log_{10}(R/\mathrm{km}),\log_{10}(L_{\gamma}/(\mathrm{erg/s})))=(8.1,49)$, B: $(7.8,50)$, C: $(9.6,51)$.}\label{fig:bench_crSi19}
\end{figure*}

\begin{figure*}[tp]
\includegraphics[width=0.5\textwidth]{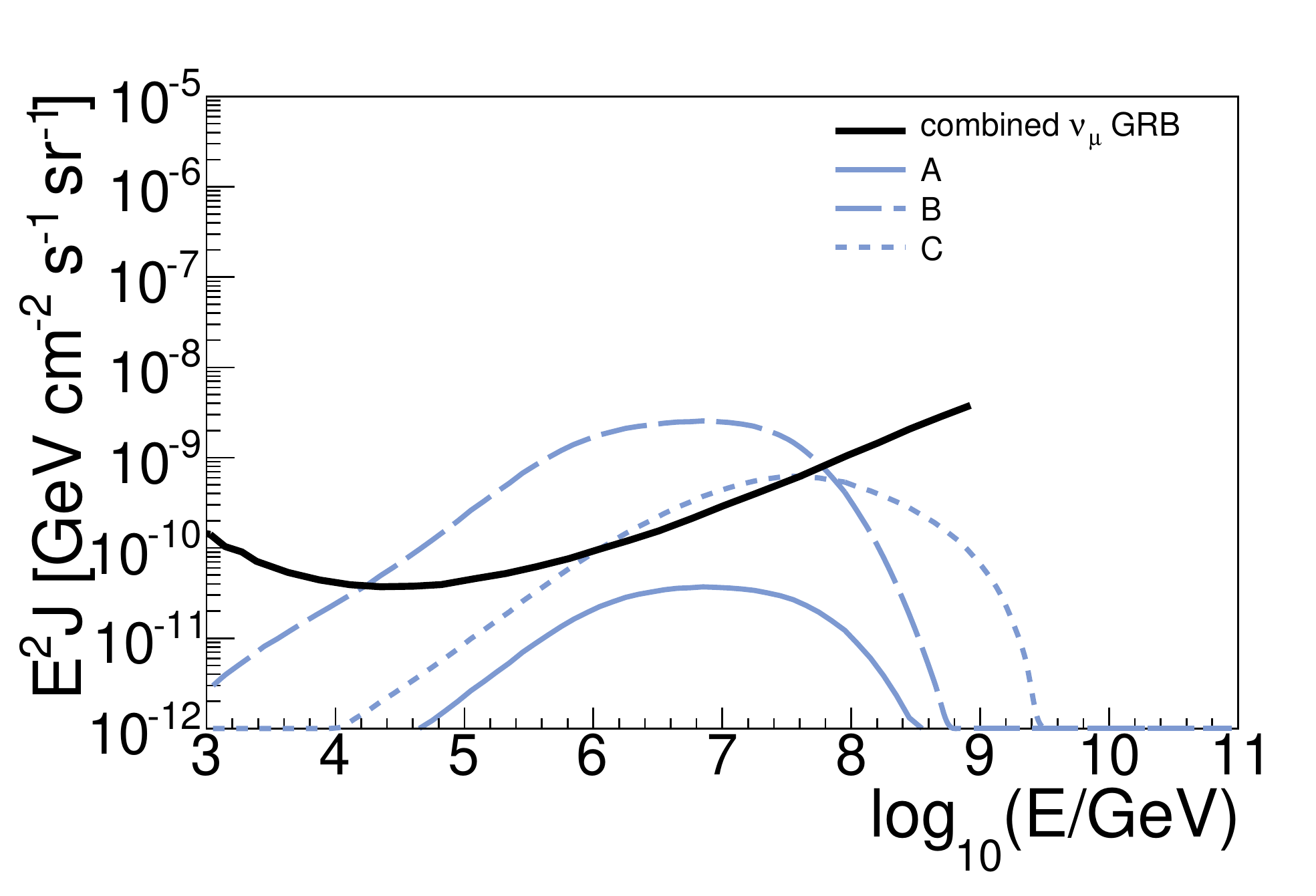}
\hspace{-0.5cm}
\includegraphics[width=0.5\textwidth]{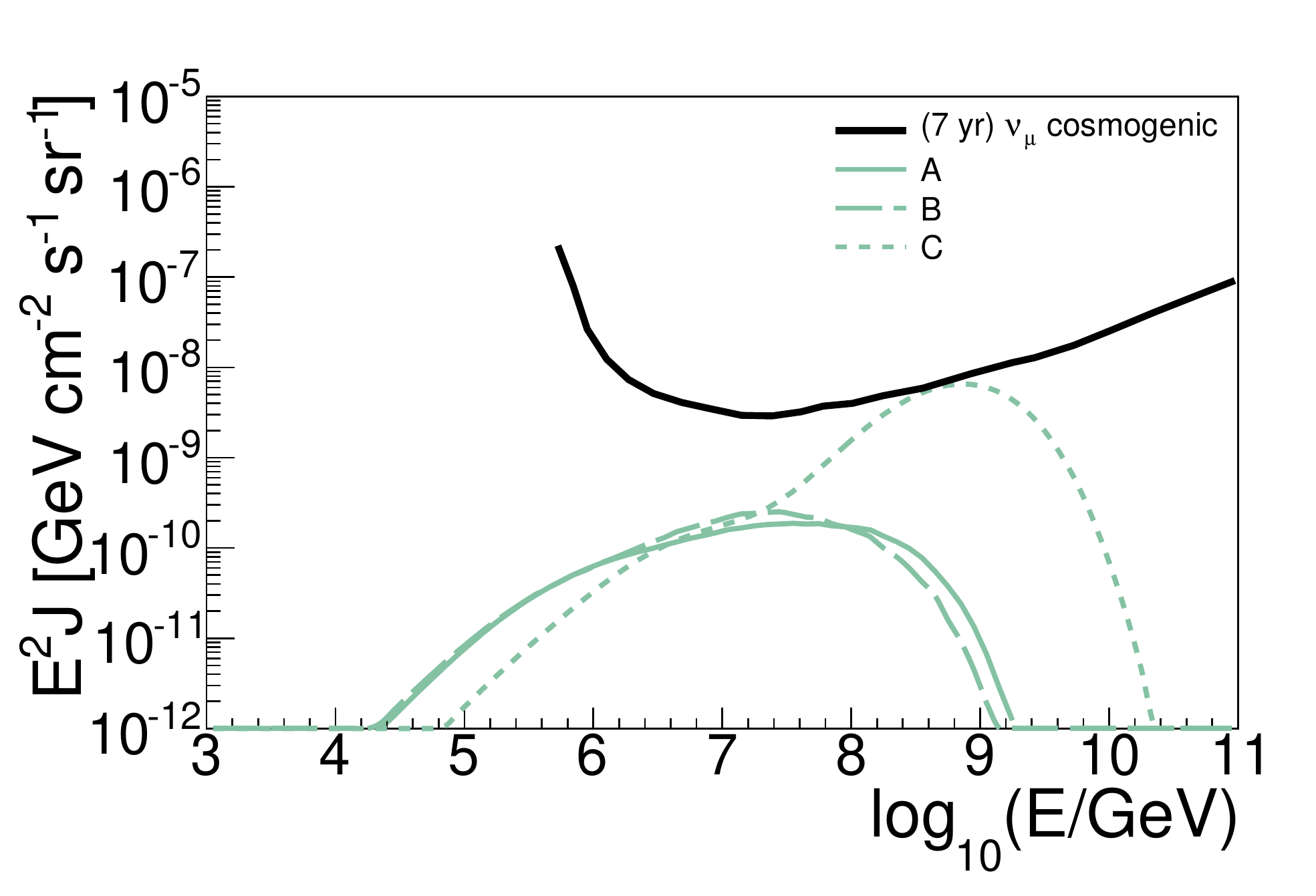}
\caption{Prompt and cosmogenic (muon flavor) neutrino spectra obtained for selected points in the parameter space in \figu{bestSi19} for the Mixed Composition Ankle Model. Here A: $(\log_{10}(R/\mathrm{km}),\log_{10}(L_{\gamma}/(\mathrm{erg/s})))=(8.1,49)$, B: $(7.8,50)$, C: $(9.6,51)$.}\label{fig:bench_neuSi19}
\end{figure*}

We also show in \figu{bench_crSi19} the energy spectra and the composition observables for other selected points in \figu{bestSi19}. Point B reproduces the cosmic ray spectrum in a reasonable way, but with a sharper transition from light to heavy masses. The left panel of \figu{bench_neuSi19} shows than enhancing the photo-disintegration with a higher source luminosity results in overshooting of the prompt neutrino limits, while the cosmogenic neutrino flux does not substantially change before reaching the Optically Thick Case. We also show point C, which represents an intermediate luminosity and a high collision radius. In this case, photo-disintegration in the source is not  efficient and results in a high maximal energy of the primary at the source (compare with \figu{nuscan}), and propagation will lead to very energetic protons at Earth. This can be seen in \figu{bench_neuSi19}, where $\sigma(X_{\mathrm{max}})$ is almost flat. The same message is given by the corresponding cosmogenic flux shown in the right panel of \figu{bench_neuSi19}.

\section{Impact of Injection Composition}
\label{sec:composition}

So far, we have tested several selected injection isotopes: $^{16}$O and $^{56}$Fe to describe the disintegration in a GRB shell as kind of extreme examples for nuclei, and $^{28}$Si for a reasonable fit to Auger data. In this section, we qualitatively discuss the dependence on the injection element ($Z$) in order to illustrate what changes to expect for different injection isotopes, why $^{28}$Si is a good example to describe Auger data, and why the best-fit parameters are found in the regions determined in the previous section. Note that we will show the results as a function of $Z$, implying that we inject the most abundant stable isotope. 

\begin{figure*}[t]
\includegraphics[width=0.49\textwidth]{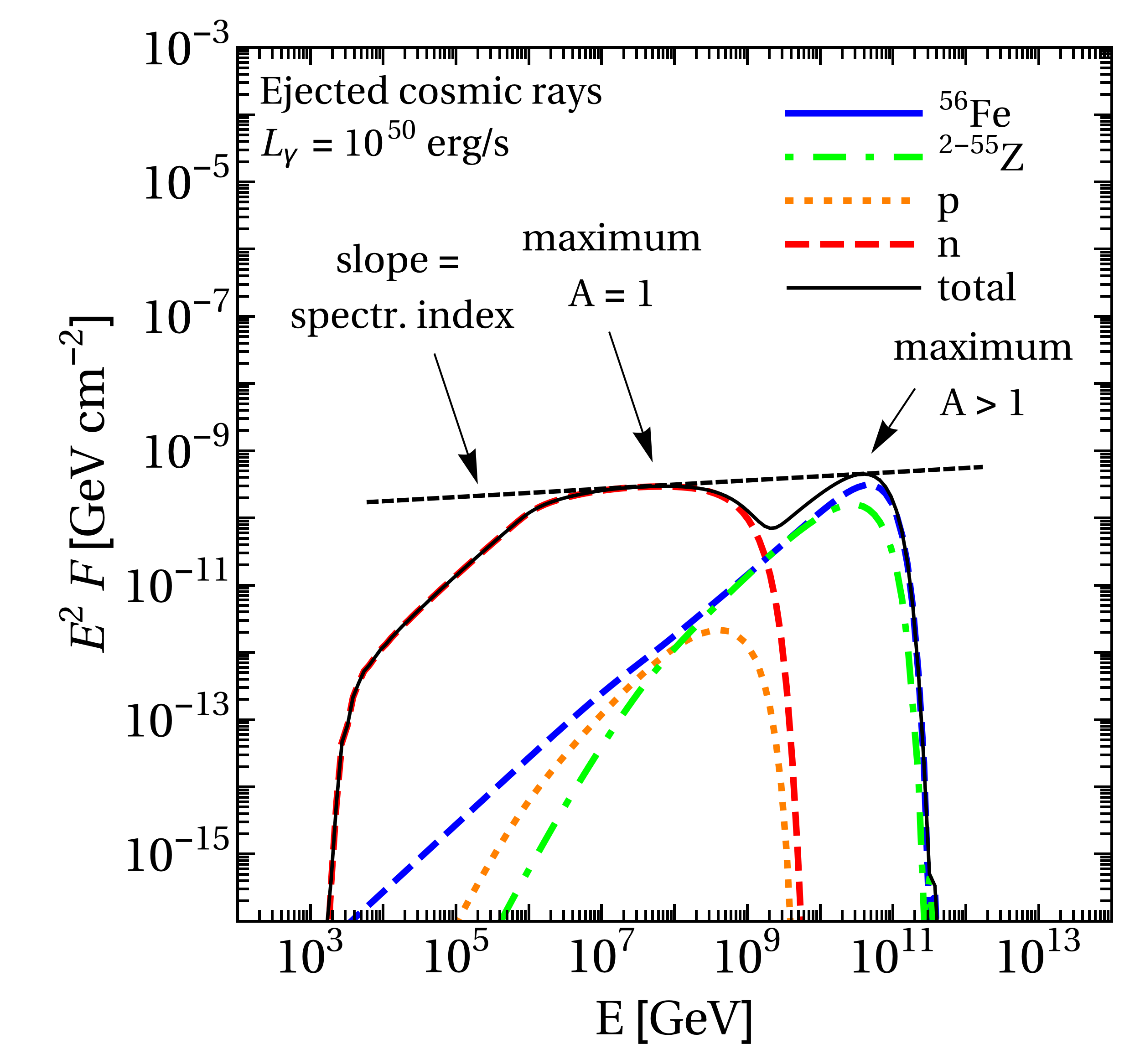}
\includegraphics[width=0.45\textwidth]{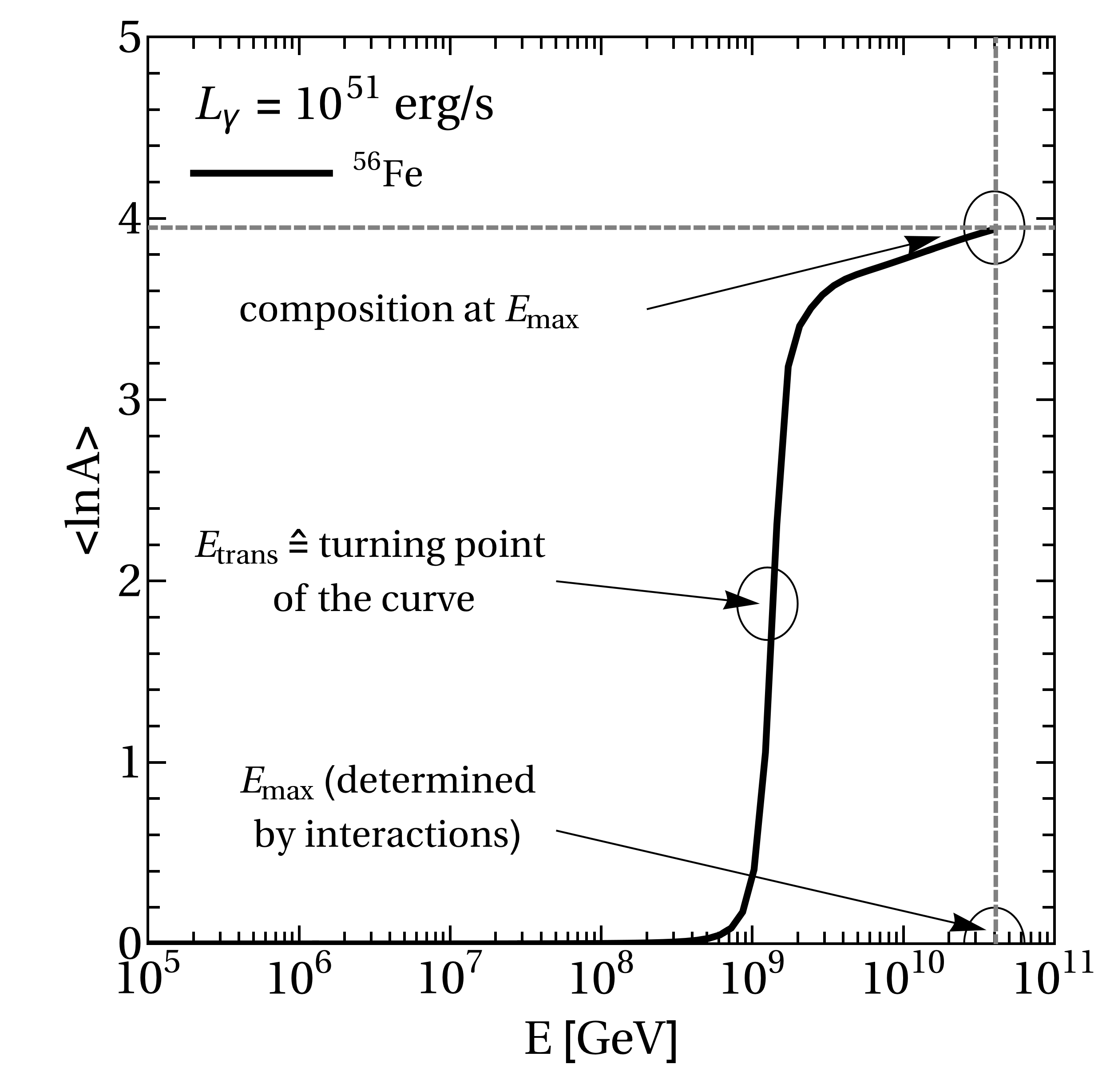}
\caption{Illustration of quantities used for qualitative analysis using one example. Left panel: ejected cosmic ray spectrum as a function of (observer's frame) energy. Right panel: ejected composition $\langle \ln A \rangle$ as a
function of (observer's frame) energy. Here adiabatic losses are taken into account, but no interactions in CMB and EBL.
The left panel illustrates how to determine the spectral index, the right panel the composition at $E_\text{max}$ and the transition energy $E_\text{trans}$. The maximum energy $E_\text{max}$ is determined by acceleration and loss processes.
}
\label{fig:methods}
\end{figure*}

We anticipate that the UHECR ejection spectrum from the sources (injection spectrum into the intergalactic medium) can be qualitatively characterized by four different estimators, see \figu{methods} for illustration:
\begin{description}
 \item[Spectral shape.] We assume that the ejection spectrum can be roughly described by a power law with a certain spectral index. Since both the neutron and charged cosmic rays contribute to the ejected spectrum, the overall spectrum  will be roughly determined by the peaks of these two contributions (see left panel).
 \item[Maximal energy.] The ejection spectrum will have a characteristic maximal energy $E_{\mathrm{max}}$. This maximal energy is determined by the equilibrium between acceleration rate and the sum of energy loss rates, see \Sec~\ref{sec:injection}.
 \item[Composition at $\boldsymbol{E_{\mathrm{max}}}$.] The composition at the maximal energy will be relevant to describe the observed composition at the highest energy, see right panel for illustration.
 \item[Transition energy to heavier composition.] If the ankle is to be described by the GRBs, the transition energy $E_{\mathrm{trans}}$ has to be in the right ballpark. It is determined from the second derivative of the composition curve, see right panel.
\end{description}
We do not discuss the cosmological source evolution as a separate parameter here, although it is well known that there is a certain degeneracy between source evolution and spectral ejection index.

\begin{figure*}[t]
\includegraphics[width=0.48\textwidth]{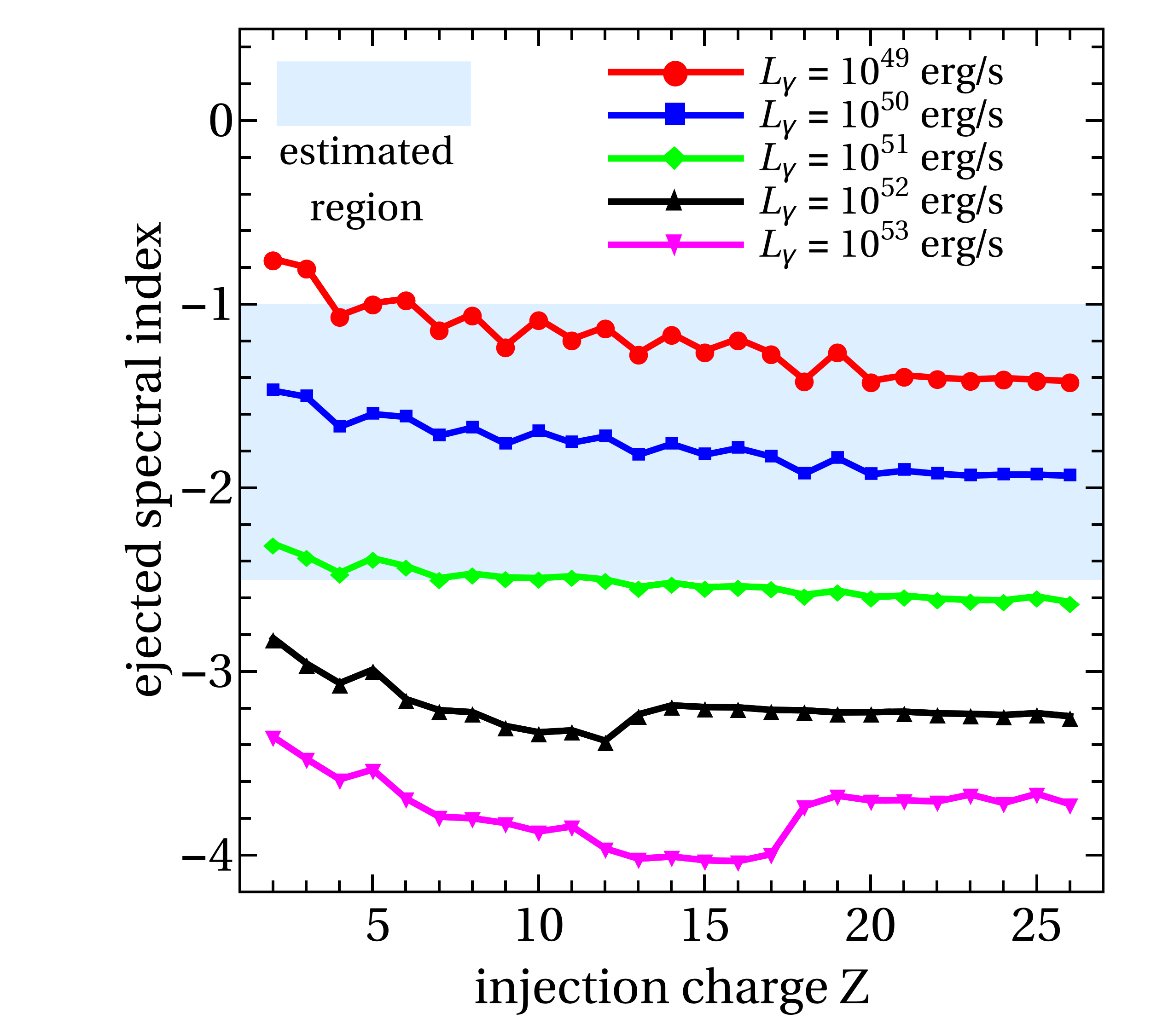}
\includegraphics[width=0.48\textwidth]{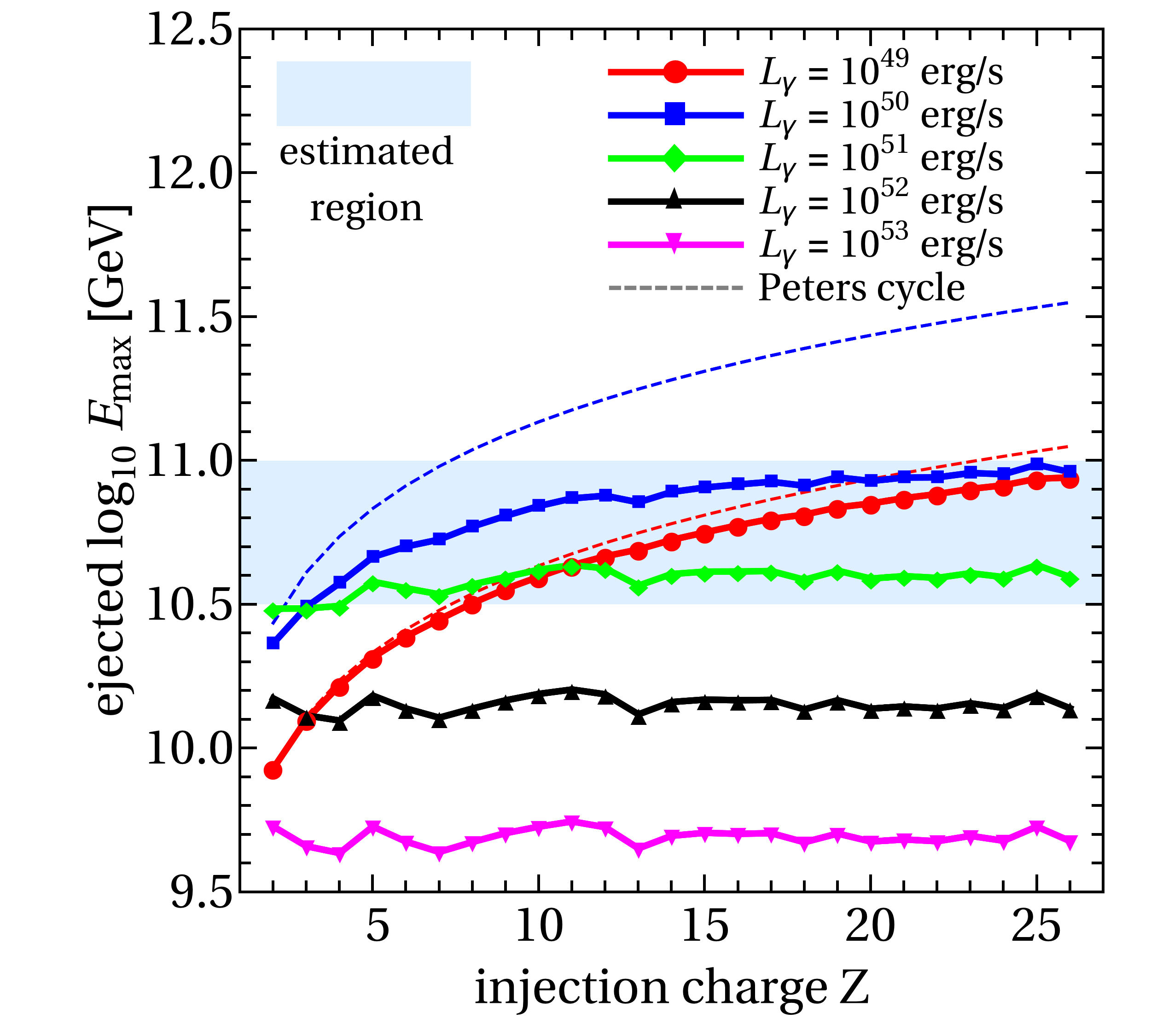}
\includegraphics[width=0.475\textwidth]{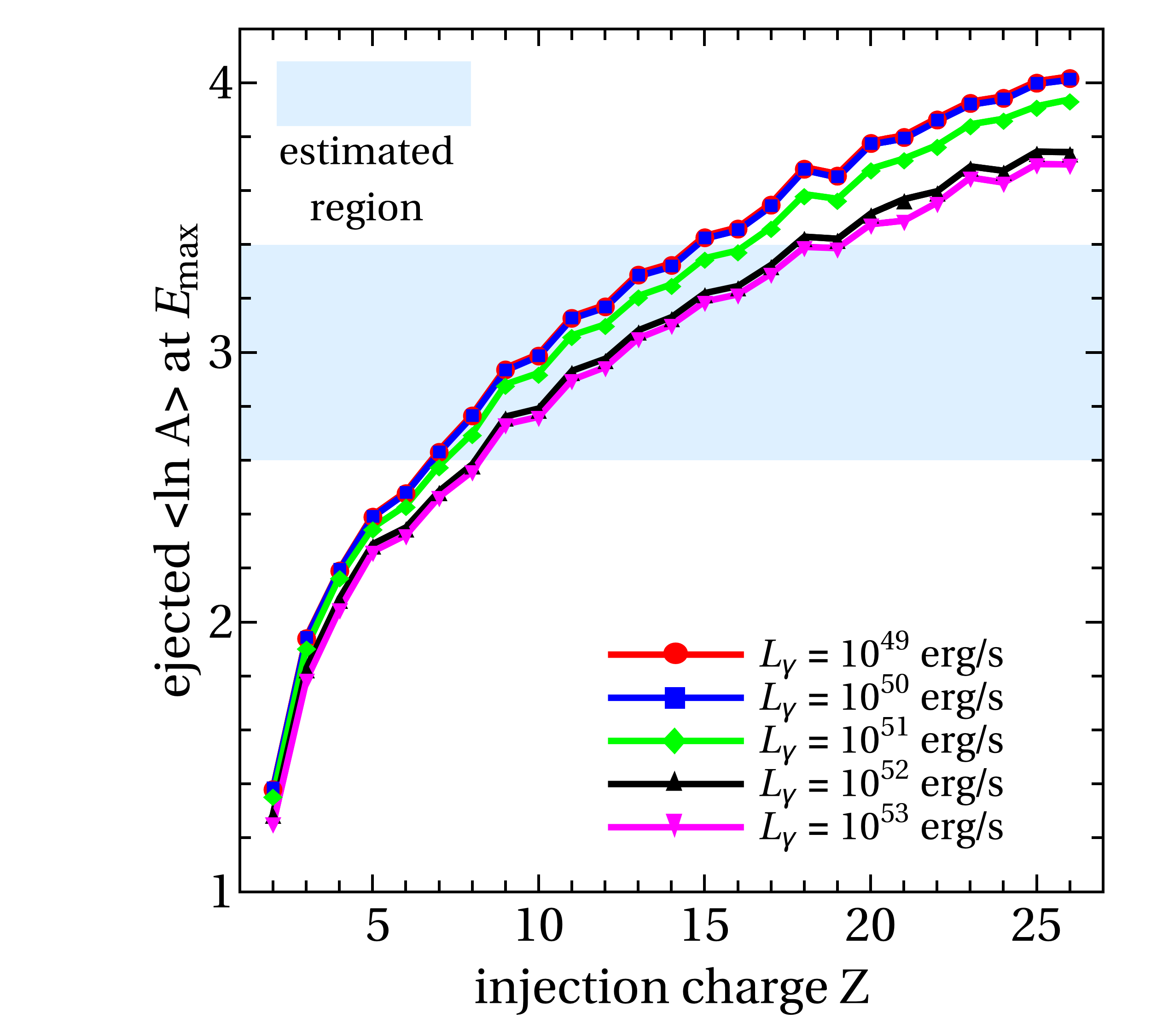}
\includegraphics[width=0.475\textwidth]{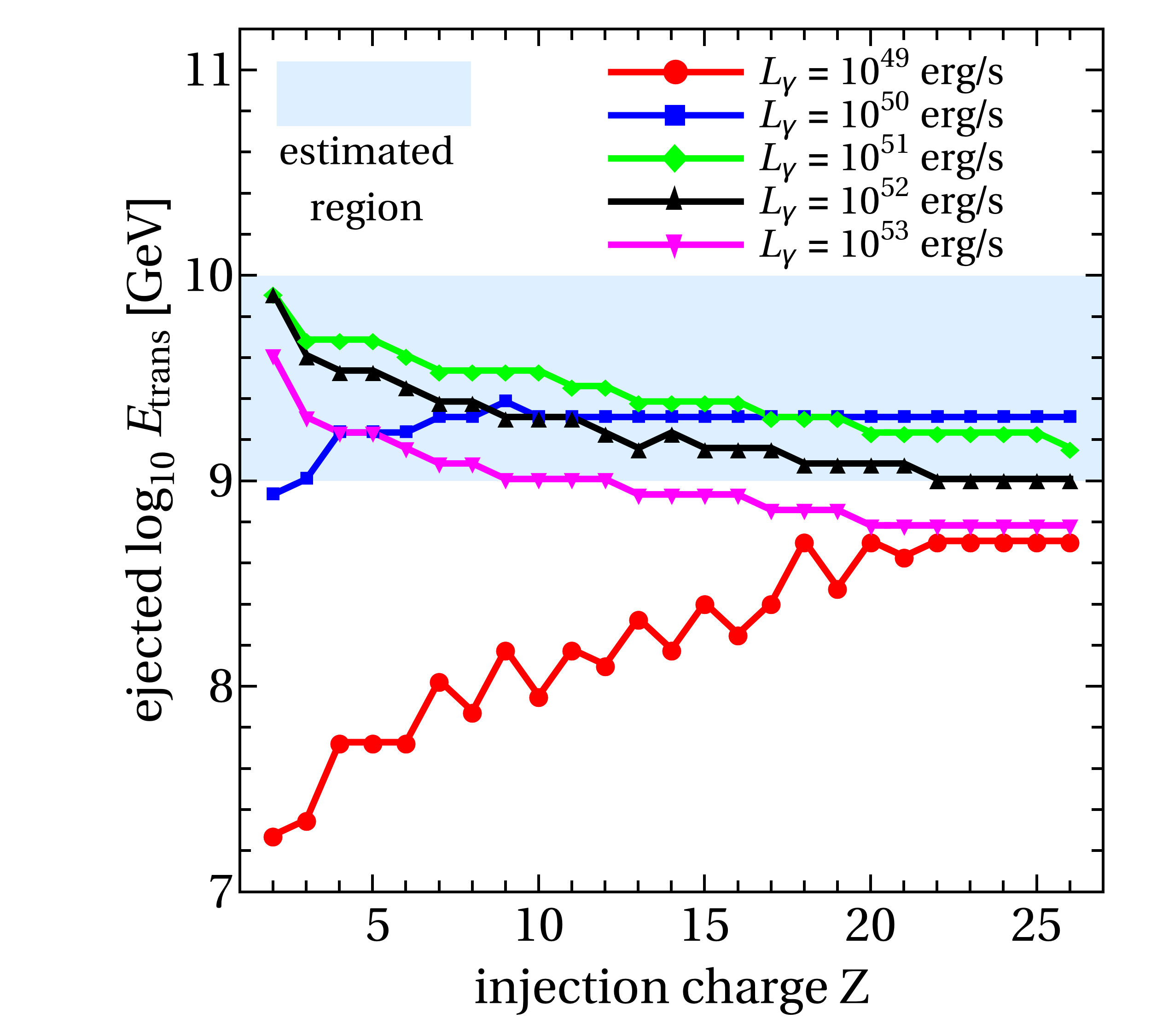}
\caption{Qualitative estimators, as defined in the main text, for the UHECR fit as a function of injection $Z$. 
The shaded regions illustrate our expectations from cosmic-ray data (top left: from~\cite{Heinze:2015hhp}, top right and bottom left: from \cite{Aab:2016zth}, bottom right: from \cite{Kampert:2012mx}. The different curves correspond to different values of $L_\gamma$, as illustrated in the legend. In the upper right panel, the maximal energy following the Peters cycle (rigidity dependent maximal energy) is illustrated for two examples. The production radius is fixed to $R  \simeq 10^{8.3} \, \mathrm{km}$ in this figure.
}
\label{fig:estimators}
\end{figure*}

The four estimators are shown as a function of $Z$ for different values of $L_\gamma$ in \figu{estimators}. The estimated preferred ranges from cosmic-ray data are shaded, \ie, we expect that the estimator value lies in the indicated band to describe data. In some of the curves, the zig-zag pattern comes from the mass pattern of the most abundant injection isotopes as a function of $Z$, see \figu{isochart} (dark blue isotopes). Note that $R$ is fixed in this figure.

From the upper left panel of \figu{estimators} one can read off that the ejected spectral index prefers $10^{49} \, \mathrm{erg \, s^{-1}} \lesssim L_\gamma \lesssim 10^{51} \, \mathrm{erg\, s^{-1}}$, the Optically Thick case results in too soft ejected spectra (see  example in \figu{proto53}). This result hardly depends on the injection composition $Z$.
The ejected cosmic rays have hard spectral indices for low GRB luminosities, as found in our fit of the Mixed Composition Ankle Model; this is in agreement with what has been found in \cite{Aab:2016zth}, where the spectra at injection are comparable to what we find here as ejected from the source. 

From the upper right panel, the maximal energy tends to be too low for large luminosities, where the maximal energy is strongly limited by $A \gamma$ interactions, while almost all other cases lead to maximal energies in a reasonable range (as long as $Z\gtrsim8$). One can also read off from this panel, when the often used Peters cycle is justified (as in \Ref~\cite{Aab:2016zth}): in that case, the maximal energy scales with rigidity. For low luminosities, corresponding to the Empty Cascade, we have found that the maximal energy is indeed limited by adiabatic losses, which is identical to this assumption. Therefore, the $10^{49} \, \mathrm{erg \, s^{-1}}$ line indeed follows the Peters cycle (see dashed curve for comparison). However, at higher luminosities (as one would typically expect for GRBs), the maximal energy becomes limited by $A \gamma$ interactions, and scales much milder with $Z$.

The lower left panel of \figu{estimators} shows the ejected composition at the maximal energy, which needs to be at least as heavy as the composition observed by Auger at the highest energy (see \eg\ the lower panels of \figu{crSi18}). While it is clear that the composition at the highest energy is somehow proportional to $Z$, the ejected $\langle \mathrm{ln} A \rangle$ depends on the distribution of secondaries in the disintegration chain if the maximal energy does not follow the Peters cycle -- which reduces the average mass for higher luminosities. Nevertheless, the dependence on $L_\gamma$ is very mild, while there is an obvious strong dependence on $Z$. From the figure, one can see that injection charges between about 7 (Nitrogen) and 14 (Silicon) may provide reasonable results. For example, for the chosen $Z=14$ in the previous section, the other estimators are relatively stable in $Z$, which means that this estimator is the only one sensitive to the injection composition in that range. 

Taking the information from the first three estimators, we can now qualitatively explain the fit range $10^{49} \, \mathrm{erg \, s^{-1}} \lesssim L_\gamma \lesssim 10^{51} \, \mathrm{erg\, s^{-1}}$ in \figu{bestSi19} for the Mixed Composition Ankle Model. The contours in the two dimensional space roughly follow the maximal energy contours (similar to \figu{nuscan}), apart from the changes due to the penalty for the overshooting at the lowest energies. The region preferred by the fit is found for intermediate values of the maximal energy, with low isotropic luminosity in order to avoid the overshooting at the lowest energies. The isolated region at high collision radius and intermediate luminosity corresponds to the highest values for the maximal energy of the residual primaries in the source, in the Empty Cascade region. 

\begin{figure*}[t]
\includegraphics[width=0.32\textwidth]{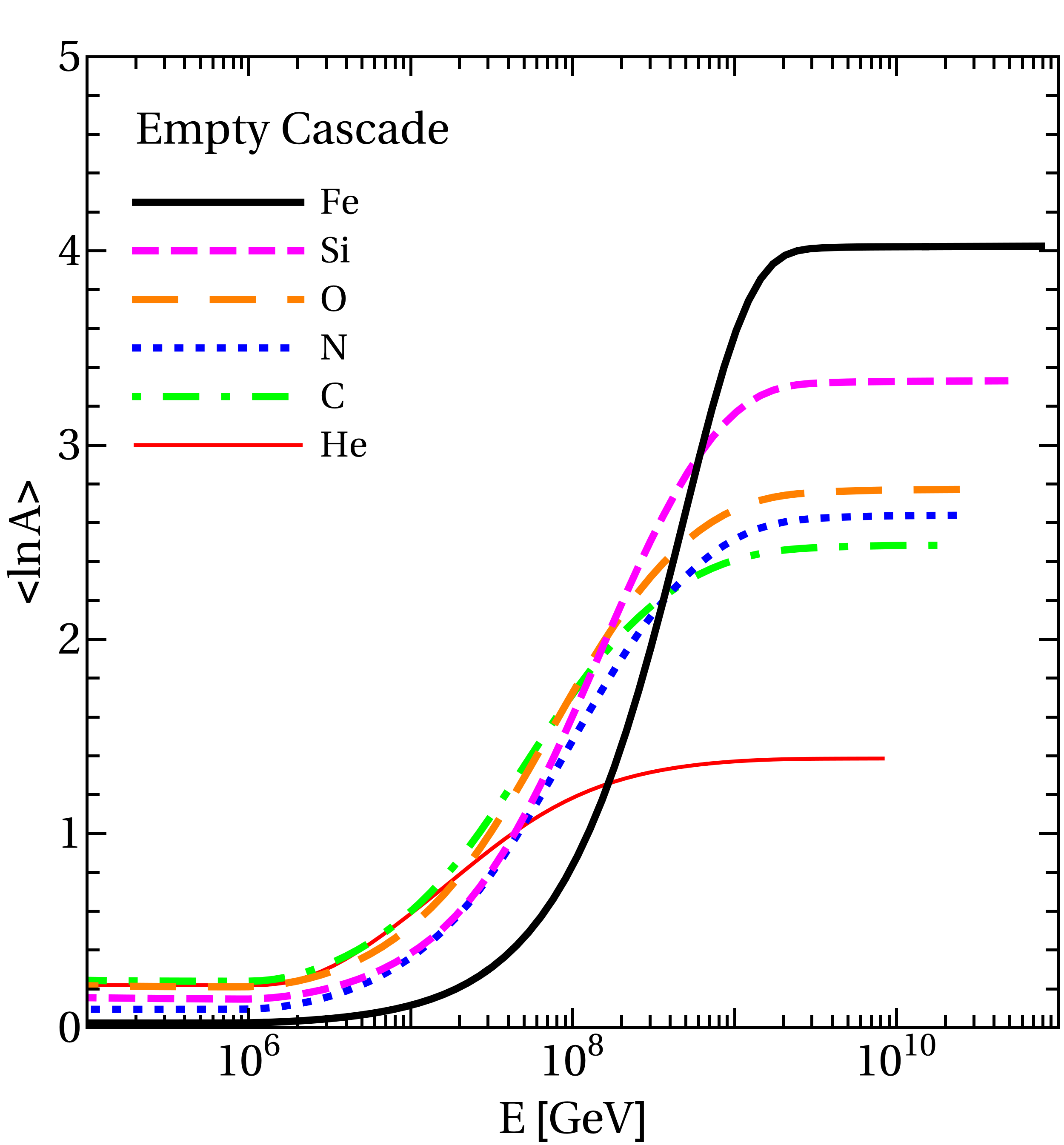}
\includegraphics[width=0.32\textwidth]{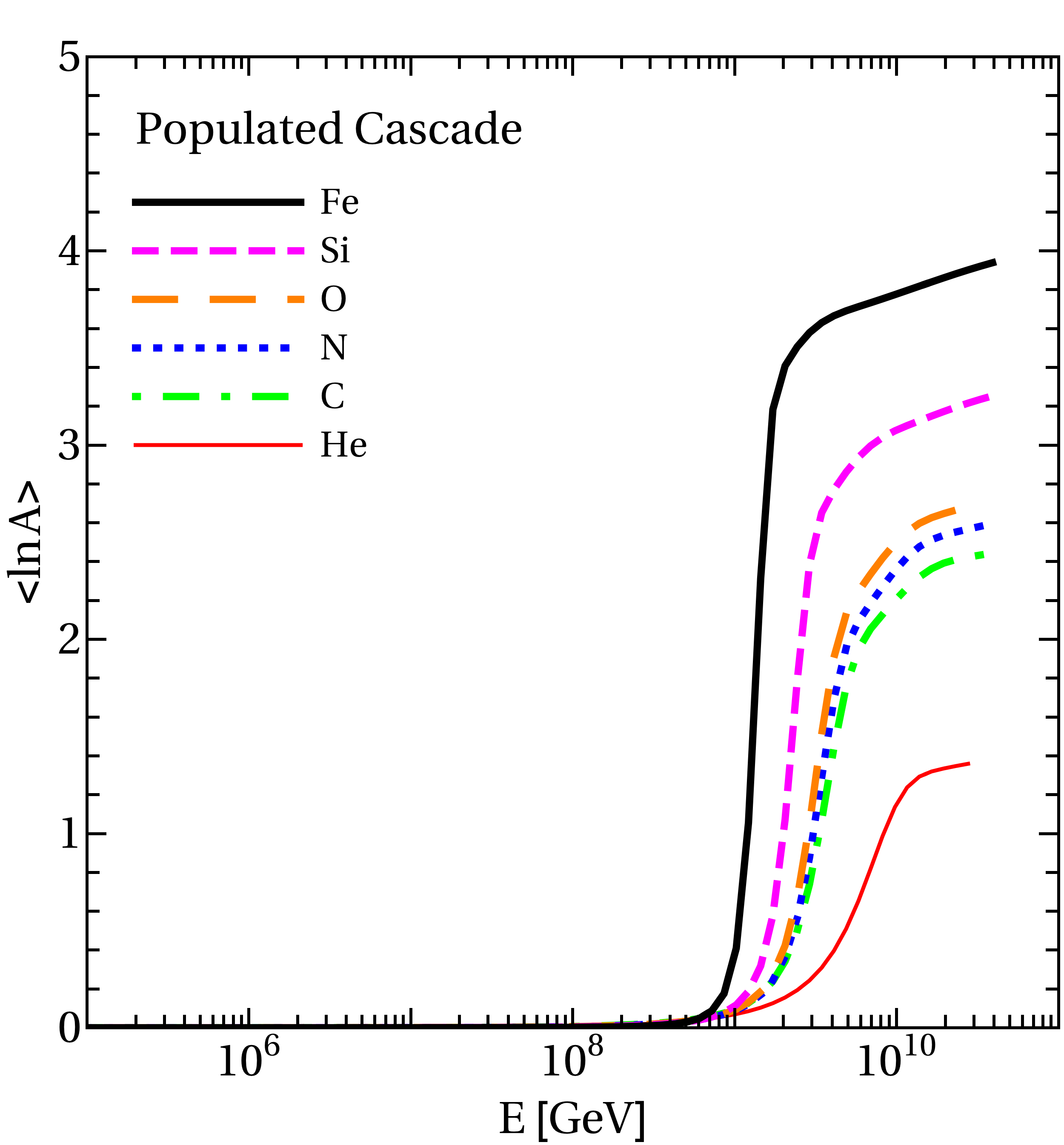}
\includegraphics[width=0.32\textwidth]{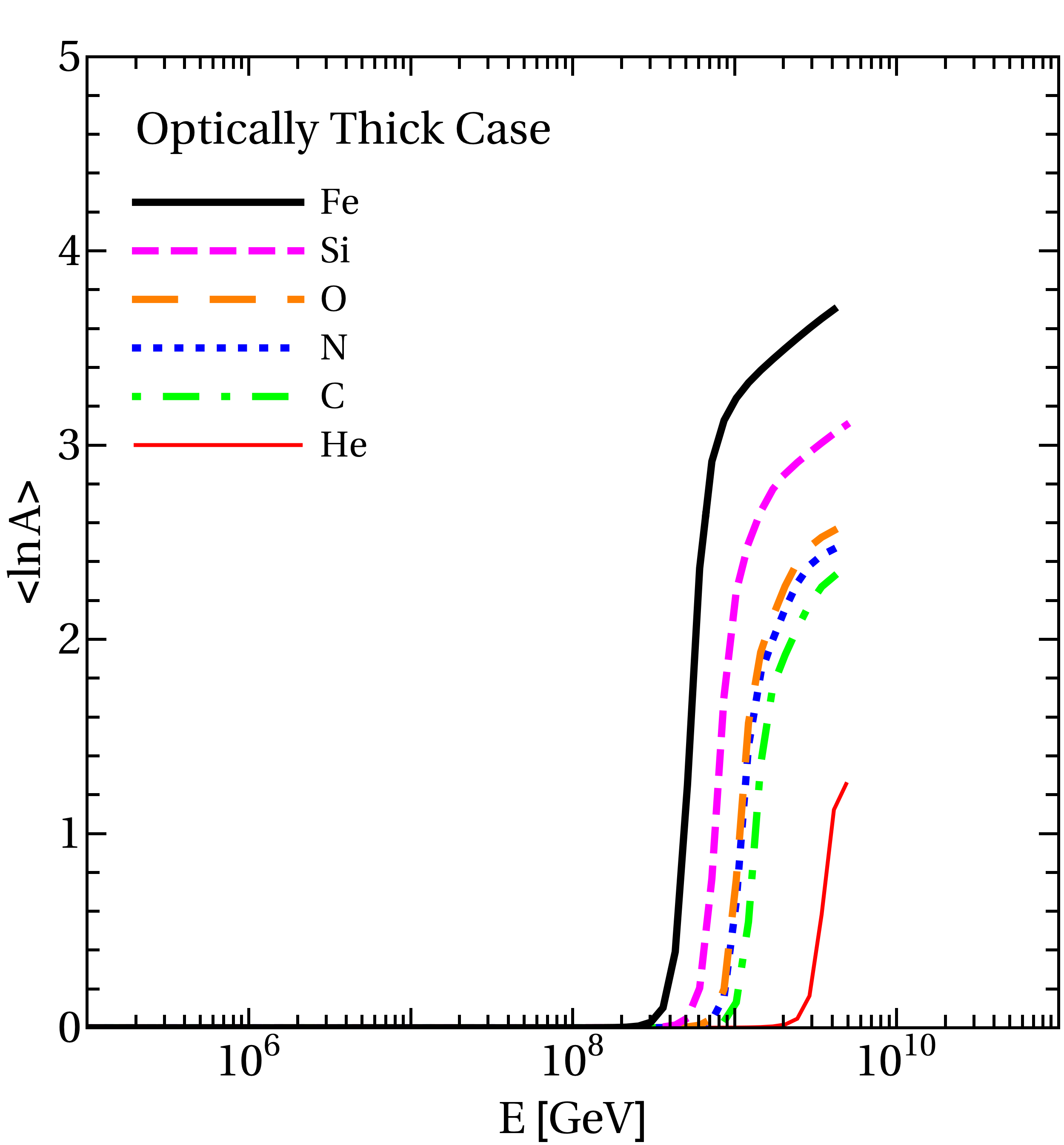}
\caption{Ejected average composition  of the cosmic rays leaving the GRB shell as a function of the energy in the observer's frame. Different luminosities correspond to the examples for different source classes in \Sec~\ref{sec:classes}. Different curves correspond to different pure injection compositions.
}
\label{fig:composition}
\end{figure*}

In the context of the Mixed Composition Dip Model, the best fit is found correspondingly to an isotropic luminosity for which the transition energy is allowed (see bottom right panel in \figu{estimators}).
We discuss this issue in somewhat greater detail in \figu{composition}, where the  $\langle \mathrm{ln} A \rangle$ is shown for our three prototypes (Empty Cascade, Populated Cascade, Optically Thick case). For the Empty Cascade, we know that the primary dominates at the highest energy (see \figu{proto49}), which leads to a relatively smooth transition at relatively low energies -- which slighlty increases with $Z$. For the Populated Cascade, the transition hardly depends on the injection composition and only slightly decreases for higher $Z$. The transition is relatively sudden and becomes smoother for higher $L_\gamma$, where the cascade populates more and more the lighter isotopes -- see Optically Thick Case. Note that the example shown in \Ref~\cite{Boncioli:2016lkt} has an $L_\gamma$ between the middle and right panels. 

Finally, it is difficult to compare the injected $Z$ to theoretical expectations, because these depend on where the jet originates, how charged cosmic rays are injected into the jet and how they are accelerated. For example, if the jet originates within a Wolf-Rayet star, there may be a significant contribution of He, O, and C. If there is a close connection with supernova explosions, a significant contribution of Si or Fe may be expected. If r-process nucleosynthesis occurs within the jet, even heavier elements may contribute, as for example studied in \cite{Metzger:2011xs}. Our approach follows the opposite direction: we trace back the observed UHECR composition into the acceleration zone and identify what is needed there to describe UHECR data. In the future, one would expect that this information will be useful for models of jet formation and injection of UHECR nuclei.

\section{Summary and Conclusions}
\label{sec:outlook}

In this study, the question if GRBs can be the sources of the UHECRs has been addressed in the multi-messenger context. While the neutrino emission from the prompt phase of GRBs has been used to constrain GRB models in stacking analyses, all these constraints have been derived for proton primaries. 
However, the UHECR composition measured by the Auger experiment indicates a dominant contribution of nuclei at the highest energies. We have, therefore, studied the UHECR paradigm for GRBs in the presence of nuclei in the sources, by taking into account the nuclear cascade in the sources with a high level of detail, and, by combining the UHECR source and propagation physics. 

As a first step, we injected nuclei with a pure composition into a GRB shell, and we have quantified the nuclear cascade, the emission of cosmic rays and the production of neutrinos. We identified three qualitatively different regimes as a function of the shell parameters, such as production radius $R$ or gamma-ray luminosity  $L_\gamma$:
\begin{description}
\item[Empty Cascade] (low $L_\gamma$ or large $R$). The source is optically thin to photo-disintegration of primary (injected) nuclei, where the nuclear cascade does not develop. The ejected cosmic rays are dominated by a hard spectrum of the residual primaries, where (as a function of injection charge) the cosmic rays follow a Peters cycle with a rigidity-dependent maximal energy. This scenario is consistent with the global fit result, recently performed by the Auger Collaboration~\cite{Aab:2016zth}.
The neutrino flux is dominated by photo-meson production off the primary nuclei; it is  relatively low because of the low target photon density.
\item[Populated Cascade] (medium $L_\gamma$ and  $R$). The source is optically thick to photo-disintegration of the primary nuclei, and the nuclear cascade can efficiently develop. The ejected cosmic rays receive a significant contribution from neutrons (disintegration products), which can easily escape because they are electrically neutral. While at the highest energies secondary nuclei produced in the disintegration chain dominate, cosmic rays from neutron decays dominate at lower energies, and the ejected overall spectrum becomes softer. Note that the ejected UHECR composition is clearly mixed even if a pure composition is injected from the acceleration zone.
The neutrino production becomes more efficient, and is dominated by photo-meson processes off secondary nuclei produced in the nuclear cascade.
\item[Optically Thick Case] (large $L_\gamma$ or low $R$). The source is optically thick to photohadronic interactions of all nucleons and nuclei. In this case, the nuclear cascade is so efficient that most energy ends up in nucleons  (protons and neutrons), which are the end of the disintegration chain. The ejected cosmic ray spectrum is strongly dominated by neutrons, and it is very difficult to reach high cosmic ray energies because nuclei will mostly disintegrate. The neutrino flux is extremely high, dominated by photo-meson production off nucleons. 
 \end{description}
We have discussed how these scenarios quantitatively depend on the shell parameters, and how our results can be easily translated into different astrophysical models. Note that the ejected UHECR spectra do not exhibit a rigidity-dependent maximal energy cutoff if the nuclear cascade is populated -- which is contrary to the assumption of many UHECR models in the literature.

Regarding the impact of different photo-meson models on the neutrino flux, we have pointed out that relatively solid predictions can be obtained if the photo-meson production has a significant contribution from nucleon-photon interactions (Populated Cascade and Optically Thick Case), for which the hadronic physics is relatively well known (such as using SOPHIA), whereas the photo-meson production off nuclei (Empty Cascade) requires further study. The expected neutrino flux in the latter case is, however, smaller.

As one of the most interesting implications for GRBs, it has been shown that the expected neutrino fluence does, for the same injection luminosity and shell parameters, hardly depend on the injection composition. 
A profound consequence from this observation is that, apart from different cosmic ray propagation effects, the neutrino stacking bounds on long-duration GRBs will apply as well if the UHECRs are, in fact, nuclei.

The next step has been the extrapolation from one GRB shell to an entire population of long-duration GRBs with identical parameters for all zones in the internal shock scenario (often referred to as ``one-zone model''). The emitted UHECRs have been propagated to Earth. It has been found that the Auger spectrum and composition can be reasonably well described even with a pure injection composition only, were two cases have been discussed:
\begin{description}
 \item[Mixed Composition Dip Model.] The same class of sources is responsible for UHECR data including the ankle. Compared to the conventional proton dip model, where the dip comes from pair production, the same feature is generated by the nucleons from disintegration (in the sources and during propagation), which pile up at energies below the ankle. 
 \item[Mixed Composition Ankle Model.] The UHECR data needs to be described only above the ankle, whereas a different population is expected to dominate at lower energy.
\end{description}
It has been demonstrated that both cases can be reasonably well described within the GRB framework. 
However, the dip hypothesis puts narrower constraints on the parameters. The protons below the ankle, often generated within the source, enhance the (prompt) neutrino flux significantly. As a consequence, the Mixed Composition Dip Model is in tension with current neutrino data for the chosen set of parameters, while the Mixed Composition Ankle Model is consistent with current experimental observations -- meaning that GRBs can be the sources of UHECR nuclei in spite of low predicted neutrino fluxes. While the results have been derived for the injection of $^{28}$Si, the dependence on the 
injection composition has been illustrated with a novel approach connecting the physics of the source with the physics of propagation.

Note that there are several limitations of our fitting procedure. First of all, the overall goodness of fit is poor in the Mixed Composition Dip Model because of the small statistical errors bars. However, we have not yet included systematic errors, such as energy scale uncertainty of the experiment. In addition, it is clear that a mixed injection composition in the source will improve the goodness of fit, which is however beyond the scope of this work. In addition, some parameters (such as the spectral index, the shape of the cutoff at the injection and the source evolution) have been fixed, which would also improve the fit. Therefore, the results in this work can be considered as ``proof of principle'' for the UHECR fit, and as "indicative" for the neutrino bounds.


 We conclude that GRBs can be the sources of the UHECRs, as we have self-consistently described the observed spectrum and composition observed by Auger even in a one-zone model with a pure injection composition into the GRB shells. However, the exclusion power of neutrino stacking bounds applies to nuclei as well, what in particularly constraints the hypothesis that the light composition below the ankle comes from the same population. If, however, the UHECR transition to the GRB contribution occurs at the ankle, GRBs can describe cosmic ray and neutrino data at the expense of relatively high baryonic loadings and obtained GRB parameters (either low $\gamma$-ray luminosities or large collision radii). 

While some of our results can be applied to multi-zone models (such as the behavior of the nuclear cascade), these models typically predict a lower neutrino flux~\cite{Bustamante:2014oka,Globus:2014fka,Bustamante:2016wpu}. A self-consistent picture has been drawn in \Refs~\cite{Globus:2014fka,Globus:2015xga} for a particular set of parameters, such as a specific injection composition mix. Dedicated parameter space studies in this context will require more efficient computing techniques, especially if the nuclear cascade is to be computed in each collision independently. In this study, we have presented the technology to perform such computations efficiently and precise enough.

{\bf Acknowledgments.}  We would like to thank Mauricio Bustamante and our colleagues from the Pierre Auger Collaboration for fruitful discussions.

This project has received funding from the European Research Council (ERC) under the European Union’s Horizon 2020 research and innovation programme (Grant No. 646623).

\appendix

\clearpage

\section{Efficient Computation of Nuclear Processes}
\label{app:nuclear}

This appendix contains details about the computational methods used for the efficient treatment of the interaction (and decay) processes, balancing performance and precision. Note that in this study, we assume that the target photons are isotropically distributed in the SRF.

\subsection{Beta Decays and Spontaneous Emission}

 The simplest approximation for the re-distribution function $p$ in \equ{repro} is the  $\delta$-function
\begin{equation}
\frac{d n_{j \rightarrow i}}{dx} (x,\sqrt{s}) \simeq  M_{j \rightarrow i} \cdot \delta(x -  \chi_{j \rightarrow i} )  \, .
\label{equ:nabsimple}
\end{equation}
The function $\chi_{j \rightarrow i}$, which depends on the kinematics of the process, describes  which (mean) fraction of the parent energy is deposited in the secondary. If a secondary nucleus is produced, a frequently applied assumption is Lorentz factor conservation, which leads to $\chi_{j \rightarrow i}  \simeq A_i/A_j$.

Using \equ{nabsimple} in \equ{prod}, one easily finds for the injection of secondary nuclei from beta decays and spontaneous emission:
\begin{equation}
 Q'_{j \rightarrow i}(E'_i) = N'_j \left( \frac{E'_i}{\chi_{j \rightarrow i}} \right) \, \Gamma'_j \left( \frac{E'_i}{\chi_{j \rightarrow i}} \right) \, \frac{1}{\chi_{j \rightarrow i}} \, M_{j \rightarrow i} \, .
\label{equ:secnuclei}
\end{equation}
Here $\Gamma'_j  = m_j/(\tau_{0,j} E'_j)$ and $\chi_{j \rightarrow i}=A_i/A_j$. For instance, for the case of $\beta^\pm$ decays, $A$ does not change, which means that $\chi_{j \rightarrow i} \simeq 1$, and $M_{j \rightarrow i}$ is the branching ratio into that channel. 

Refined computations for the neutrino spectrum of beta decays from relativistic ions can be, for example, found in the context of long-baseline experiments, so-called $\beta$-beams, see \Ref~\cite{BurguetCastell:2003vv}. For the neutrino injection, we use \equ{secnuclei}  with a peak value $\chi_{j \rightarrow \nu}$ directly from the peak of the re-distribution function extracted from \Ref~\cite{BurguetCastell:2003vv}, related to the $Q$-value (obtained from the mass difference between parent and daughter nucleus). 
\
\subsection{Photo-disintegration}

The  interaction rate  $\Gamma'_j$ for photo-hadronic interactions in an isotropic target photon field is given by
\begin{eqnarray}
\Gamma'_j (E'_j) & = & \int d \varepsilon' \int\limits_{-1}^{+1} \frac{d \cos \theta'_{A \gamma}}{2} \, (1- \cos \theta'_{A \gamma}) \cdot \, \nonumber \\
& & \cdot n'_\gamma(\varepsilon') \, \sigma^{\mathrm{abs}}_j(\epsilon_r) \, .
\label{equ:Agamma}
\end{eqnarray}
Here $n'_\gamma(\varepsilon')$  is the photon density as a function of photon energy $\varepsilon'$ and the pitch angle between the photon and proton momenta $\theta'_{A \gamma}$, $\sigma^{\mathrm{abs}}_j(\epsilon_r)$ is the absorption  cross section for species $j$, and 
\begin{equation}
\epsilon_r=\frac{E'_j \, \varepsilon'}{m_A} (1 - \cos \theta'_{A \gamma})
\end{equation} is the photon energy in the parent rest frame (PRF) in the limit $\beta'_A \approx1$.\footnote{
Note that  $\epsilon_r$ corresponds to the available center of mass energy $\sqrt{s}$ of the interaction, as
$s = m_A^2 + 2 \, m_A \, \epsilon_r $.}

 It is convenient to write the interaction rate in the form of an integral over the photon density $n'_\gamma(\varepsilon')$ as 
\begin{equation}
\Gamma'_j(E'_j) = \int\limits d \varepsilon' \, n'_\gamma(\varepsilon') \,
f_j\left( y'  \right) \, ,
\label{equ:irates}
\end{equation}
where $y' \equiv (E'_j \varepsilon')/m_A$ corresponds to the ``typical'' center-of-mass energy, and
\begin{equation}
 f_j(y') \equiv \frac{1}{2y'^2} \int\limits_0^{2y'} d \epsilon_r \, \epsilon_r \, \sigma^{\mathrm{abs}}_j(\epsilon_r)  \, 
\label{equ:f}
\end{equation}
is an integral over the cross section (which is, by definition, zero below the threshold); it can be interpreted at pitch-angle averaged cross section.
Here the idea is that the function $f_j(y)$  can be pre-computed (it only depends on the cross section, and the integral takes care of the pitch angle averaging), and that the interaction rate can be obtained in a single integral in \equ{irates}. 

For the secondary nuclei injection, we re-write \equ{prod} (using \equ{nabsimple}) as
\begin{eqnarray}
Q'_{j \rightarrow i}(E'_i) & = & N'_j \left( \frac{E'_i}{\chi_{j \rightarrow i}} \right) \frac{m_{A_j}}{E'_i} \cdot \label{equ:prodsimple2}
 \\
& & \cdot  \int dy' \, n'_\gamma \left( \frac{m_{A_j} \, y' \, \chi_{j \rightarrow i}}{E'_i} \right) \,  g_{j \rightarrow i}(y') \,  \nonumber ,
\end{eqnarray}
with $\chi_{j \rightarrow i} \equiv A_i/A_j$ and a function 
\begin{equation}
 g_{j \rightarrow i}(y') \equiv \frac{1}{2y'^2} \int\limits_{\epsilon_{\mathrm{th}}}^{2y'} d \epsilon_r \, \epsilon_r \, \sigma^{\mathrm{abs}}_j(\epsilon_r)  \, M_{j \rightarrow i}(\epsilon_r) \, .
\label{equ:g}
\end{equation}
Here it is taken into account that the secondary multiplicity strongly depends on the center-of-mass energy  (or $\epsilon_r$) with pre-computed functions $g_{j \rightarrow i}$, while the re-injection can be still obtained by a single integral \equ{prodsimple2}.  For the case of a constant target photon spectrum (such as in this work), not even a single integral is needed.

\begin{figure*}[tp]
\includegraphics[width=\textwidth]{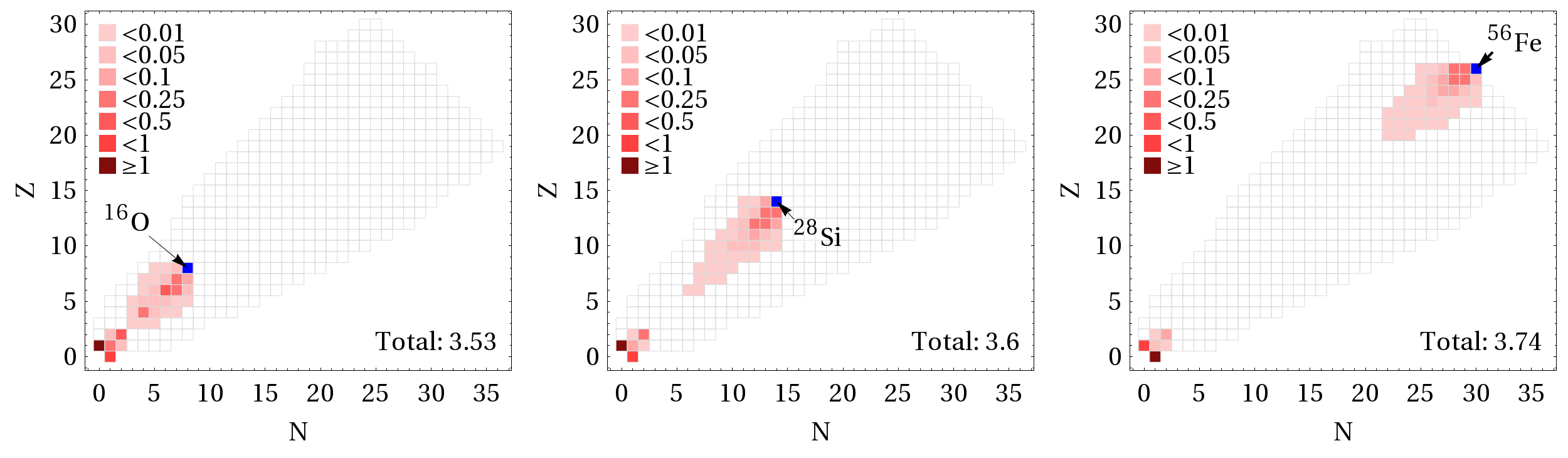}
\caption{Secondary multiplicities for the disintegration of $^{16}$O (left), $^{28}$Si (middle) and $^{56}$Fe (right), marked in blue, for the TALYS disintegration model used in this work. The color coding refers to the pitch-angle averaged number of secondary nuclei $g_{j \rightarrow i}(y')/f_j(y')$ produced on average for $y'=50 \, \mathrm{MeV}$. The total (average) number of secondary nuclei and nucleons produced per interaction is shown in the lower right corner of each panel.}
\label{fig:dismodel}
\end{figure*}

The values for $f(y')$ and $g_{j \rightarrow i}(y')$ are produced from the disintegration models (such as TALYS), as described in the main text.  Note that $\sigma^{\mathrm{abs}}_j(\epsilon_r)  \, M_{j \rightarrow i} = \sigma^{\mathrm{incl}}_{j \rightarrow i}(\epsilon_r)$ (total inclusive cross section), which can be extracted from the nuclear models. Comparing \equ{g} with \equ{f}, we can read off that the quantity $g_{j \rightarrow i}(y')/f(y')$ describes the secondary multiplicity as a function of $y'$, including the pitch angle averaging in the isotropic target photon field. We illustrate this quantity for three different examples in \figu{dismodel} for  $y'=50 \, \mathrm{MeV}$ (which is slightly above the GDR, because the model is more sophisticated there). In all cases, nucleons are produced in each integration process, as well as light nuclei. The residual nuclei tend to be populated along the main diagonal. However, compared to one-dimensional (one isobar) disintegration models, such as the Puget-Stecker-Bredekamp model~\cite{Puget:1976nz,Khan:2004nd}, the isotope chart will be populated in two dimensions: along the main diagonal, and perpendicular to it in terms of unstable isobars. The importance of using a sufficiently sophisticated disintegration model has been addressed  in \Ref~\cite{Boncioli:2016lkt}.

\subsection{Photo-meson Production}

The  interaction rate for photo-meson production can be obtained as in \equ{irates}. The main difference to disintegration, however, is that the re-distribution functions of the secondary pions should be taken into account. The approach presented here is a further advancement of \Ref~\cite{Hummer:2010vx}.

If the re-distribution function for the secondaries (especially the pions) is to be described by \equ{repro}, it turns out to be difficult to avoid double integrals in the injection function, such as over nucleus and photon energy. These, however, are one of the bottlenecks for efficient computations, and brute-force sampling methods of the interaction models frequently do not take this into account. The idea  in \Ref~\cite{Hummer:2010vx} for $p\gamma$ interactions was basically to discretize one of these integrals into a small number of ``interaction types'' (in their language), where the splitting into $t$-channel production, resonances, and multi-pion production was physics-motivated. These interaction types had different characteristics in terms of their multiplicities and inelasticities, and were evaluated similarly to \equ{prodsimple2}.  We propose a different method here, which will allow for automatic definitions of the interaction types for many isotopes, and which will even outperform  \Ref~\cite{Hummer:2010vx} in terms of efficiency and precision -- based on similar principles.

We re-write one of the integrals into an $x$-integral, and we define  $x$-dependent ``interaction types'' in the language of \Ref~\cite{Hummer:2010vx} by discretizing that integral. We find that for $T$ such interaction types the injection of secondaries is given by
\begin{eqnarray}
Q'_{j \rightarrow i}(E'_i)  & = &  \sum\limits_{k=1}^{T} \Delta \tilde x_k  \, N'_j \left( \frac{E'_i}{10^{\tilde x_k}} \right) \frac{m_{A_j}}{E'_i} \times \, \label{equ:proddistildex3}
 \\
& \times & 
 \int d y' \, n'_\gamma \left( \frac{m_{A_j} y' 10^{\tilde x_k}}{E'_i} \right)  \,   h_{j \rightarrow i}(\tilde x_k,y') \, , \nonumber
\end{eqnarray}
where $\tilde x=\mathrm{log}_{10}(x)$, and where we define a new re-distribution function
\begin{eqnarray}
 h_{j \rightarrow i}(\tilde x_k,y') & \equiv & \frac{1}{2y'^2} \int\limits_{\epsilon_{\mathrm{th}}}^{2y'} d \epsilon_r \, \epsilon_r \, \frac{d\sigma^{\mathrm{incl}}_{j \rightarrow i}}{d \tilde x}(\tilde x_k,\epsilon_r)  \, .
\label{equ:h}
\end{eqnarray}
Here we identified  $\sigma^{\mathrm{abs}}_j  \, dn_{j \rightarrow i}/d \tilde x$ with the  (differential) inclusive cross section $d\sigma^{\mathrm{incl}}_{j \rightarrow i}/d \tilde x$. The idea is similar to the photo-disintegration: \equ{proddistildex3} only contains a single integral, to be summed over several discrete values $\tilde x_k$. The information on $h$ can be directly compiled from the inclusive cross sections once an appropriate splitting in terms of $x$ is defined. 

Comparing \equ{h} to \equ{g} (and the corresponding re-injection functions), we note that
\begin{equation}
 g_{j \rightarrow i}(y') = \sum\limits_{k=1}^T \Delta \tilde x_k \,  h_{j \rightarrow i}(\tilde x_k,y') \, , \label{equ:gh}
\end{equation}
which implies that the re-distribution function for the secondaries has to add up to yield the secondary multiplicity -- and the interaction types should be chosen accordingly. The simplest example, frequently used in the literature, is to pick $T=1$ and $dn_{j \rightarrow i}/d \tilde x(\tilde x_1,y') = M_{j \rightarrow i}$ with $\tilde x_1=\log_{10} \chi_{j \rightarrow i}$, in which case \equ{proddistildex3} reduces to \equ{prodsimple2}. It is clear that using several such interaction types with different values of $\tilde x$ will lead to more precise results, at the expense of computation time linearly growing with $T$.

Compared to \Ref~\cite{Hummer:2010vx}, one can identify their interaction types $M_{j \rightarrow i}^{\mathrm{IT}} f^{\mathrm{IT}}(y') \leftrightarrow \Delta \tilde x_k \, h_{j \rightarrow i}(\tilde x_k,y')$ with ours (compare \equ{proddistildex3} to Eq.~(28) in \Ref~\cite{Hummer:2010vx}).  Our splitting into interaction types with corresponding $\tilde x$-values is performed  as a function of the average interaction energy $y'$, which allows for energy-dependent multiplicities. Since the contributing physical processes change as a function of interaction energy, both approaches lead, in principle, to similar results.

\begin{figure*}[tp]
\includegraphics[width=0.8\textwidth]{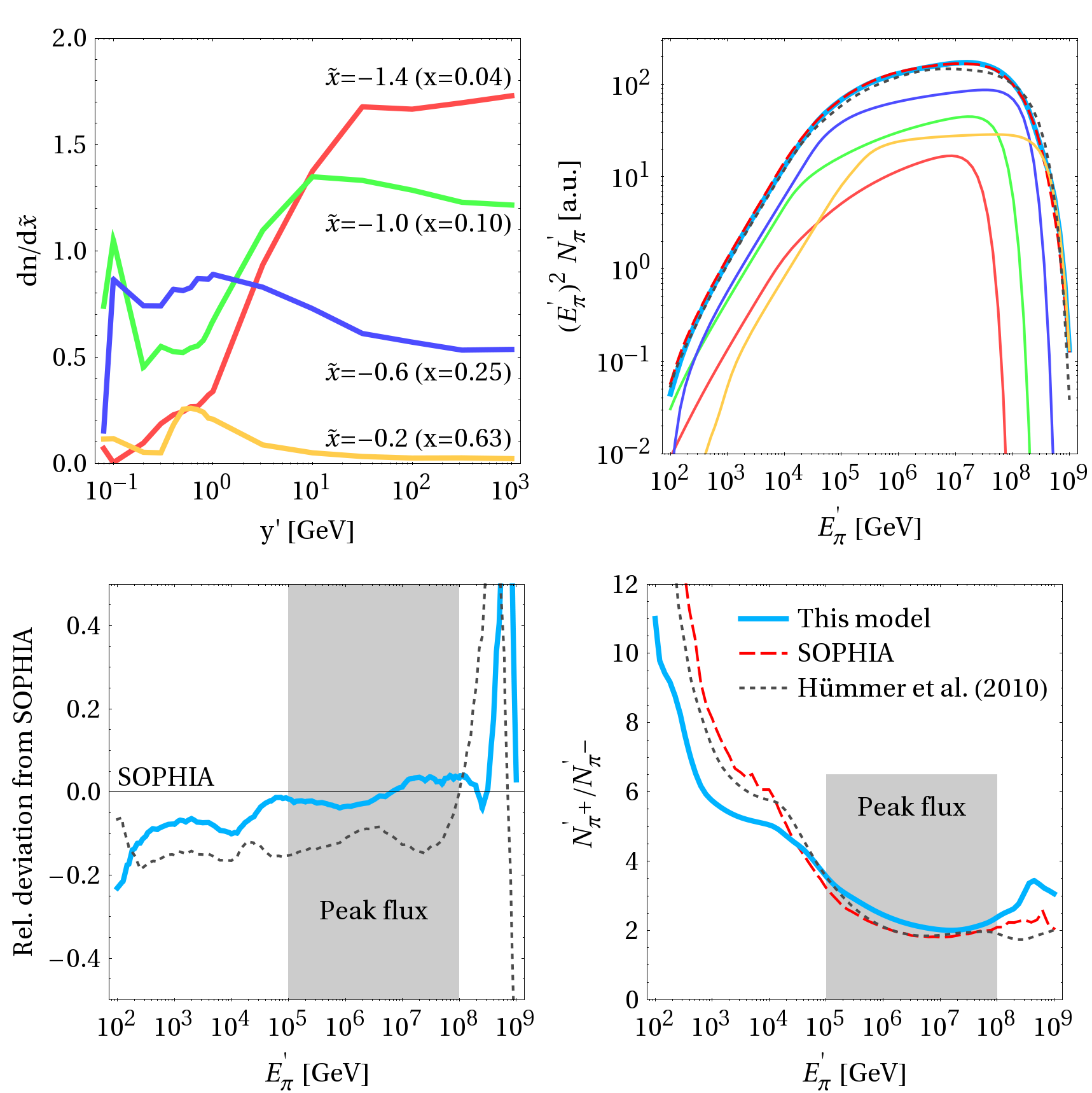}
\caption{Illustration of the new interaction model implementation (based on SOPHIA) for photo-meson production of $\pi^+$ from $p\gamma$ interactions for the GRB benchmark from H{\"u}mmer et al. (2010)~\cite{Hummer:2010vx}: re-distribution functions (upper left panel), spectral contributions (upper right panel), relative deviation from SOPHIA (lower left panel), and charged pion ratio (lower right panel). See main text for details.}
\label{fig:newimodel}
\end{figure*}

Let us test this at the example of $p\gamma$ interactions using SOPHIA~\cite{Mucke:1999yb}. First of all, we use the mean value theorem on \equ{h}, which yields 
\begin{equation}
h_{j \rightarrow i}(y') \simeq f_{j \rightarrow i}(y') \, \frac{dn_{j \rightarrow i}}{d \tilde x}(\tilde x_k, \langle \epsilon_r \rangle)
\end{equation}
 with an appropriate choice $0 \le \langle \epsilon_r \rangle \le 2y'$. It turns out that  $\langle \epsilon_r \rangle \simeq  y'$ is a good approximation (the ``typical center-of-mass energy''). In that case the re-distribution function  can be easily extracted from SOPHIA without modifications assuming an isotropic target photon spectrum and appropriate values of $\varepsilon'$ and $E'_p$ for a flat target photon spectrum (re-call that $y'=\varepsilon' \, E'_p / m_p$). 

We have tested several different splittings in $\tilde x$, and found that using $T=4$ with $\tilde x_1=-1.4$, $\tilde x_2=-1.0$, $\tilde x_3=-0.6$,  and $\tilde x_4=-0.2$ (for fixed $\Delta \tilde x=0.4$) gives reasonable results, while being computationally inexpensive. We illustrate the results for this interaction model and its comparison to SOPHIA in \figu{newimodel}. The upper left panel illustrates the distribution function $dn/d\tilde x(\tilde x_k,y')$ for different values of $\tilde x$. The typical assumption for photohadronic interactions is that pions take about 20\% of the proton energy, corresponding to $x \simeq 0.2$.  From the figure, one can easily see that the re-distribution funcion for $x=0.25$ peaks at around the $\Delta$-resonance (a few hundred MeV). For higher energies, where multi-pion production dominates, smaller values of $x$ yield larger pion multiplicities. An important message of the figure is captured by the $y'$-dependence of the re-distribution function, which is different from the usually shown $\tilde x$-dependence (for fixed interaction energy): The functions describe the injection contributions from different (pitch-angle averaged) center-of-mass energies for fixed values of $\tilde x$ -- which correspond to the ratio of daughter and parent energies. 

The upper right panel of \figu{newimodel} shows the contributions from the different $\tilde x$-types (same color coding), and the total (light blue). One can easily see that the main contribution comes from $\tilde x \simeq 0.25$,  while low (high) energies receive some contribution from smaller (larger) $\tilde x$-values. The total spectrum matches the one produced with SOPHIA  (red dashed) very well and even better than \Ref~\cite{Hummer:2010vx} (dotted, model Sim-B therein). The better relative reproduction of the spectrum is also documented in the lower left panel of \figu{newimodel}, whereas \Ref~\cite{Hummer:2010vx} reproduces the charged pion ratios somewhat better at low energies (see lower right panel). In summary, we obtain a precision somewhat better than \Ref~\cite{Hummer:2010vx}, with only four instead of 23 interaction types, \ie, the new method is more than a factor of five faster. We have also tested alternatives with 8 and 17 $\tilde x$-values, which however do not provide a significant gain of precision. 

The implementation of the superposition model for nuclei is described in the main text, where $\tilde x_k^A = \tilde x_k^p-\mathrm{log}_{10} A$ can be easily found from the one for protons/neutrons. This model does not yet exploit the strength of our method in full, which is to be discussed in future publications.

\section{Effect of Minimum Photon Energy Cutoff}
\label{app:ephmin}

\begin{figure*}[tp]
\includegraphics[width=0.49\textwidth]{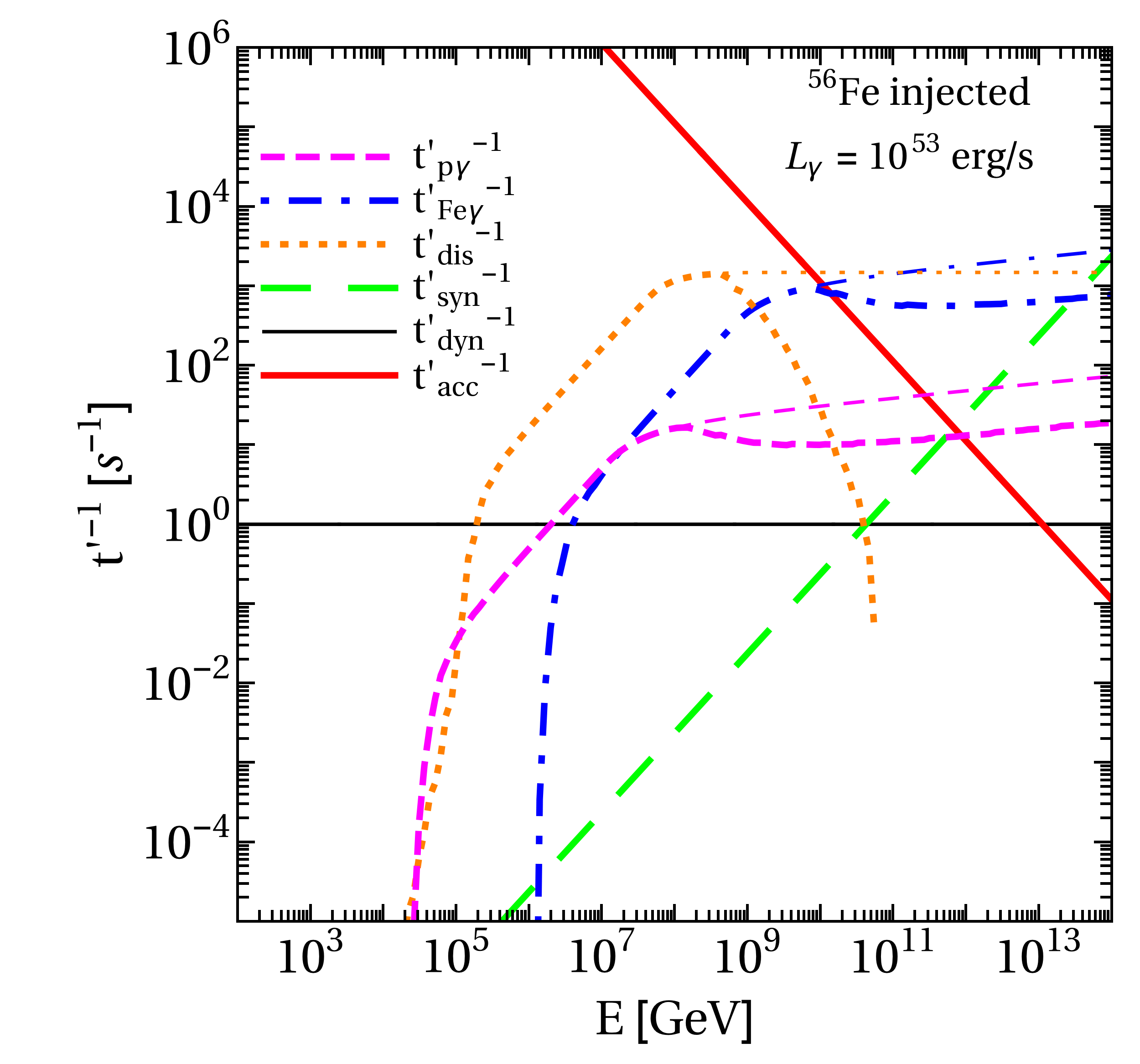}
\includegraphics[width=0.49\textwidth]{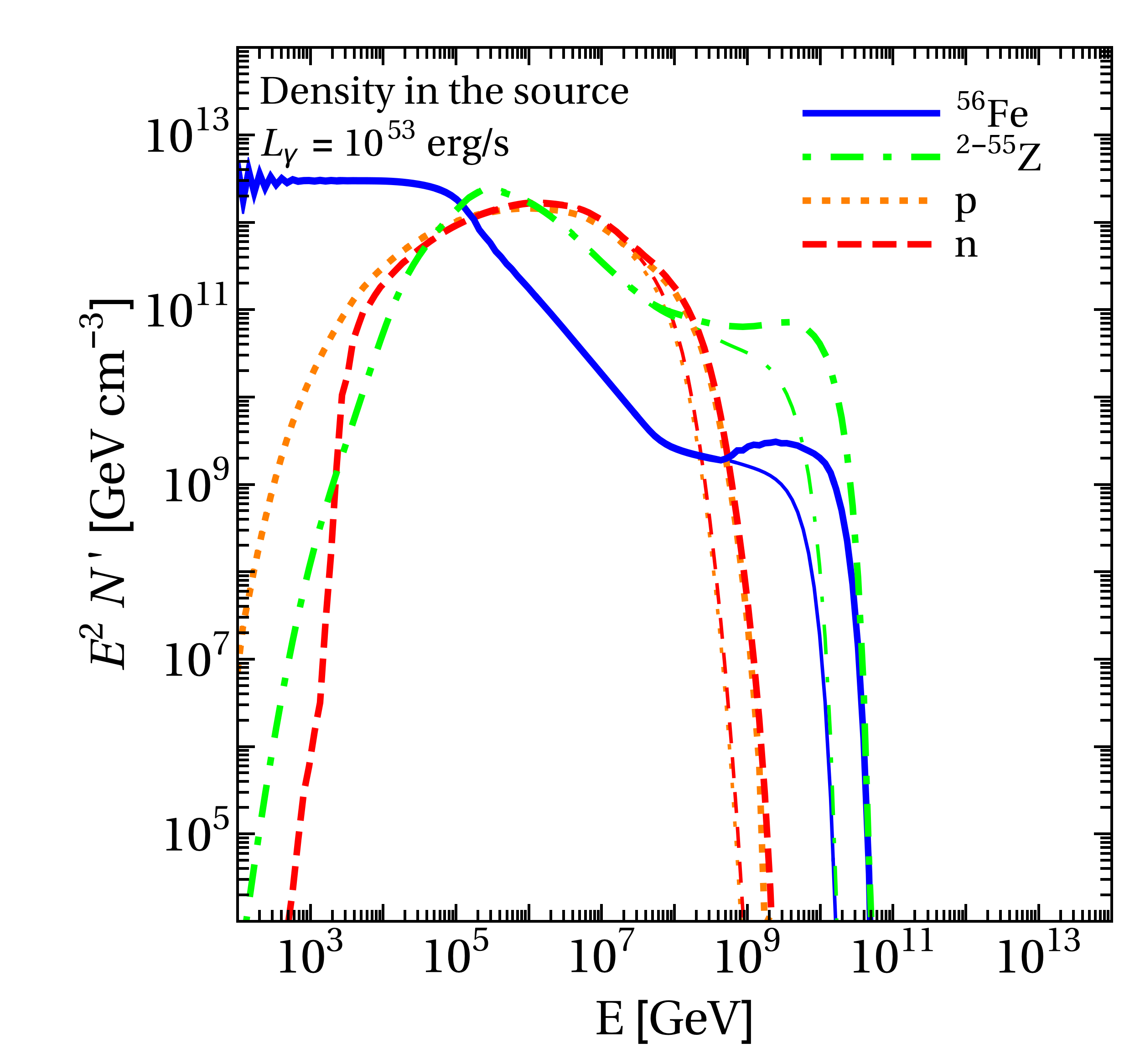}
\includegraphics[width=0.49\textwidth]{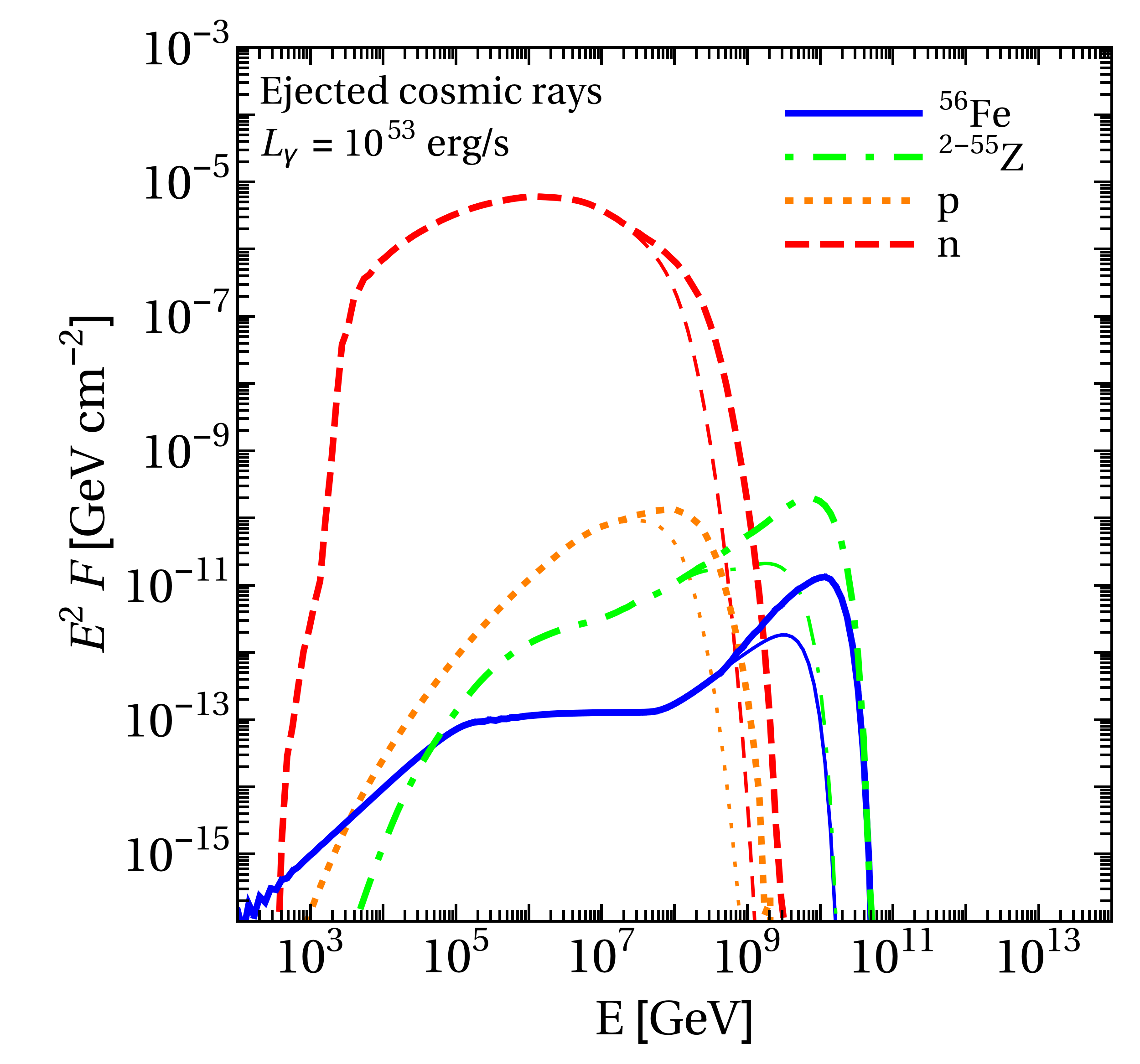}
\includegraphics[width=0.49\textwidth]{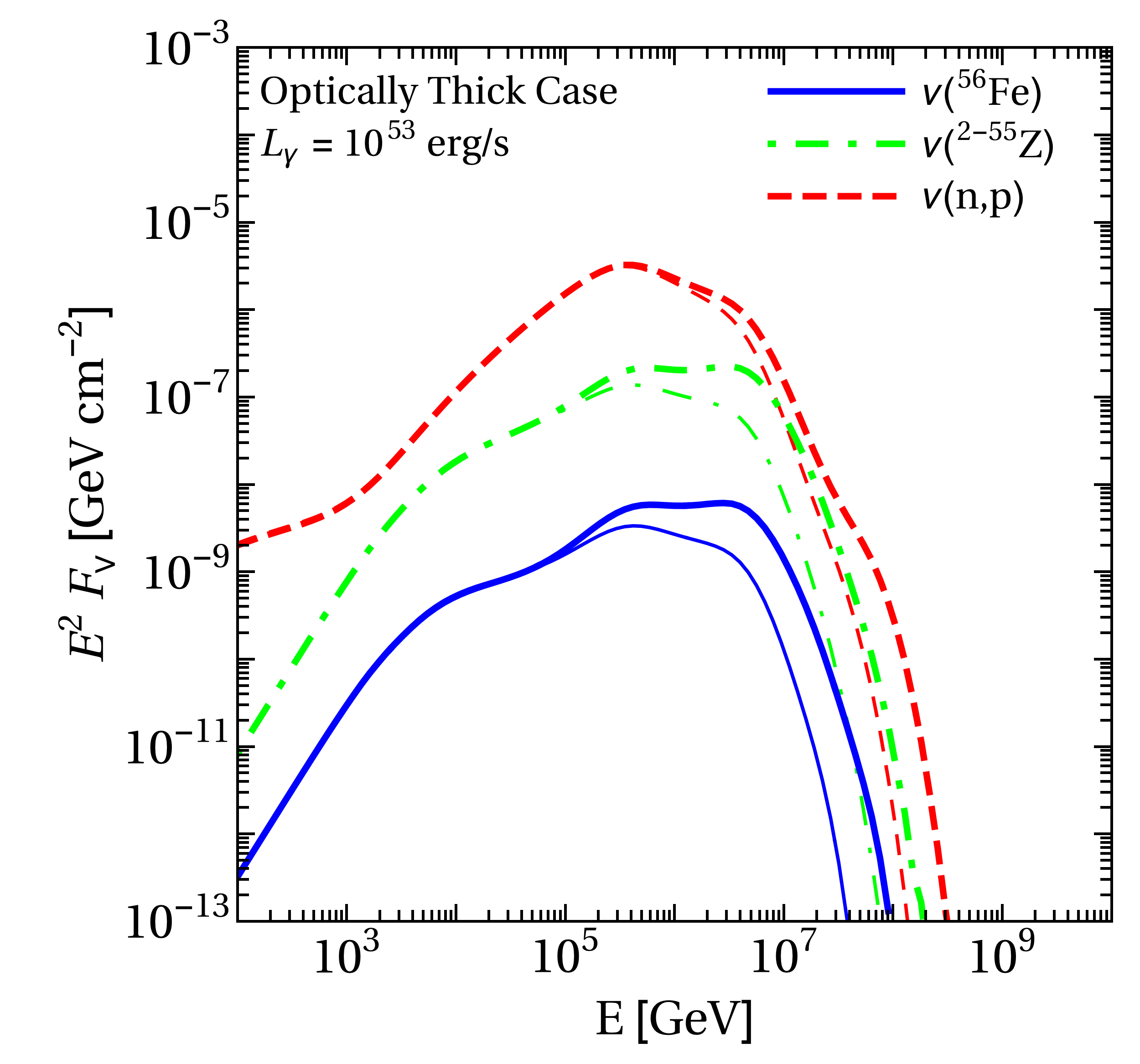}
\caption{Effect of a minimal photon energy cutoff at $\varepsilon'_{\gamma,\text{min}} = 0.1 \varepsilon'_{\gamma,\text{br}}$ on the example in the Optically Thick Case shown in \Figs~\ref{fig:proto53}~and~\ref{fig:nuflux}. Thin curves show the original plots, and thick curves the result with cutoff.
}
\label{fig:ephmin}
\end{figure*}

In the main text of this study, we assume that the minimal photon energy is low enough such that UHECRs will always find an interaction partner to produce the giant dipole resonance. For UHECR with a $\gamma$-factor of $10^{10}$ in the SRF, that means that $\varepsilon'_{\gamma,\mathrm{min}} \lesssim 10^{-3} \, \mathrm{eV}$ to produce the GDR.  If, however, the low-energy photon spectrum cutoff is at higher energy, as it may come from synchrotron self-absorption~\cite{Wang:2007xj,Murase:2008mr}, the disintegration at the highest energies will be dominated by the photo-meson regime. Let us choose one extreme example here, \ie, $\varepsilon'_{\gamma,\text{min}} = 0.1 \varepsilon'_{\gamma,\text{br}} \simeq 100 \, \mathrm{eV}$, such that photo-disintegration will cease about one order of magnitude beyond the break. 

A comparison between this case (thick curves) and our standard assumption (no such cutoff, thin curves) is shown in \figu{ephmin}. From the rate plot (upper left panel), it is clear that the ceasing disintegration rate leads to slightly higher maximal energies, which are re-covered in the upper right and lower left panels for the densities in the source and the ejected spectra, respectively. The effect of the maximal energy on the neutrino spectra (lower right panel) is small because the peak flux, coming from disintegration products (nucleons), is hardly affected.  Therefore we expect slight modifications of the UHECR fit in this case, whereas the neutrino flux predictions hardly change. Since the UHECR output is actually somewhat higher, the tension with the prompt neutrino flux, which scales with the baryonic loading, will be somewhat released.


\section{Optically Thick Proton-Neutron System}
\label{app:optthick}

\begin{figure*}[tp]
\includegraphics[width=0.49\textwidth]{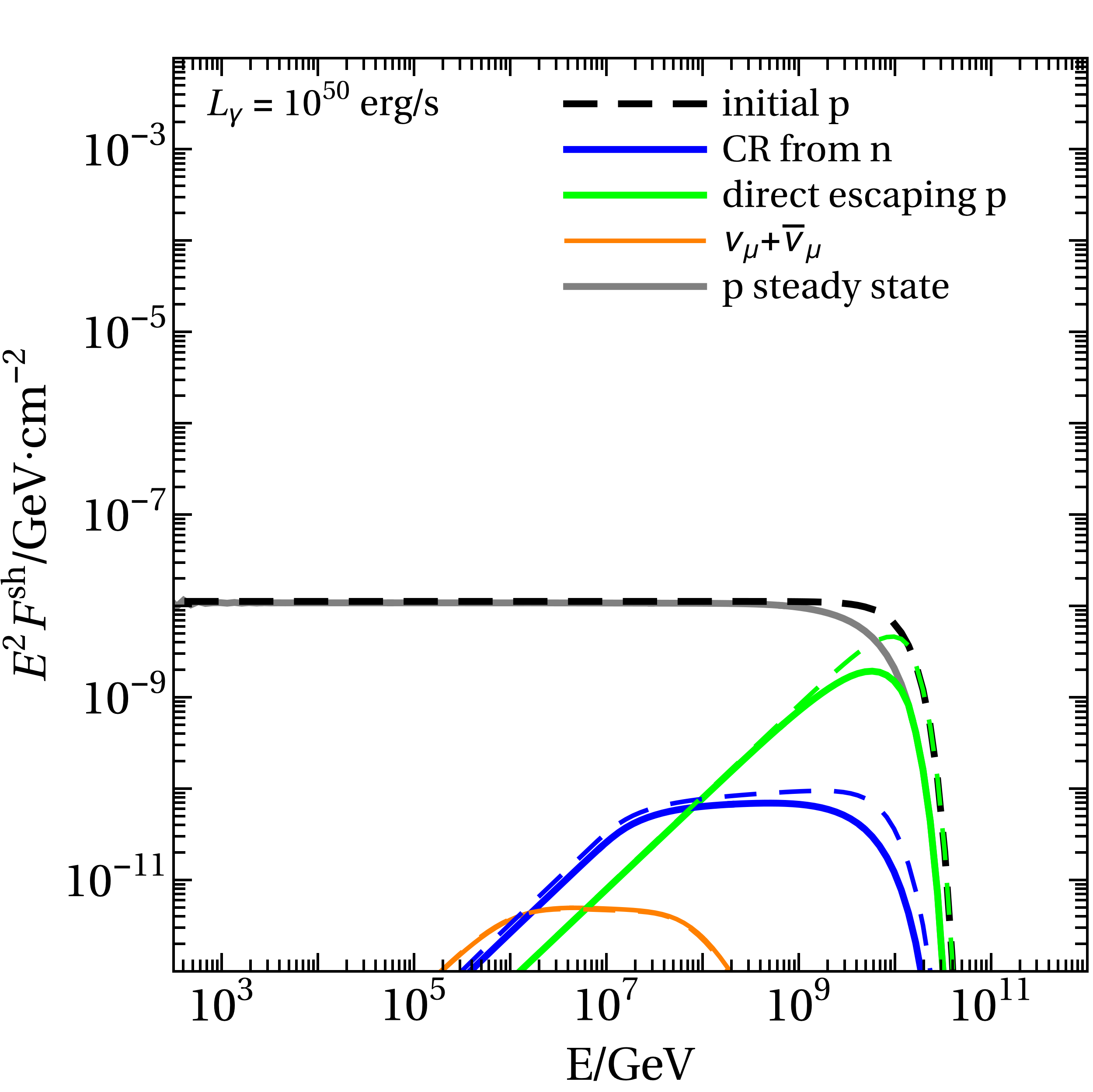}
\includegraphics[width=0.49\textwidth]{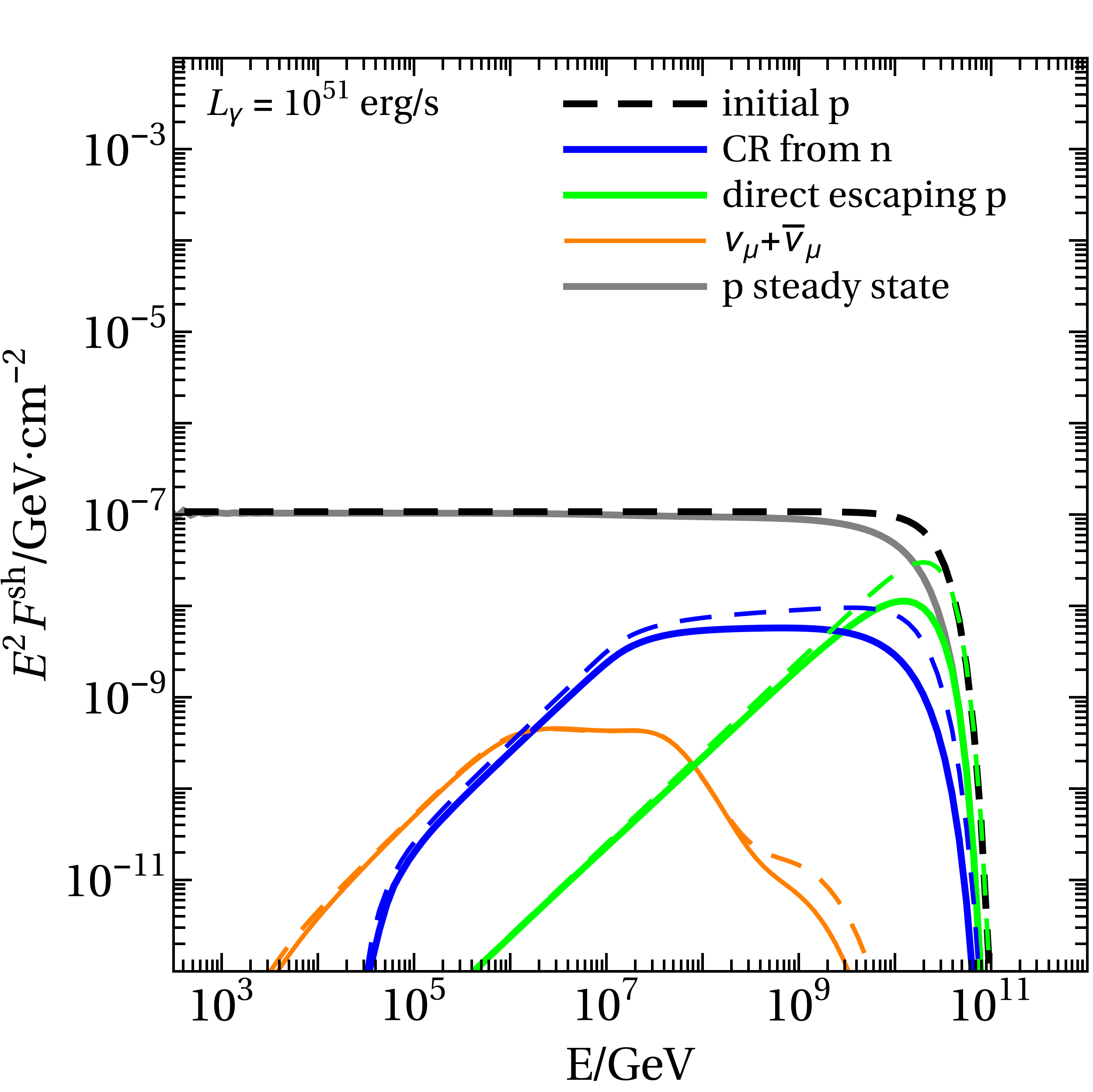}
\includegraphics[width=0.49\textwidth]{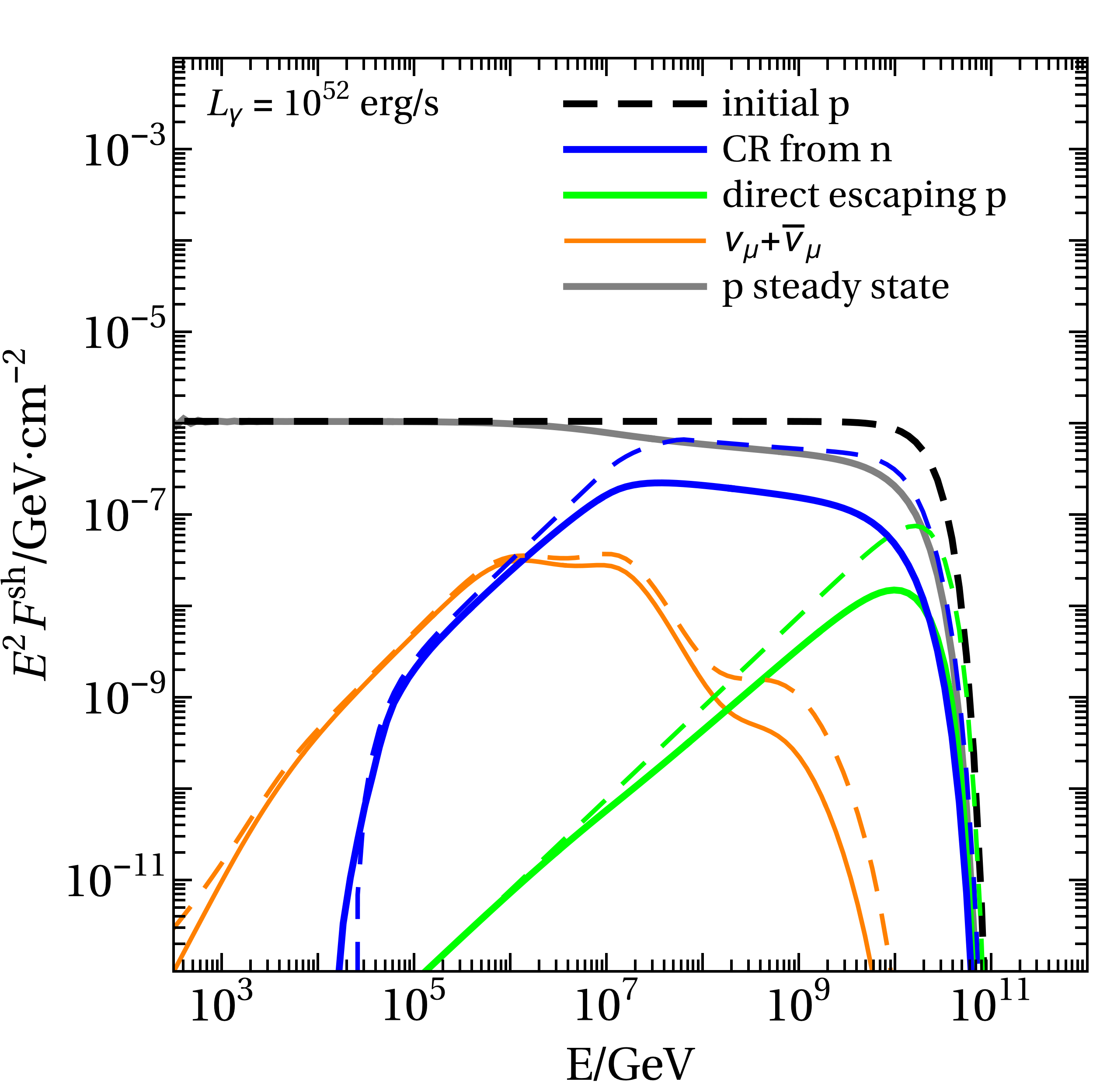}
\includegraphics[width=0.49\textwidth]{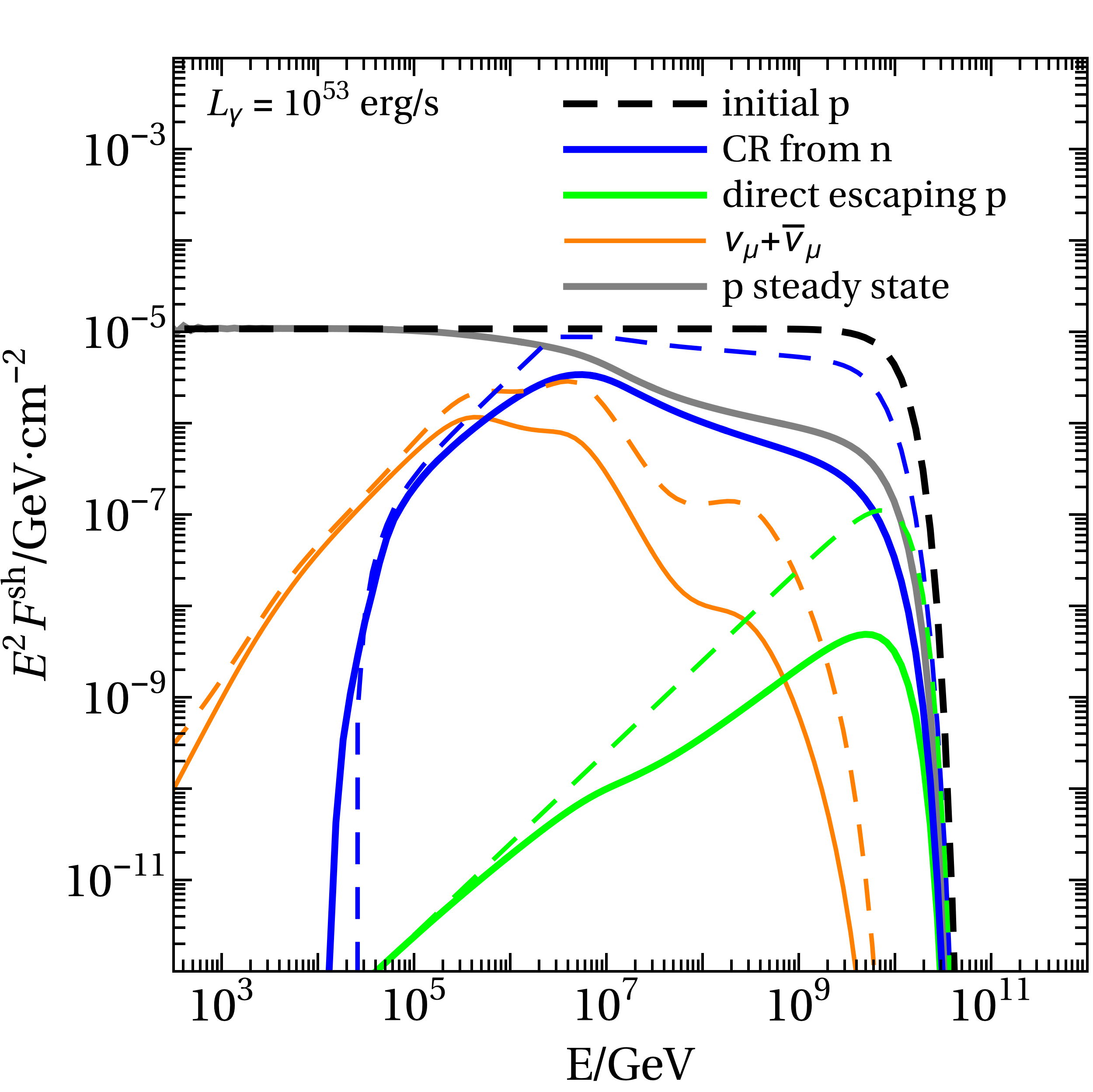}
\caption{Comparison of optically thin (upper row) and thick (lower row) proton-neutron systems for pure proton injection. Ejected particle fluences per shell are shown as a function of the energy in the observer's frame for different luminosities. The dashed curves refer to the model used in \Ref~\cite{Baerwald:2013pu}, while the solid lines refer to the model used in this work. See main text for details. 
}
\label{fig:optthickness}
\end{figure*}

In \figu{optthickness} we discuss the impact of optical thickness to photohadronic interactions in a coupled proton-neutron system (with proton injection only). 

An effective treatment of the optically thick case was performed in \Ref~\cite{Baerwald:2013pu}, where the proton spectrum $N'_p$ was assumed to be a power law with a cutoff at the maximal energy given by the dominant radiation process; see thick black curves in \figu{optthickness}. The neutron injection $Q'_{n, \mathrm{inj}}$  was then computed using the proton and photon densities. Since the neutrons are stopped by photohadronic interactions over the respective mean free path,  it was assumed that only neutrons produced at the edges of the shells (within the mean free path of the photohadronic interactions) can escape; see dashed blue curves. This is numerically equivalent to computing the steady state for the neutrons from $N'_n \simeq Q'_{n, \mathrm{inj}} \, t'_{p \gamma}$, assuming that it is dominated by the photohadronic interaction rate at the peak of the spectrum. Then one uses  $Q'_{n,\mathrm{esc}}=N'_n \cdot t'^{-1}_{\mathrm{dyn}}$, \ie, neutrons from the whole region can escape over the free-streaming timescale -- which we also do in this work. In that case, one has $Q'_{n,\mathrm{esc}}=Q'_{n,\mathrm{inj}} \,  t'^{-1}_{\mathrm{dyn}}/t'^{-1}_{\mathrm{p \gamma}}$, which reproduces the assumption in \Ref~\cite{Baerwald:2013pu}. This approach however neglects that the proton (steady) spectrum itself deforms in the case of high optical thicknesses, and that there is substantial feeddown of particles to lower energies by multiple interactions. 

In this study, we treat the proton-neutron system as time-dependent fully coupled partial differential equation system, computing the steady state explicitly, as illustrated in \figu{pde} (left panel). The  photohadronic energy losses are treated discretly, \ie, the protons and neutrons escape with the photohadronic rate, and are partially re-injected at lower energies and the other species. In the optically thin (to photohadronic interactions) regime, this method yields results similar to earlier works.  Compare the solid and dashed curves in the upper panels of \figu{optthickness}, which shows a comparison between the escape spectra computed with the method in \Ref~\cite{Baerwald:2013pu} (dashed curves) and this study (thick solid curves). In these cases, the optical thickness to photohadronic interactions is lower than one, which means that there is hardly any feedback from the neutrons to the protons, and the proton spectrum is hardly affected by photohadronic interactions. There are small differences, which come from the slightly different effective maximal energy (see curves ``p steady state'', showing $N'_p/t'_{\mathrm{dyn}}$, compared to ``initial proton'' curves). The reason is that we add an adiabatic cooling term for the protons in this study, which produces exactly the same result as an equivalent escape term for a power law spectrum (\cf, initial proton and steady proton spectra), but modifies the shape of the cutoff (see \equ{master}: a cooling term is sensitive to the derivative of $N'_p$ at the cutoff). 

There are however differences in the optically thick regime, see \figu{optthickness} for two high luminosity cases (lower panels). One important difference is that we normalize the injection luminosity with the baryonic loading  (see \equ{Anorm}), as compared to the steady state density in \Ref~\cite{Baerwald:2013pu} (which needs to be powered by a correspondingly higher injection luminosity). The proton steady state is then computed explicitly, and, especially at the highest energies, suppressed by the photohadronic interaction rate. For $L=10^{52} \, \mathrm{erg/s}$, the difference between the old and new methods is moderate because the optical thickness is moderate (around three). The new method implies a slightly lower normalization (because of the injection luminosity normalization) and a slightly stronger depletion of the neutron escape spectrum at higher energies (because of multiple interactions). 

For extremely large luminosities $L=10^{53} \, \mathrm{erg/s}$ (optical thickness: 36), the protons (and neutrons) from multiple interactions in fact pile up close to the pion production threshold, whereas the escape spectra at the highest energies are heavily suppressed -- which leads to softer escape spectra. The neutron density $N'_n$ follows the proton density $N'_p$, because the proton-neutron system becomes fully coupled by forward and backward reactions in the case of high optical thicknesses. This means that there is an impact on the ejected cosmic ray flux shape in the optically thick regime, whereas one can easily see from the figure that the neutrino flux is less affected. In that case, the spectral neutrino peak comes from the photo-pion production close to the threshold, where the model-dependent impact is smaller. While the change of normalization reduces the flux somewhat, the additional production channel from neutron interactions, which we take into account here, partially compensates for that.


\end{document}